# Recent progress in voltage control of magnetism: Materials, mechanisms, and performance

Cheng Song,[a,b,*] Bin Cui,[a,†] Fan Li,[a,b] Xiangjun Zhou,[a] Feng Pan[a,b]

[a] Key Laboratory of Advanced Materials (MOE), School of Materials Science and Engineering, Tsinghua University, Beijing 100084, China.

[b] Beijing Innovation Center for Future Chip, Tsinghua University, Beijing 100084, China.

## Abstract

Voltage control of magnetism (VCM) is attracting increasing interest and exciting significant research activity driven by its profound physics and enormous potential for application. This review article aims to provide a comprehensive review of recent progress in VCM in different thin films. We first present a brief summary of the modulation of magnetism by electric fields and describe its discovery, development, classification, mechanism, and potential applications. In the second part, we focus on the classification of VCM from the viewpoint of materials, where both the magnetic medium and dielectric gating materials, and their influences on magnetic modulation efficiency are systematically described. In the third part, the nature of VCM is discussed in detail, including the conventional mechanisms of charge, strain, and exchange coupling at the interfaces of heterostructures, as well as the emergent models of orbital reconstruction and electrochemical effect. The fourth part mainly illustrates the typical performance characteristics of VCM, and discusses, in particular,

[*] Corresponding author. Tel: +86-10-62781275; fax: +86-10-62771160
E-mail address: songcheng@mail.tsinghua.edu.cn

[†] Present at Max-Planck Institute of Microstructure Physics, Weinberg 2, D-06120 Halle, Germany

its promising application for reducing power consumption and realizing high-density memory in several device configurations. The present review concludes with a discussion of the challenges and future prospects of VCM, which will inspire more in-depth research and advance the practical applications of this field.

*Abbreviations:*

2-DEG two dimensional electron gas; AAIM, 1,3-diallylimidazolium; ABIM, 1-allyl-3-butylimidazolium; AEIM, 1-al-lyl-3-ethylimidazolium; AFM, antiferromagnetic; AHE, anomalous Hall effect; ALD, atomic layer deposition; $\alpha$, magnetoelectric coupling coefficient; BFO, $BiFeO_3$; BTO, $BaTiO_3$; CFO, $CoFe_2O_4$; DEME, N,N-diethyl-N-(2-methoxyethyl)- N-methylammonium; DOS, densities of states; EDL, electric double layer; EELS, electron energy loss spectroscopy; EMIM, 1-ethyl-3-methylimidazolium; FE, ferroelectric; FE-FET, ferroelectric field effect transistor; FET, field-effect transistor; FM, ferromagnetic; GMR, giant magnetoresistance; HAADF, high angle annular dark field; HM, heavy metal; $H_C$, coercive field; $H_{EB}$, exchange bias field; IL, ionic liquid; $\kappa$, dielectric constant or permittivity; $K_U$, magnetocrystalline anisotropy; LAO, $LaAlO_3$; LCMO, $La_{1-x}Ca_xMnO_3$; LSMO, $La_{1-x}Sr_xMnO_3$; $\lambda$, screening length; MA, magnetic anisotropy; MAE, magnetocrystalline anisotropy; MFM, magnetic force microscopy; MFTJs, multiferroic tunnel junctions; MOKE, magneto-optic Kerr effect; MPPR, N-methyl-N-propylpiperidinium; MR, magnetoresistance; MRAMs, magnetic random access memories; MTJ, magnetic tunnel junction; $\boldsymbol{m}$, magnetic moment; $M_S$, saturated magnetization; NFO, $NiFe_2O_4$; PCMO, $Pr_{1-x}Ca_xMnO_3$; PEEM, photoemission electron microscopy; PEO, polyethylene oxide; PLD, pulsed laser deposition; PMA, perpendicular magnetic anisotropy; PMN-PT, $PbMg_{1/3}Nb_{2/3})O_3$-$PbTiO_3$; PTO, $PbTiO_3$; PZT, $PbZr_{1-x}Ti_x)O_3$; $P$, polarization; $P_{down}$, polarization downward; $P_{up}$, polarization upward; $\Phi$, potential barrier; $R_{Hall}$, Hall resistance; SHE, spin Hall effect; SOT, spin–orbit torque; SQUID, superconducting quantum interference device; SRO, $SrRuO_3$; STEM, scanning transmission electron microscopy; STO, $SrTiO_3$; STT-MRAM, spin transfer torque magnetic random access memory; $\boldsymbol{\sigma}$, spin polarization direction; TER, tunnel electroresistance; TFSI, bis-(trifluoromethylsulfonyl)imide); TMPA, N,N,N-trimethyl-N-propylammonium; TMR, tunnel magnetoresistance; $T$, temperature; $T_C$, Curie temperature; $\boldsymbol{\tau}_{DL}$, damping-like torque; $\boldsymbol{\tau}_{FL}$, field-like torque; VCM, voltage control of magnetism; VCMA, voltage control of magnetic anisotropy; $V_G$, gate voltage; $V_O$, oxygen vacancies; XANES, x-ray absorption near edge structure; XAS, x-ray absorption spectroscopy; XLD, x-ray linear dichroism; XMCD, x-ray magnetic circular dichroism; YIG, $Y_3Fe_5O_{12}$; YMO, $YMnO_3$;

## Contents

## 1. Introduction

The control of magnetism and spin phenomena, which corresponds to switching between the basic "0" and "1" signals in information technology, has been intensely pursued during the past few decades [1–4]. It has generally been accepted that a magnetic field is the only means to switch magnetization and to maintain unchanged the magnetic behaviors of magnetic materials once they have been prepared [4]. The use of cumbersome magnets or coils occupies a large amount of space and entails serious energy consumption, especially when taking the remarkable trend in miniaturization of magneto-electronics into account [4–6]. Thus, there is a pressing need to employ nonmagnetic means to switch and modulate magnetism. Compared with other nonmagnetic routes, such as strain, doping, current, and light, etc., a voltage has been proven to be able to manipulate magnetism with a combination of advantages, including low power dissipation, reversibility, nonvolatility, high speed, and good compatibility with the conventional semiconductor industry [7–13].

In fact, Maxwell’s equations first reveal that the two independent phenomena of magnetic interaction and electric charge motion are intrinsically coupled to each other [14]. The thought of utilizing electric fields to control magnetism could date back to the 1960s [15]. In 2000, Ohno *et al*. [16] demonstrated the tuning of saturated magnetization and Curie temperature by an electric field in a diluted magnetic semiconductor (In,Mn)As. A large number of experimental works and theoretical investigations on the modulation and switching of magnetism have emerged in recent years, driven both by an urge to understand the mechanisms involved and by the demand for better performance, which is shown by the increasing number of publications every year as depicted in Fig. 1. The research regarding the interaction between charge and spin has developed into the exciting area of voltage control of magnetism (VCM). This field shows great potential to impact magnetic data storage, spintronics, and high-frequency magnetic devices, especially taking into account recent initiatives in internet of things, big data, artificial intelligence, and cloud computing.

Inspired by studies on magnetic semiconductors, researchers began investigating the voltage control of magnetic metals. Although it was previously thought that is was difficult to realize a large enough electric field effect in metals due to the short screening length [17], the manipulation of magnetism with an electric field effect was observed in ultrathin FePt and FePd owing to the large electric field-induced modulation of surface charges through ionic liquid gating [18]. Relevant investigations were soon expanded to ferromagnetic metal/ferroelectric oxide bilayers, such as Ni/BTO and CoFe/BFO, as well as Ta/CoFeB/MgO sandwiches [19–21]. The latter are the fundamental heterostructures for the current read–heads in high-density hard disks and the emerging magnetic random access memory (MRAM) technologies [3]. Since ultrathin metals are commonly used for electrical modulation, the interface of ferromagnetic metal/dielectric gating profoundly affects the efficiency of VCM. Thus the role of the interface in the VCM was extensively discussed in previous publications [22–24].

On the other hand, complex oxides have drawn increasing attention in recent years, because the strong couplings between lattice, charge, spin, and orbital degrees of freedom provide a broad playground for a variety of exotic properties tuned by various electrical methods including VCM [25]. Thanks to the high permittivity ($\kappa$) and the low carrier density ($n$), oxide thin films often display a much larger screening length ($t_S = (\kappa\hbar^2/4me^2)^{1/2}(1/n)^{1/6}$, $\hbar$ is the Planck constant, $m$ and $e$ are the electron mass and charge, respectively), guaranteeing a strong electric field effect for the manipulation of magnetic properties. The typical examples of VCM are found in multiferroic-based oxide bilayers, such as LSMO/PMN-PT, PZT/LSMO and LSMO/BFO [26–28]. These systems were commonly involved in inverse magnetoelectric coupling [29,30], a revival in multiferroic materials from the 1980s. That is, changing the magnetic properties ($\Delta M$) by applying an external electric field ($\Delta E$), $\Delta M = \alpha\Delta E$, where $\alpha$ is magnetoelectric coupling coefficient [30]. It is worth pointing out that inverse magnetoelectric coupling could be considered a typical VCM behavior to some extent when ferromagnetic/ferroelectric bilayer films are used. In recent years, the magnetic properties of ferromagnetic films, including saturated magnetization ($M_S$), coercive field ($H_C$), magnetic anisotropy (MA), and Curie temperature ($T_C$), etc., have been effectively manipulated by electrical means [24,31,32].

In terms of the mechanisms embedded in VCM, there are various origins. First,

the modification of charge carrier densities is thought to play a crucial role in ferromagnetic semiconductors and oxides [33,34]. An induced or spontaneous electric polarization at the interface of the dielectric/magnetic layers results in carrier accumulation or depletion, leading to a variation in the magnetic exchange interaction and anisotropy. In magnetic metals, the electric field manipulates the carrier density and the resultant Fermi level position at the interface, which determines the magnetic anisotropy [24,35]. The progress in ferromagnetic/ferroelectric heterostructures has brought about significant innovation in the mechanisms for VCM, where the strain effect is commonly involved. As known as the inverse piezoelectric effect, an external electric field would change the lattice of ferroelectric crystals when the ferroelectric switching occurs. The strain induced by the lattice variation is then transferred to the adjacent magnetic layer, causing modulations of magnetic properties through magnetostriction [36]. Once the ferroelectric layer is replaced by a multiferroic film (e.g., BFO) with a combination of ferroelectric and antiferromagnetic features, the exchange coupling mechanism provides another route for the electrical modulation of the spin arrangement in the ferromagnetic layer [37,38].

The textbook-style mechanisms referring to charge, lattice, and spin degrees of freedom mentioned above have been widely summarized in previous reviews [30,31,39,40]. Interestingly, recent advances in the characterizations of cutting-edge electronic structures and in dielectric materials indicate that two emergent mechanisms, orbital reconstruction and the electrochemical effect, are responsible for VCM in some systems. For example, the interfacial modulation of Mn orbital occupancy in LSMO/BTO heterostructures alters the $T_C$ and magnetoresistance of LSMO [41]. The situation turns out to be dramatically different when the inert oxide gate is replaced by an ionic conductor (e.g., $GdO_x$) with high $O^{2-}$ mobility or an electrolyte; the ion motion and the subsequent reversible redox in the magnetic medium driven by a gate voltage leads to a variation of magnetic anisotropy and saturated magnetization [42,43].

More than one and a half decades have already passed since the first experimental report on VCM in (In,Mn)As. The focus is gradually shifting not only toward the excavation of in-depth physical mechanisms, but also toward the improvement of application-oriented performance. One of the most promising examples of nonvolatile memory, spin-transfer torque magnetic random access memory (STT-MRAM) is at the heart of information storage and spintronic devices [6,10]. The present STT-MRAM chips with the storage volume of 64 Mb have shown extensive applications, including aerospace industry and automobile industry. Very recently, Everspin began planning production of 256 Mb and 1 Gb products [44], and SK Hynix and Toshiba also announced 4 Gb STT-MRAM in 2016 International Electron Devices Meeting [45]. Such a scale of storage volume is comparable to that of DRAM (dynamic random access memory), and even higher than that of SRAM (static random access memory). These achievements promote this promising memory design to the multi-billion dollar market for persistent memory in the field of storage devices and servers, which could be used in the internet of things, big data, artificial intelligence, and cloud computing, etc. Nevertheless, to realize magnetization switching by spin–transfer torque (STT), a high current density ($10^6$ A/cm$^2$) passing through each magnetic tunnel junction is persistently needed [6]. As it is generally accepted that the threshold current of semiconductors is at $10^5$ A/cm$^2$, the operation of magnetization switching in MRAM with a lower density current is particularly urgent. It is considered to be fundamentally transformative once an external voltage can assist or achieve magnetization switching. This interesting phenomenon was observed in CoFeB/MgO/CoFeB junctions, where the magnetization behaviors and tunneling magnetoresistance were manipulated by voltage pulses associated with much smaller current densities of $10^4$ A/cm$^2$ [46], and even a low switching energy of ~6 fJ/bit [47,48]. In this scenario, voltage-controlled magnetic anisotropy (VCMA) coefficient is employed to evaluate the energy efficient writing [49,50]. Moreover, the multiferroic tunnel junction with the ferromagnetic and ferroelectric orders exhibits four distinct resistance states with a combination of electric and magnetic fields, which provides a promising way toward the realization of high-density memory.

On the other hand, the use of pure voltage to realize deterministic nonvolatile 180 $^o$ magnetization switching is highly sought after, not only driven by an understanding of fundamental physics, but also motivated by an eagerness for information storage with low power consumption. To replace the magnetic field with an electric field in a magnetization reversal, the electric field-controlled exchange coupling and interactions between the ferroelectric–ferromagnetic domains are utilized in CoFe/BFO [51] and CoFeB/IrMn/PMN-PT [52]. Note that spin–orbit torque (SOT) provides an alternative way to achieve current-induced magnetization switching through a public line. A three-layer structure consisting of heavy metal/ferromagnetic metal/oxide is conventional in the study of SOT, where the spin Hall effect from heavy metals (such as Pt and Ta) and/or the Rashba effect from the inversion asymmetry of the interfaces contribute to the spin–orbit torque and the effective switching field [53–55]. Recently, the control of SOT in both metallic heterostructures and topological insulators was reported with the purpose of decreasing the critical switching current, which is an extension of VCM [56–58].

Although large amount of significant works have been done in the area of VCM, it is still a developing and energetic research topic with a lot of open questions in the performances, fundamental mechanisms, and practical applications of VCM. Firstly, the magnetic medium, dielectric gating, and their interface profoundly affect VCM with different behaviors and efficiencies. Note that the type of dielectric gating not only affects the manipulation efficiency, but also determines the intrinsic features of electrical modulation [59–62]. Even more ambiguously, opposite modulation tendencies are commonly observed in very analogical ferromagnetic metal systems. For instance, it is reported that positive gate voltage enhances the perpendicular magnetic anisotropy [22,24], whereas it does the opposite in other works [23]. Thus, there is a pressing need to sort out electrical modulation at the category level to address the question of how the magnetic medium and dielectric gating affect VCM. Secondly, the mechanisms responsible for magnetization modulation by electric fields are hard to separate completely in experiments [32,63]. For example, the carrier density modulation and strain effect are commonly mixed together in ferromagnetic/ferroelectric bilayers, because the ferroelectric polarization always causes both charge accumulation/depletion and a simultaneous strain effect [64,65]. Moreover, from the perspective of fundamental research, further development of VCM relies heavily on a thorough elucidation of the underlying mechanism, which has provoked studies for several novel mechanisms in brand new systems during the last few years. Thirdly, VCM is proposed to establish the next generation nonvolatile memory technologies with ultralow power consumption, high density, high speed, and a good compatibility with the semiconductor industry. To switch the magnetization or reduce the current density in current-induced magnetization reversal by external voltage are promising ways for the further reducing of the power consumption in spintronic devices.

Our review aims to supply a comprehensive summary of the burgeoning field of VCM and illuminate its distinct mechanisms. After this brief introduction in Section 1, we will illustrate the materials for the magnetic medium and dielectric gating, as well as various properties manipulated by the electric field in Section 2. Considering that the three conventional mechanisms cannot satisfy the rapid development of VCM in different novel systems, the latest mechanisms are added and discussed in Section 3. Subsequently, in Section 4 we discuss the recent focus on the promising application of VCM in spintronics. Finally, the challenges and future prospects for voltage control of magnetism are presented. It should be clarified that it is an almost impossible mission to depict all the studies carried out worldwide due to the fast pace of research in this area, and therefore some publications will inevitably be missed.

## 2. Film materials for voltage control of magnetism

Materials used for voltage control of magnetism can be classified into magnetic and dielectric materials according to their roles. An electric field forms via the dielectric materials under an applied external voltage and, subsequently, the performance of the magnetic materials is modulated under the electric field. To combine these two materials together for VCM, four types of device configurations are generally adopted as shown in Fig. 2: FET (field effect transistor) type, where the dielectric layer lays on the magnetic layer; BG (back gating) type, where the dielectric layer lays beneath the magnetic layer as a back grid; MTJ (magnetic tunnel junction) type, where the dielectric layer is sandwiched by two magnetic layers as a barrier); and, the Nano (nanostructure) type, where dielectric and magnetic materials are grouped together in the form of a nanostructure. FET-type VCM devices are developed from traditional FETs in the semiconductor industry, which usually utilize electric field-induced carrier density variation to manipulate the magnetic properties in semiconductor, oxides, and ultrathin ferromagnetic metal. In BG-type devices, piezoelectric materials or multiferroic materials, e.g. PMN-PT or BFO, are generally used as the bottom grid to control the magnetic metals and oxides through strain transfer or exchange coupling. In the MTJ-type devices, it is convenient to manipulate the magnitude and even a sign of magnetoresistance via the external voltage. In contrast, nano-type devices are based on some ferromagnetic-ferroelectric oxides systems, which can form self-assembled or artificial nanostructures, demonstrating a remarkable VCM due to the enhanced magnetoelectric coupling.

Note that magnetic metals dominate in the information storage devices, both in hard disk and the emerging MRAM. Since VCM was first realized in diluted magnetic semiconductors, taking advantage of the electric field effect on magnetism-related carrier density, magnetic materials have been extended from initial semiconductors to metals and then oxides, associated with a deeper insight and an extended understanding of the underlying mechanisms. At the same time, with the increased demand for a larger electric field effect, many efforts have been made to replace the traditional normal dielectric materials with either ferroelectric materials with a spontaneous polarization or electrolytes with an electric double layer to enhance the effect. As the materials vary, the mechanism underlying the VCM differs, involving five distinct embedded mechanisms: i) modulation of charge carrier densities (Charge), ii) strain effect (Strain), iii) exchange coupling or exchange bias (EB), iv) orbital reconstruction (Orbital), and v) electrochemical redox (Electrochemistry). In consequence, we mainly focus on the character of magnetic materials and dielectric materials in this section, accompanied by a brief introduction of the corresponding mechanisms in some comprehensive tables, which can help in the selection and design for VCM.

## 2.1 Magnetic medium

Almost all magnetic materials have the potential to be controlled by electric fields. For example, macroscopic magnetic properties of ferromagnetic films, including MA, $H_C$, $M_S$, exchange bias field ($H_{EB}$), $T_C$, and magnetoresistance (MR) have been effectively manipulated by electrical means, as shown in Fig. 3. In addition, domain wall movement under an external magnetic field can be manipulated by voltages as well. According to the conductive character and chemical composition, magnetic materials can be classified into magnetic metals, semiconductors, and oxides. Magnetic metals, which have a wide application in the information technology industry, show the advantages of strong magnetization, high $T_C$, simple fabrication, low cost, and possible perpendicular magnetic anisotropy (PMA) [1,2]. Nevertheless, the short screening length in metals due to the high conductivity seriously limits the VCM [24,66]. Compared with metals, semiconductors show a remarkable superiority in large screening length, but an inferior one in small magnetic moments and low $T_C$ [2,67], while oxides exhibit high thermostability and abundant physical properties [25,31,68], which would supply more opportunities for voltage control.

### 2.1.1 Magnetic metals

Taking advantage of the superiority of robust magnetization, high $T_C$, and possible PMA, many attempts have been made in magnetic metals to achieve VCM experimentally and theoretically, involving magnetic metals or alloys such as Fe, Co, Ni, Co/Ni, Co/Pd, Fe-Ga, FePt, FePd, CoFe, CoPd, NiFe, and CoFeB. The robust magnetization and high $T_C$ guarantee the performance at room temperature while the properties of PMA show great application potential in the information technology industry. Although the short screening length due to the high conductivity of metals limits the electric field effect, a breakthrough was achieved via enhancement of the electric field with the introduction of ferroelectric (FE) materials and electrolytes as the dielectric materials. At the same time, an effective control of magnetism by external voltages has been observed recently in some AFM metals (e.g., IrMn and FeMn) and metals with an FM–AFM transition (e.g., FeRh), which enriches the metallic system for VCM [69–71]. To summarize this research in magnetic metals, the device structure, magnetoelectric coupling coefficient ($\alpha$), temperature ($T$), coupling mechanism, and corresponding results of voltage control are listed in Table 1.

Achieving an effective control of magnetism in metals via the electric field is mainly realized through three device types, that is, FET-type, BG-type, and MTJ-type devices. The FET type with the dielectric layer laid above the magnetic layer and the BG type with the dielectric layer embedded under the magnetic layer are commonly used in designs to attain a large electric field via the electrolyte and ferroelectric layers, respectively. At the same time, the MTJ type with a current-perpendicular-to-plane structure (the current is perpendicular to the plane) is often considered in topics on VCM as well, which show an exclusive potential for high-density storage.

In FET-type devices with the dielectric layer on the surface, a large electric field can be easily applied on the magnetic layer with the help of an electrolyte (particularly an ionic liquid). Thus, through many attempts, various performance behaviors of magnetic metals have been successfully modulated by electric fields, with the breakthrough in the short screening length limitation. By immersing the device in an electrolyte, magnetocrystalline anisotropy and coercivity in ultrathin FePt and FePd of 2 nm is modulated by an electric field [18]. With the gate voltage applied through an ionic liquid (IL), voltage control of FM metals was observed in similar structures with MgO/Co/Pt [61], $HfO_2$/FeMn/Co/Pt/$[Co/Pt]_4$ [69], $HfO_2$/IrMn/Co/Pt/$[Co/Pt]_4$ [70], $HfO_2$/Ni/Co/Pt [22], and $HfO_2$/$[Co/Ni]_n$/Pt [62] heterostructures with PMA, where the magnetic properties were modified, including magnetic moments, $T_C$, coercivity, magnetic anisotropy, and exchange bias. In addition, an electric field could also be directly applied on the oxide-capping layer to gain an effective manipulation of the magnetism. For example, magnetic anisotropy in MgO/Fe systems is controlled through spin–orbit interaction, where the hybridization of Fe 3*d* orbitals and O 2*p* orbitals or the Fermi surface can be modulated under the electric field [23,72], while in MgO/FeCo and $GdO_x$/Co, the magnetization, coercivity, magnetic anisotropy, and domain are modulated under the electric field, due to the change in the oxidation states of the interfacial magnetic metals [42,43,73,74].

In contrast to the FET structure, in BG-type devices, the dielectric layer is embedded under the magnetic layer, where FE or piezoelectric materials usually serve as the dielectric layer to provide an electric field, with the magnetic metal as a top electrode. Voltage control of magnetism in BG-type devices was first realized in CoPd/PZT, demonstrating a dependence of the polar Kerr rotation on the external electric field [75]. After that, the voltage control of magnetic anisotropy (VCMA) and moment was achieved in many works based on FM/FE heterostructures, e.g., Fe/BTO, Co/PMN-PT, Ni/BTO, and CoFeB/PMN-PT [19,59,76,77]. With the advances in characterization methods, the reversible switching of the magnetic domain in FeGa/BTO using a static (DC) electric field was directly observed by *in-situ* Lorentz microscopy [78], while the electrical modulation of domain behavior, that is, out-of-plane magnetization, was demonstrated in Ni/BTO by XMCD and PEEM [79]. Recently, research interests have tended toward magnetization switching by external voltages, which have been realized in the heterostructure of Co/PMN-PT and $CoFeB/AlO_x/CoFeB$/PMN-PT MTJ [59,80]. In addition, exchange bias can also be tuned by voltage with an AFM layer introduced into the heterostructure, CoFe/BFO/SRO/PMN-PT, CoFe/BFO, $[Co/Pt]/Cr_2O_3$, $[Co/Pd]/Cr_2O_3$, and $NiFe/YMnO_3$ [20,60,81–85], etc. Besides FM metals, VCM has been attained in an FeRh alloy with an FM–AFM transition near room temperature, showing a variation in transition temperature and giant electroresistance [71,86].

Since the two device types mentioned above are generally in a current-in-plane configuration, VCM realized in a current-perpendicular-to-plane MTJ-type device has received much attention for high-density storage applications. In the MTJ-type device, the MTJ part serves as a pseudocapacitor and most attempts have been focused on the switching of the magnetization by electric fields, which can be classified into two schemes, that is, changing the coercivity or magnetic anisotropy, respectively. Based on the first scheme, a magnetization reversal induced by the electric field in the CoFeB/MgO/CoFeB MTJ has been observed according to the junction resistance under a small assistant magnetic field, due to the reduced coercivity of CoFeB under the electric field [46]. On the other hand, taking advantage of the electric field effect on magnetic anisotropy, magnetization can be reversed by the voltage pulse in a Fe/MgO/FeCo MTJ, where the magnetic anisotropy and related switching of the FeCo layer are modulated by the external voltage while the magnetization of the Fe layer is stable under electrical stimulation [87,88]. Since the electric field pulse temporarily arranges the easy axis of the free layer in-plane, the magnetization of the free layer under an external fixed perpendicular magnetic field is switched. It is noteworthy that with the increasing thickness of the MgO barrier in the MTJ device, the switching energy of the electrical switching remarkably decreases [47,48]. It promises a decrease in energy consumption when combined with the magnetization switching induced by the current via the spin-transfer torque (STT), which shows tremendous potential for compact memory and integrated circuits with lower energy consumption [89].

To guarantee a fine magnetic metal layer for VCM, the magnetic metals or alloys are prepared by different methods, such as magnetron sputtering and electron beam deposition. Magnetron sputtering is a thin film deposition method that is widely used in the industry with advantages including high efficiency, yield, and low-cost production [90,91]. It can meet the deposition requirement of most magnetic metals and alloys. In contrast, electron beam deposition with a high background vacuum and the largest energy density of $10^9$ W/cm$^2$, demonstrates a superiority in the deposition of highly pure metals and binary oxide films [92]. It is noteworthy that with the advances in sputtering and electron beam deposition technology, preparation of magnetic metals with PMA has been realized, including a number of ultrathin FM metal alloys, such as $L1_0$-ordered (Co,Fe)-Pt alloys, Co/(Pd,Pt) or Co/Ni multilayers, and CoFeB/MgO [10,22,93,94]. The realization of VCM in most of these systems brings the promise of a bright future for a new generation high-density nonvolatile information storage and logic devices, where a fine thermostability and a low critical current are in demand for current-induced magnetization switching.

### 2.1.2 Magnetic semiconductors

A magnetic semiconductor is prepared by doping a transition-metal element (e.g., Mn and Co) into a nonmagnetic semiconductor (e.g., GaAs, InAs, $TiO_2$, and ZnO) [9,67], which behaves common semiconductor characteristics and magnetic properties at the same time. When compared with magnetic metals, FM semiconductors have a larger screening length, guaranteeing a high efficiency in the voltage control of magnetic layers. It is well known that FM semiconductors show a strong magnetism dependence on the hole concentration, as described by the *p*-*d* Zener model [95]. Thus, an effective manipulation of magnetism can be attained via the extraction and injection of hole carriers in the FET channel under positive and negative gate voltages, accompanied by the suppressed and enhanced $T_C$ or magnetic moment, respectively. FM semiconductors used for VCM are summarized in Table 2.

We focus first on (In,Mn)As and (Ga,Mn)As, which are the most well-known FM semiconductors. Voltage control of the magnetic phase transition in FM semiconductors was first observed in (In,Mn)As, as reflected by the modulation of an anomalous Hall effect (AHE) [16]. Subsequently, the $T_C$, magnetic moment, magnitude, and sign of the AHE coefficient of (Ga,Mn)As were changed by the electric field in a series of works [96–104]. Nevertheless, the low intrinsic $T_C$ somehow limits the application of (In,Mn)As and (Ga,Mn)As, which also inspired much research aimed at the improvement of $T_C$ in these systems through various means such as the proximity effect [105].

Due to the limitation of the low intrinsic $T_C$, studies on the influence of the electric field on magnetic properties have involved numerous magnetic semiconductors, for example, groups IV, II–VI, III–V, topological insulators, etc. [106–115]. After 2000, many studies in experiment and in theory revealed that the $T_C$ of some *p*-type FM wide-bandgap semiconductors, such as ZnO, $TiO_2$, and GaN, could be enhanced to higher than room temperature [9,95], which further inspired a new wave of research. Since ferromagnetism in diluted magnetic oxides like Co:ZnO originates from the defects in the system such as oxygen vacancies ($V_O$) [116], the AHE and magnetic phase transition of Co:ZnO and Co:$TiO_2$ can be manipulated by gate voltage applied through $SiO_x$ and IL, respectively [109,111]. Additionally, the saturation magnetization and coercivity field of $Zn_{0.95}Co_{0.05}O$ in the Pt/$Zn_{0.95}Co_{0.05}O$/Pt is reversibly controlled, taking advantage of the resistive switching under the electric field [110]. It is noteworthy that, although diluted magnetic oxides are oxides in chemical composition, we attribute them to FM semiconductors in light of their semiconductor conductive character.

In order to achieve VCM in semiconductor-based devices, ultrathin films with a high crystalline quality are usually needed. In general, narrow gap semiconductors, like (In,Mn)As, (Ga,Mn)As, and topological insulators, are usually prepared by molecular beam epitaxy, which is well known for an epitaxial growth of ultrathin films. On the other hand, diluted magnetic oxides with a wide gap, such as Co:ZnO, are commonly deposited via magnetron sputtering or pulsed laser deposition.

#### 2.1.3 Magnetic oxides

Magnetic oxides, displaying a multitude of new physics and potential applications, have drawn a great deal of interest in research since the investigation of the half-metal oxides, which are promising for the realization of the large tunnel magnetoresistance effect with high stability and multiple functions (e.g., multiferroics). Since magnetic oxides, especially perovskite oxides, have a good coherence with FE oxides such as BTO, PZT, and PMN-PT in the crystal structure, which guarantees the preparation of high-quality epitaxial FM/FE heterostructures, VCM in magnetic oxides is mainly achieved in the form of the FET, MTJ, and BG types with FE oxides serving as the dielectric layer [7,117]. In addition, spinel magnetic oxides ferrites can also be used as the magnetic layer. It is worth mentioning that multiferroic oxides, for example, BFO, exhibit both antiferromagnetic and ferroelectric properties at the same time, making it possible for multiferroic oxides to work as the antiferromagnetic and dielectric layer simultaneously except for the single role of the dielectric layer in FM/FE heterostructures. The research on VCM in magnetic oxides is summarized with the corresponding data in Table 3.

As a model for magnetic oxides, manganites lie at the heart of this research in light of the fact that their $T_C$ is close to room temperature, for example, 370 K for LSMO ($x$ = 0.33). Meanwhile manganites have a high spin polarization to 95% and rich magnetoresistance, for example, tremendous magnetoresistance, anisotropic magnetoresistance, and planer Hall magnetoresistance [68,118–120]. Among various manganites, most studies are focused on LSMO, LCMO, and PCMO for the realization of VCM. Because the origin of the magnetism is the double exchange in the $Mn^{3+}$-O-$Mn^{4+}$ chain, where the charge transfer calls for an identical magnetic structure in two Mn ions [117], the variation in the doping level and corresponding $Mn^{3+}/Mn^{4+}$ ratios can be used to modulate the double exchange, which promises an effective electric-field manipulation of magnetism via the charge mechanism. Taking LSMO as an example, a series of magnetic phases appears with the variation in the doping level [121,122], that is, LSMO changes from an insulating phase with seriously suppressed magnetism to a good ferromagnet and then an antiferromagnet, as the Sr doping varies from $x < 0.16$ to $x = 0.16$–$0.5$ and then $x > 0.5$ [8]. For example, in FET-type PZT/LSMO and BTO/LSMO heterostructures, $T_C$, magnetization, and magnetic anisotropy are modulated by the electric field based on the accumulation and deletion of carriers near the interface, which is controllable via FE polarization switching [27,34,65,123]. Similarly, in multiferroic tunnel junctions such as Co/PZT/LSMO and LSMO/BTO/LCMO/LSMO, the tunnel barrier height and the transition between different magnetic phases are manipulated via the variation in carrier density under the electric field, reflected in the four-state memory under a combination of magnetic and electric fields [124,125].

In addition, the electronic phase dependence of manganites on the strain provides an alternative for VCM in manganites, with the strain mechanism playing a dominant role [126]. For some FM manganites like LSMO, the in-plane tensile strain favors the in-plane orbital occupancy, resulting in the in-plane magnetic easy axis for even *A*-type antiferromagnets with oppositely aligned ferromagnetic planes of the {001}, while the compressive strain stabilizes an occupancy of the orbitals out of plane, inducing the perpendicular easy axis for even *C*-type antiferromagnets with oppositely aligned ferromagnetic planes of the {110} [127,128]. In this way, an electrically induced strain gained by the inverse piezoelectric effect-induced magnetostriction could be used to alter the strain-related magnetism in BG-type devices [64]; for example, in LSMO/PMN-PT, the magnetic moment follows the strain modulation in FE crystals under an external electric field [26,129].

Compared with perovskite manganites, spinel magnetic oxides ferrites, e.g., CFO, NFO, and $Fe_3O_4$ (or $Zn_{0.1}Fe_{2.9}O_4$), usually show a $T_C$ much higher than room temperature and a large resistivity, making them appropriate for various applications such as information storage, spintronics, and high-frequency chips [37,130]. Similar to the case of a perovskite heterostructure, the magnetization and magnetic anisotropy have been effectively controlled under the electric field in the various heterostructures of ferrimagnetic spinel/FE perovskite oxides, such as CFO/PMN-PT, NFO/PZT, $Fe_3O_4$/CFO/PZT, $Fe_3O_4$/PMN-PT, and $Zn_{0.1}Fe_{2.9}O_4$/PMN-PT [131–138]. Besides the heterostructure devices, ferrimagnetic spinel/FE perovskite oxide composites with self-assembled or artificial nanostructures (i.e. Nano-type) also attract significant attention as a unique system for VCM, where the ME coupling is thought to be relatively larger than that in the heterostructure devices [139,140]. In general, composite nanostructures could be classified into two groups: (i) a 0–3 structure that is usually designed with a magnetic nanostructure implanted in a piezoelectric matrix;

(ii) a 1–3 structure, such as single-layer self-assembled structures and some artificially constructed nano-rods embedded in a piezoelectric matrix [141]. Numerous investigations of FM spinel/FE perovskite systems have been reported with different combinations between the FE perovskites (BTO, BFO, PTO, and PZT) and the spinel phases (CFO and NFO) [130,142–145].

Multiferroic oxides, as a special type of magnetic materials, usually demonstrate an interaction between the antiferromagnetic and ferroelectric properties, promising a controllable magnetism under an electric field in a single layer instead of the commonly used FE/FM oxide multilayers. As one of the most well-known multiferroics, a large number of works have been carried out on BFO, to realize an effective control of magnetism involving the antiferromagnetic domain and the spin wave by electric field-induced ferroelectric switching [146–149]. Similarly, voltage control of the magnetic structure was demonstrated in multiferroic $HoMnO_3$ [150].

Generally, magnetoelectric composites are prepared by either the PLD or sol–gel method. The PLD assisted by reflection high-energy electron diffraction guarantees the epitaxial growth of the high-quality FM/FE oxide heterostructures with a precise control of the interface and thickness in the atomic-layer scale, constructing the cornerstone of voltage control of FE/FM heterostructures [117]. Compared with the delicate PLD method, sol-gel is more appropriate for the preparation of thicker film from tens of nanometers to tens of micrometers with a higher efficiency and a lower cost [141].

## 2.2 Dielectric gating medium

Dielectric materials in the devices for VCM act primarily as a medium for the application of an external voltage or electric field, calling for a high insulation of the materials. According to the working principles and states, dielectric materials can be divided into three classes, those are, normal dielectric and high-$\kappa$ materials, FE materials, and electrolytes. With the advantage of simple fabrication, the normal dielectric and high-$\kappa$ materials can produce a static electric field on the magnetic material via the separation of the center of the positive and negative charges under an external voltage, which is volatile in the absence of an electric field. In contrast, the FE materials stand out in terms of the high insulating behavior and the remnant polarization, which guarantees a nonvolatile manipulation effect. Meanwhile, the electrolytes (e.g., ionic liquids), as a novel dielectric material, can generate a rather large electric field by the movement of charged ions and resulting formation of electric double layer at the interface of dielectric/magnetic layers.  We summarize the performance characteristics of various dielectric materials in Table 4. It is also noteworthy that some VCM devices based on resistance switching could work without dielectric materials like the FM semiconductor Co:ZnO [110].

### 2.2.1 Normal dielectric and high-$\kappa$ materials

For the realization of VCM, a large-magnitude electric field is usually expected in light of the corresponding significant modulation of carrier density and magnetic properties. Because the change in carrier density under the gating effect in a capacitance structure is related to the value of $CV_G/e$, where $C$ is the capacitance per unit area, $V_G$ denotes the gate voltage, and $e$ stands for the electron charge, an apparent control of FM performance can be achieved by increasing the value of $C$ for the case with the same voltage. The capacitance of a normal dielectric material is $C = \kappa\varepsilon_0/d$, where $\kappa$ and $\varepsilon_0$ denote the relative permittivity and the vacuum permittivity, respectively, while $d$ marks the dielectric-layer thickness. [151]. Hence, it is necessary to increase the $\kappa$ and reduce the $d$ of the dielectric layer for obtaining a larger $C$ and a remarkable VCM consequently.

As the most common dielectric material among modern semiconductors, silicon dioxide ($SiO_2$) was used early on in VCM. Nevertheless, the relative dielectric constant (permittivity, $\kappa$) of $SiO_2$ is only 3.9 [152], which seriously limits the electrical manipulation effect. Hence, some high-$\kappa$ materials such as MgO, $Al_2O_3$, $HfO_2$, and $ZrO_2$ are used in FET-type devices as the dielectric layer between the electrode and FM materials to enlarge the electric field [152]. On the other hand, MgO and $Al_2O_3$ could also serve as the tunneling barrier in MTJ-type devices, thus providing a good opportunity for the electrical manipulation of practical TMR [94,153].

To avoid the deterioration of dielectricity due to a poor crystalline quality associated with a possible electric leakage, magnetron sputtering and atomic layer deposition (ALD) are commonly used to prepare the dielectric materials [23]. Magnetron sputtering favors a high-quality interface between the FM and dielectric materials, which is especially important for voltage control of the delicate interfacial magnetism, while ALD can offer dense dielectric layers, with the advantage of a high breakdown field and a small leakage current [154].

### 2.2.2 Ferroelectric crystals and films

Ferroelectric materials have been widely used in VCM, taking advantage of the spontaneous electric polarization, the inverse piezoelectric and magnetostriction effect, and the possible multiferroic property with the AFM moment, which just favors an effective VCM via carrier density modulation, strain effect, and exchange coupling, respectively.

Since opposite polarization states of the FE layer will accumulate or deplete the carriers in the FM layer near the interface of the heterostructure, the polarization reversal controlled by external voltages can be used to modulate the magnetic properties related to carrier density [34]. Compared with normal dielectric materials, FE materials (e.g., BTO, PZT, and PMN-PT) could dramatically enlarge the permittivity by two orders of magnitude [155], resulting in a much more apparent change in carrier density as shown in Table 4. It is noteworthy that the FE polarization shows a hysteresis loop similar to the FM magnetization when sweeping the electric field below FE $T_C$. Hence, a remanent polarization in the FE material after withdrawing the voltage promises a nonvolatile electrical manipulation in the FE/FM heterostructure [156].

At the same time, utilizing the inverse piezoelectric and magnetostriction effect in FE materials, which means the emergence of an electric field-induced lattice variation, strain-related magnetism in the FM materials above can be manipulated by electric fields, such as the magnetic moment, anisotropy and even magnetization switching [26,59,76]. Among the large quantities of present FE materials, PMN-PT relaxor ferroelectrics have been widely used for strain-mediated VCM due to their excellent ultrahigh strain and piezoelectric behavior. Since the generally used in-plane piezo strain along the [100] direction in (001) PMN-PT crystal shows volatile butterfly-like behavior lacking a remanent strain, i.e., the strain state disappears after the voltage is withdrawn, the voltage control of the magnetization is volatile with only one magnetization state at the zero field [26]. Nevertheless, with the combined action of the absence of MAE in CoFeB and the 109° FE domain reversal in PMN-PT, an electric field-controlled loop-like magnetization was observed, which shows the representative nonvolatile characteristic [77]. In addition, by altering the normal direction of PMN-PT from (001) to (011), the in-plane piezo strain can also exhibit a loop-like behavior, which guarantees nonvolatile electrical manipulation of magnetism [157]. A similar loop-like behavior was also observed in PZT [158,159].

In addition, some FE systems show multiferroic properties with two or more ferroic-order parameters. Some typical multiferroic materials like BFO and YMO are both FE and AFM [146,160], while $Cr_2O_3$ has a uniaxial AFM spin structure and linear magnetoelectric effect with $\alpha$(263 K) = 4.13 ps/m [161]. Considering that the electric field can alter the antiferromagnetic domain, inducing the variation in the magnitude and polarization of exchange bias field [20,28,60,81], VCM in such a multiferroic/FM heterostructure is achieved on the basis of the interfacial exchange interaction between the FM and AFM orders.

Considering the quality of the sensitive properties of FE oxides, a high crystalline quality is usually in demand to attain a good insulator and a fine FE polarization, especially when the thickness of the FE layer becomes much small. In addition, a good interface between the FM layer and the FE layer would enhance the effect of VCM. In this view, the thin films of FE oxides are usually prepared by PLD to guarantee good quality and performance. Sometimes, a melting method is also used when a FE oxide crystal is needed, such as PMN-PT, which is generally used as a single-crystal substrate.

### 2.2.3 Electrolytes

Electrolytes, known for the separation of cations and anions to opposite electrodes under an electrical voltage, have drawn a great deal of attention for applications in VCM, including ionic liquids with a distinctive electric double layer (EDL) and solid-state electrolytes with high oxygen mobility. The electric double layer enormously enhances the magnitude of the electric field at the interface, successfully breaking through the limit of the short screening length in many systems, while the high oxygen mobility associated with a large quantity of transferable oxygen ions favors an effective control of magnetism by electric fields via an electrochemical process.

An ionic liquid is a typical electrolyte material, which first shows various potential applications in electrical- or electrochemical-related devices such as capacitors and batteries, etc. [162–164]. Compared to conventional electrolyte solutions such as organic and aqueous solutions, whose concentration cannot remain constant due to the volatility when exposed to air, ionic liquids show superiority as dielectric materials for FET-type devices (Fig. 4a), for their high freezing point near room temperature. At the same time, the low molecular weight of ionic liquids makes them much more conductive than conventional electrolytes, while the high polarization ensures a large electric field for VCM. Furthermore, a high frequency characteristic has been demonstrated in the system of ionic liquids, where a fast operation speed on the order of up to MHz was observed [165,166], despite the well-known slow responses of the polymer-dielectric FETs owing to high resistivity and slow polarization relaxation [167].

For applications in VCM, the common cations in ionic liquids include DEME, EMIM, MPPR, AAIM, AEIM, ABIM, TMPA, $KClO_4$, and $CsClO_4$, while the common anions include TFSI, PEO, and $BF_4$. [168]. The physical properties of these materials relevant to VCM are also summarized in Table 4. During VCM, cations and anions of ionic liquids are driven separately to the gate electrode and channel by a gate voltage, as shown in Fig. 4b and c. Consequently, an electric double layer forms on the surface of the electrodes, where sheets of negative and positive charges couple with each other. The opposite charges come from the ions in the ionic liquid and the electrons or holes in the bottom magnetic layer. Compared with normal dielectric materials, the EDL behaves at very large $C$ of ~10 $\mu F/cm^2$ in that the distance between the two charged plates, as denoted by $d$, is close to the size of the ions ($C = \kappa\varepsilon_0/d$). Hence, a dramatically large variation in carrier density $\Delta n_S$ ($\Delta n_S = CV_G/e$) can be realized, which is at a level of $10^{15}$ $cm^{-2}$ [165,169,170]. These high-density carrier injections with the help of EDL are dramatically larger than those of the $SiO_2$ dielectric layer [171]. In comparison, considering the commonly used inorganic-dielectric FET where the $SiO_2$ gate dielectrics as thick as 300 nm behave as a capacitance around 10 $nF/cm^2$, a typical modulation of charge can be merely ~$10^{13}$ $cm^{-2}$ [168], while the FET using FE gate dielectrics can support a carrier density change of $10^{14}$ $cm^{-2}$ [172].

It is noteworthy that, along with the voltage control of two-dimensional electron gas, superconducting properties, and the metal–insulator transition via EDL, [173–177], voltage control of magnetism, such as coercivity, magnetic anisotropy, and magnetic phase transition by EDL, has been achieved in many FM metals and oxides [18,61,70,109,121,169,178]. Although it is well known that positive and negative gate voltage will inject and extract electrons from the system, the fundamental physical understanding of EDL charging in VCM is still under intense debate, because both electrostatic doping and redox (i.e., oxygen ions or vacancy migrations) have been found in VCM via EDL. Some preliminary studies suggest that the mechanism of EDL gating is sensitive to the operation temperature, frequency, oxygen concentration, magnitude of $V_G$, and humidity, etc. [62,178,179], which further complicates the understanding.

In comparison to ionic liquids, gadolinium oxide ($GdO_x$) is an alternative electrolyte, which can be used for VCM. $GdO_x$ is a kind of solid-state electrolyte with a high oxygen mobility, which can serve as a reservoir of oxygen ions under an electric field. Hence, with the positive and negative electric field in an FET device, the oxygen ions in $GdO_x$ can be driven toward or away from the interface between $GdO_x$ and the FM material, to control the interfacial oxidation of FM materials. In this way, the magnetism related to oxidation states can be modulated under the electric field, such as magnetization, magnetic anisotropy, coercivity, and the magnetic domain of PMA Co [42,43,74,180]. Due to the slow migration speed of ions in solids, external heat is usually needed to reduce the operational time of the electric field application [42,43].

To introduce electrolytes into VCM, electrolytes are usually combined in the form of an FET-type device. In general, the droplet of ionic liquid is placed directly on the magnetic oxides to obtain a significant manipulatory effect for oxygen ions or vacancies [121,169,178]. Nevertheless, in some situations, the insertion of an oxide layer between the FM metals and the ionic liquids is adopted to avoid an irreversible chemical reaction due to the direct contact [22,61,70]. Interestingly, in light of the high freezing temperature, a frozen ionic liquid is sometimes used to deliberately fix the ion positions and the corresponding electric field of the EDL at low temperature, promising a long-term electric field effect and a nonvolatile VCM. Note that, a gate voltage can only be applied effectively when the electrolyte is still in the liquid state and the electric field effect in the frozen electrolyte is rather difficult to be changed. In comparison, the deposition of the solid electrolyte $GdO_x$ used for VCM was usually achieved via reactive sputtering in previous works [180], where a metal Gd target was used under an argon/oxygen gas mixture atmosphere.

## 3. Mechanisms under voltage control of magnetism

The mechanism for voltage control of magnetism depends on the choice of magnetic and dielectric materials, the thicknesses of thin films, the crystal orientations, and the operational mode of the electric field. Consequently, there are many possible mechanisms primarily due to the diversity of factors mentioned above. In our review, the common mechanisms are divided into five types that is, carrier modulation, strain effect, exchange coupling, orbital reconstruction, and electrochemical effect. Among them, the first three mechanisms are textbook-style mechanisms referring to charge, lattice, and spin degrees of freedom, respectively, which are widely utilized to explain the classic VCM phenomena. Recent advances in cutting-edge characterizations of electronic structure and in dielectric materials indicate that two emergent mechanisms of orbital reconstruction and electrochemical effect are responsible for VCM in some systems. Various magnetic behaviors such as magnetic anisotropy, magnetization intensity, exchange bias, magnetoresistance, and Curie temperature can be manipulated based on these mechanisms as shown in Fig. 5. We will introduce the work in VCM based on the various mechanisms and cover it in detail and discuss the interactions and characterizations of these different mechanisms.

## 3.1 Carrier modulation

When the magnetic properties of the heterostructures are intimately linked to the carrier density, changes in the carrier doping level will significantly modulate their magnetic properties. Because the modulation of carrier density under the electric field is common and inescapable, magnetic metals, semiconductors, and oxides can be manipulated by the electric field through this charge mediated mechanism.

### 3.1.1 Modulation of carrier density in ultrathin metals

In ferromagnetic metallic systems, the change in magnetism is often related to the carrier density, that is, the density of the itinerant electrons in metals [18,61,151,181]. Since there exists a strong screening effect in metals, the influence of the electric field cannot deeply penetrate into the bulk of the metal. However, a large electric field can be used to modulate the carrier density and electron occupancy in ultrathin metal systems, which have a high surface-to-volume ratio, and therefore it can control the magnetism. With the assistance of liquid electrolytes, the large electric field induced by EDL shows an extensive modulation effect in ultrathin FePt and FePd, as displayed in Fig. 6a and b [18]. The surface charge is altered by the high voltage applied and the magnetic properties of the entire film are modified. The unpaired *d* electrons with an energy close to the Fermi level in 3*d* metals act as the free carriers that compose the modulated surface charge, and they also dominate the essential magnetic properties. The change in carrier density directly affects the magnetocrystalline anisotropy ($K_U$) and the corresponding $K_U$ energy (MAE) via the variation in the number of 3*d* electrons [23,182]. Taking into account the different MAE dependence on the band-filling in FePt and FePd, the same applied voltage leads to an opposite change in the MAE in the two systems, and thus the coercivity [183,184]. Besides, for 3*d* metals such as Fe, Co, and Ni, the change in carrier density has a prominent modulation effect on the magnetic anisotropy [185,186]. In a bcc Fe/MgO(001) junction where Fe is only a few atomic layers, an electric field smaller than 1 MV/cm could induce a large modification of the magnetic anisotropy by changing the orbital occupancy of Fe-3*d* close to the MgO layer [23]. The application of voltage could affect the energy of the $3z^2{-}r^2$ orbital and the corresponding electron occupancy in $3z^2{-}r^2$, $xy$, and $x^2{-}y^2$ orbitals. Then the magnetic anisotropy is tuned owing to the spin–orbit coupling. The applied voltage causes the shift in the Fermi level, which also changes the relative electron occupancy in the different orbitals and the consequent magnetic anisotropy.

At the same time, the Curie temperature in metallic ferromagnetic multilayers is closely connected to the carrier density. The change in MAE due to the voltage-controlled carrier density and electron occupancy can drive the $T_C$ modification in ultrathin Fe and Co films, as shown in Fig. 6c and d [61,151,181]. The first-principles calculations predict that owing to the spin-dependent screen effect and *p*-*d* hybridization in metals, the applied electric fields can alter the spin-spiral formation energy and Heisenberg exchange parameters [187,188]. The integrated exchange interaction energy suggests that in different systems a positive electric field can lead to an increase (Co/Pt and Co/Ni) or a decrease (Fe) in $T_C$. However, it should be noted that charge-mediated VCM in a metallic system is strongly limited by the small screening thickness, which makes the charge mechanism a minority in voltage control of FM metals.

The work above focuses on magnetic modulations in the FM layer. However, it is also found that once a FE polarization is induced under the electric field, the interfacial spin of the FE layer can be tuned as well. According to theoretical calculations, a significant interfacial magnetoelectric (ME) coupling can be induced by a modulation of the FM/FE interfacial covalent bonding upon reversal of the FE polarization in the heterostructures of Fe/BTO. [189]. Large modulations of both the Fe and Ti magnetic moments in the Fe/BTO heterostructure have also been demonstrated theoretically [189–191]. In order to detect this interfacial ME coupling, X-ray resonant magnetic scattering were performed on Fe/BTO (1.2 nm)/LSMO tunnel junctions, revealing that the Fe layer governs the interfacial magnetism of Ti [192], which is corroborated by first-principles calculations [189,190,193].

#### 3.1.2 Modulation of carrier density in diluted magnetic semiconductors

The most prominent feature for a diluted ferromagnetic semiconductor is that its ferromagnetism is correlated to hole-mediated magnetic interaction, as described by the *p*–*d* Zener model. This intrinsic feature makes it possible for the DMS to be used as an outstanding model for VCM, based on the FET configuration. Consequently, the pioneering work on the observation of VCM is realized in (In,Mn)As as shown in Fig. 7 [16]. A negative gate voltage enhances the hole density and leads to an increase in the FM interaction among Mn ions, accompanied by enhanced magnetic moments and Curie temperature, whereas a positive counterpart does the opposite. A similar behavior was also observed in (Ga,Mn)As [103]. More interestingly, magnetic anisotropy, which is extensively used in information storage, has been shown to depend on the hole density in (Ga,Mn)As films. Hence, as the gate voltage decreases from a positive value to a negative one, the hole concentration is greatly reduced, corresponding to a uniaxial easy axis switching from $[\bar{1}10]$ to [110] [99].

To directly monitor the field effect on the magnetic moment and ordering temperature, SQUID was also employed to determine the voltage dependent magnetization of (Ga,Mn)As in a quantitative way. Note that the ultrathin (Ga,Mn)As films with a thickness of 3.5 nm and a Curie temperature below 25 K are specially chosen here, because quantum critical fluctuations in the local hole density would expand the capability of electrical modulation [33]. Moreover, to realize the nonvolatile electrical modulation, a ferroelectric gated FET configuration is developed, where the thickness of the (Ga,Mn)As is set to 7 nm and the P(VDF-TrFE) (polyvinylidene fluoride with trifluoroethylene) ferroelectric gate prepared at a low temperature is used rather than a FE oxide layer [194]. Then both the hysteresis loop and Curie temperature of (Ga,Mn)As were tuned due to the modulation of carrier density.

For diluted ferromagnetic oxides, voltage control of magnetism can be traced back to 2005, in the reversible modulation of saturated magnetization and coercivity of Co:$TiO_2$ films by the polarization of the PZT ferroelectric gate [195]. Utilizing the technology of ionic liquid gating, the anomalous Hall effect of Co:$TiO_2$ was demonstrated to be correlated to the carrier density with application of a gate voltage of several volts, thus demonstrating the critical influence of electron carriers on room-temperature magnetism in the samples [106]. Recently, the concept of VCM was extended to the vigorously pursued magnetic topological insulators (TIs), e.g., $Cr_{0.15}(Bi_{0.1}Sb_{0.9})_{1.85}Te_3$ films [196]. The transport of magnetic TI was modulated by the electric field effect via the $SrTiO_3$ back gate, which also serves as the substrate, associated with the emergence of the quantum anomalous Hall effect. The TI shows a large spin Hall angle, defined as the ratio of the spin Hall conductivity to the electrical conductivity, providing a source of spin current for the switching of magnetic films attached to it [197,198]. This topic will be further discussed in Section 4.3. Compared with metals with high conductivity, the electric field effect can penetrate into a thicker magnetic semiconductor layer due to the larger screening thickness, resulting in a more obvious modulation in magnetic properties. We also note that almost all the voltage controls of magnetism in semiconductor materials are based on the charge modulation. The reason for the simple mechanism in a magnetic semiconductor is closely related to the origin of magnetism in the semiconductor, that is, the holes mediate the magnetic interaction.

### 3.1.3 Modulation of carrier density in ferromagnetic oxides

Compared with FM metals and semiconductors, the dielectric layers used in FM oxide-based heterostructures are usually ferroelectric (FE) materials. The spontaneous or induced polarization will result in the accumulation of electrons or holes at the interface between the dielectric and magnetic layers and corresponding magnetic properties modulations. For a FE material like PZT, the electron or hole modification is on the order of $10^{14}$ $cm^{-2}$, which is dramatically higher than what can be achieved utilizing $SiO_x$ as the dielectric layer [199]. Furthermore, such an effect is nonvolatile after withdrawing the electric field. The ferroelectric FET has been used to modulate various physical properties such as superconductivity [200,201] and metal–insulator transitions [172] for a long time. In the last 10 years, the voltage controls of magnetism in ferroelectric FET were reported both in first-principles calculations and in experiments.

The doped manganite supplies an ideal arena for VCM because of its rich electronic phase for different chemical doping levels [8,202]. The significant ME coupling based on the change in hole carrier density was theoretically proved in the $La_{1-x}A_xMnO_3$/BTO (001) system, where the *A* ion could be Ca, Sr, or Ba and the doping level *x* was set to be 0.5, close to the FM and AFM phase-transition point [189]. The variation of the holes in $La_{0.5}A_{0.5}MnO_3$ during the polarization switching of the BTO, which occurs at the interface, will result in a transition between FM and AFM phases. For different FE polarization states, an increase in holes favors the AFM state while a decrease in holes stabilizes the FM state. Additionally, another theory based on the double exchange interaction between two orbitals was used to simulate the influence of interfacial hole or electron accumulation on the phase transition of manganites during the FE polarization switching. This model also found a remarkable ME coupling and bipolar resistive switching during the polarization reversal [203].

Charge-meditated VCM in ferromagnetic and ferroelectric oxide heterostructures was first observed by Molegraaf *et al.* [27] in PZT (250 nm)/LSMO (4 nm). MOKE signals measured at 100 K demonstrate that the dependence of saturation magnetization in LSMO on the external electric field behaves as a hysteretic loop, as displayed in Fig. 8a, where the reversal of magnetic moment in LSMO is relevant to the switching of FE polarization in PZT. Due to the remnant polarization of the FE layer, the magnetic moment modulation is stable even after the applied electric field is removed, suggesting a nonvolatile electrical manipulation via the FE gate control. At 100 K, the magnetoelectric coupling coefficient is calculated to be $0.08–0.62\times10^{-8}$ s/m, while magnetoelectric response is enhanced with the increase in temperature and reaches the maximum value of $1.35\times10^{-8}$ s/m at around 180 K [204]. It is very interesting that the peak of $\alpha$ located at a temperature near the magnetic $T_C$, indicates that a larger electric field effect can be gained in light of the transition between ferromagnetic and paramagnetic phases in manganites.

The modification of carrier density under different electric fields is demonstrated using X-ray absorption near edge spectroscopy (XANES) in Fig. 8b [34] and electron energy loss spectroscopy (EELS) [205] in Fig. 8c. The main finding in Fig. 8b is the observation of an energy shift in the Mn XANES of around +0.3 eV upon reversing the PZT polarization from the depletion to the accumulation state, suggesting that the valence state of Mn becomes higher. Figure 8c maps the local variations of the Mn valence in nanoscale in the LSMO/PZT/LSMO heterostructure by EELS. A magnetic asymmetry at the LSMO/PZT interface was observed, which depended on the local PZT polarization and led to gradual modulation of moments from the interface to the bulk of the thin film. In the top LSMO, the EELS peak position of the sample increases gradually to nearly ~0.2–0.3 eV in the region of 2.5–3 nm near the interface, while the LSMO near the bottom interface shows a delicate fluctuation in the EELS peak position of ~0.1 eV. These results for chemical valences in manganites under different FE polarization states strongly demonstrate the role of carrier density variation in the VCM in these systems.

Actually, not only the valence, but also the orbital occupancy of manganites is thought to be modulated by the electric field, collectively resulting in modification of the magnetic moment. In Fig. 8d, LSMO in the accumulation state behaves as stronger in-plane $x^2$–$y^2$ orbitals, which will dramatically weaken the origin of magnetism in manganites, i.e., the double exchange effect [34]. This modulation in orbital occupancy will lead to a variation in the magnetic moment of around (4+3+0.2)/12 $\mu_B$/Mn between different polarization states, which is in qualitative agreement with the first-principles calculations in LSMO [189,206–209]. To analyze the role of orbital occupancy in VCM, we will discuss the details in Section 3.4.

Note that the effect of VCM is highly related to the chemical compositions and thicknesses of manganite films. LSMO is a hole-doped magnetic oxide whose magnetic properties is sensitive to a slight variation in the doping level. LSMO with different doping levels $x$ = 0.20, 0.33, and 0.50, and FE PZT are used to understand the influence of chemical composition on VCM [123]. When $x$ is 0.20, the $T_C$ value is very sensitive to the doping level with a slope $dT_C/dx$ of 1300. The slopes for the case of $x$ = 0.50 and $x$ = 0.33 are reduced to 800 and 300, respectively. Thus, FE polarization switching should have a larger effect on the magnetic properties of LSMO ($x$ = 0.20) at the phase boundary than that of LSMO at the center of the FM phase, which can be demonstrated by their resistivity dependence on the temperature as shown in Fig. 9a and b. Except the doping level, the electric field effect also relies strongly on the distance of the FM layer from the FE/FM interface as can be seen in Fig. 9c. The differences of carrier density in different polarized $e_g$ orbitals will dramatically be reduced with the increasing distance from the FE/FM interface and become negligible after the forth unit cell. This conclusion is also supported in other theoretical work [203,210]. Hence, for these voltage controls of magnetism based on charge modulation, the ultrathin FM oxides at the boundary between two electronic phases are usually used.

The modifications of carrier density under FE reversal are also adopted to explain the variation of the magnetic properties in $La_{0.85}Ba_{0.15}MnO_3$ (10 nm)/PZT [211] and BTO (19 nm)/LSMO (10–50 nm)/STO (001), where the magnetic moments are studied by MOKE and SQUID [65]. For the $La_{0.85}Ba_{0.15}MnO_3$/PZT device structure, different hysteretic curves of $La_{0.85}Ba_{0.15}MnO_3$ are observed at room temperature when the polarization of PZT are switched [211]. Nevertheless, the magnetization is enhanced in the depletion state, which is opposite to the observation in the LSMO/PZT system [27,34]. It also indicates that there should be a novel origin for the modulation magnetic property in the $La_{0.85}Ba_{0.15}MnO_3$ channel layer. A linear ME coupling is demonstrated in the STO/SRO/STO heterostructure by first-principles calculations [212]. The voltage controls of magnetism based on the charge modification were widely observed or calculated in BTO/LSMO [65,213], BFO/LCMO [214], BTO/SRO [212,215], BTO/$Fe_3O_4$ [131], and $CrO_2$/BTO/Pt [216], *etc*.

The carrier density modulation is one of the most popular mechanisms for VCM in oxides because the polarization switching in the FE layer is bound to the carrier density modulation in the adjacent FM oxide layer. For example, the $T_C$ and magnetic moment can be significantly changed under the modulation of the electric field. The magnetic properties dependence on the electric field tracks the polarization variation in the FE layer as the electric field varies. In voltage controls of magnetism based on charge modification, the nature of the ME effect can be divided into three types: (i) the spin imbalance is increased under the electric field, which contributes to the modification of magnetic moments [77]; (ii) the electronic phase of the magnetic layer is modulated by the electric field, resulting in the ferromagnetic–paramagnetic or ferromagnetic–antiferromagnetic phase transition [34,217]; and (iii) the magnetic anisotropy is modulated due to the various density of states near the Femi level under different polarizations [23,218]. However, the origins of the magnetic property variations under electric fields can be complex sometimes, because there can be more than one type of ME effect in the actual materials.

### 3.1.4 Modulation of carrier density in non-magnetic oxide heterostructures

With the advance of technologies in thin film growth, the delicate manipulation of complex oxide heterostructures at an atomic level becomes achievable and gives rise to a large number of exotic states and unexpected performance characteristics at the interface. The most famous representative is the modulation of the two-dimensional electron gas (2-DEG) and the interfacial magnetism in the insulating nonmagnetic LAO/STO heterostructure [219–224]. Such a novel magnetic state and 2-DEG is closely related to the interfacial charge transfer from LAO to STO, which also provides an ideal opportunity for VCM. Both FET [225] and BG [226] types of devices could be used to finish this task, taking into consideration the fact that the interfacial magnetism is surrounded by two insulting layers. The transport measurements have proved the significant perpendicular magnetoresistance in this system, which can be switched from positive to negative as the voltages change from negative to positive, as shown in Fig. 10a. This voltage-controlled magnetoresistance is thought to be relevant to the sharp enhancement of a Rashba spin–orbit coupling with the increasing external voltage [201].

By applying different voltages on the top of the LAO/STO interface, the dependence of interfacial magnetism on the carrier density at the interface can be characterized by MFM as shown in Fig. 10b [220]. As the external voltage gradually changes from –4 V to 0 V, the perpendicular magnetic domain structure is weakened and spin aligns in plane. The finding of electrically controlled magnetic signals at the LAO/STO interface promises a novel path for applications in spintronics. Despite some attempts for an effective VCM in this fancy system, the origin of the magnetic state still remains unclear, and various explanations such as electrically controlled spin-wave propagation and detection, spin-polarized transport, and spin-torque transfer are proposed but not demonstrated.

Except the 2-DEG system, Grutter *et al.* [227] found a novel magnetic layer at the $CaRuO_3/CaMnO_3$ interface, which could be manipulated by external voltages (Fig. 11a). When a −400 V bias voltage was applied on the $CaMnO_3/CaRuO_3$ heterostructure, an increase of magnetic moment to 2.86 $\pm$ 0.4 $\mu_B$ per Mn is observed (Fig. 11b), which is ascribed to the enhancement of the double-exchange interaction with more electron transfers from the $CaRuO_3$ to $CaMnO_3$. In general, the interfacial magnetism in the nonmagnetic heterostructure is caused by the charge transfer (from LAO to STO or from $CaRuO_3$ to $CaMnO_3$). Thus, the electric field provides an effective route to confine the movement of electrons and the corresponding magnetism at the interface. The phenomena and physics at the interface of dissimilar oxides have attracted considerable attention in the field of condensed matter over the past 10 years. The utilization of electric fields to manipulate these interfacial phenomena (e.g., magnetism) is significant for revealing the in-depth physical mechanism and promoting this area in application.

### 3.2 Strain effect

Strain engineering is an ideal arena for controlling magnetism [119,228,229]. It is generally recognized that the ME coupling effect through strain transfer from the ferroelectric layer to the ferromagnetic layer will induce a remarkable modulation in magnetic properties. Such a type of electric-field manipulation of magnetism has been widely demonstrated in ferromagnetic metals and oxides (such as Co, CoFeB, LSMO, and CFO) prepared on a ferroelectric layer or substrates (such as PMN-PT, PZT, and BTO). Under the voltage applied to the FE materials, the lattice or shape of the FE layer is modulated through an inverse piezoelectric effect. Subsequently, the deformation of the FE layer will generate a strain, which will transfer to the proximate magnetic materials, resulting in modulations of $H_C$, magnetic anisotropy, even magnetization switching, mediated by the inverse magnetostrictive effect. In the section, we will focus on the strain-mediated VCM in FM or AFM metals and FM oxides.

### 3.2.1 Strain mediation in ferromagnetic metal/ferroelectric oxide bilayers

With the advantages of high $T_C$, flexibility, and easy production, ferromagnetic metal systems attract a great deal of attention in VCM. Different from those mediated by charge modulation, the voltage controls of magnetism based on strain in metallic materials are much more effective without the limitation of the small screening length. The voltage control of ferromagnetic metals was realized experimentally in a large number of systems such as Fe/BTO [19], Fe-Ga/BTO [78], CoFe/PMN-PT [230], CoFeB/PMN-PT [77,80], Co/PMN-PT [59,231], CoPd/PZT [75], and Ni/BTO [76]. Using phenomenological treatments, Pertsev [232] and Nan [233] reported that the magnetic properties of FM thin films is elastically coupled with FE layers through a ME coupling based on strain mediation, suggesting that the spin reorientation within the film plane as well as between the in-plane direction and the out-of-plane direction can be realized with the help of an electric field. In this way, the magnetism involving the magnetic anisotropy, coercivity, magnetic moment, and the magnetization reversal can be controlled by a gate voltage.

We first focus on the modulation of magnetic moments under external voltages.

In the voltage control of FM metals based on strain effect, the PMN-PT relaxor ferroelectrics have been widely used as the FE layer [234]. Normally, the in-plane piezostrain of PMN-PT with a (001) orientation shows a symmetric and volatile butterfly-like behavior lacking in remanent strain, and the voltage-dependent magnetization follows the strain variation in a butterfly-like shape [235]. Thus, the strain state and changes in magnetism under an electric field cannot be retained after removing it, which is unacceptable for information storage. In 2012, Zhang *et al.* [77] demonstrated a nonvolatile electrical manipulation of magnetism in CoFeB/(001) PMN-PT (Fig. 12a) that differed from the previous work. A loop-like behavior in the magnetization dependence on the electric field can be seen clearly with an electric field reversal between +8 and –8 kV/cm, and a relative variation of the magnetization about 25% is observed. The percentage of net 109° domain switching is found to be 26%, which is quantitatively comparable to the 25% relative change in magnetization (Fig. 12b and c), suggesting that the nonvolatile electrical manipulation of magnetism depends strongly on the 109° domain switching of PMN-PT.

According to the analysis above, the FE domain structure plays a vital role in nonvolatile VCM. However, the origin of the net 109° domain switching in PMN-PT (001) substrates, which may be closely related to some defects introduced during crystal growth [236], still remains elusive and desperately needs more investigation. Another approach to realizing nonvolatile VCM is employing unipolar poling voltage in PMN-PT with the (011) orientation [157,237]. If the electric field is swept circularly from positive to negative but does not exceed the negative coercivity along the $[01\bar{1}]$ direction in this substrate, the electric field dependence on polarization as well as the strain shows a hysteresis.

$BaTiO_3$ is another frequently used FE substrate in VCM. Sahoo *et al.* [19]

investigated the temperature-dependent magnetic moment of Fe which is grown on BTO. The structural variation in BTO will generate different strain states in the FM layer, resulting in a remarkable modulation of magnetic moment and coercivity in the Fe layer. In a similar system, modifications of the Fe spin structure are imaged by MFM, which display different domain structures with BTO in different temperature and polarization states [238,239]. In these studies, a change in the magnetization based on strain modulation is demonstrated under an electric field or the temperature variation. For the strain modulation, the domain structure at different external electric fields and temperatures plays a critical role. The coupling between ferroelectric and ferromagnetic domain structures was reported in CoFe thin film grown on BTO crystal. The FM domain pattern follows the FE stripe domain switching under the electric field, which not only clearly demonstrates the role of strain in the ME coupling, but also provides a promising way to realize the electrical-write FM domain structure [240].

Despite the control of magnetic moments, many attempts have been focused on the realization of tunable magnetic anisotropy under electric fields through ME effects in a FM/FE heterostructure because it was thought to be the first step toward magnetization switching controlled by gate voltages. The magnetic anisotropy (MA) of Fe was first found to be different on BTO with different structures at different temperatures: a tetragonal-phase BTO favored a fourfold symmetry in the MA of Fe, while the BTO in orthorhombic phase stabilized a twofold symmetry [241]. In the following, the magnetic anisotropy of FeGaB films was directly manipulated by the strain transferred from bottom PZN-PT (011) substrates under an electric field at room temperature (Fig. 13a) [242]. In Fig. 13b, the magnetization is measured along [100] with the electric field applied along [011]. As the electric fields were enhanced from 0 to 6 kV/cm, the saturation fields in hysteretic curves sharply increased from 1 to 70 mT. Here, the strain should pay for the enhanced difficulty in magnetization reversal, because the external electric field would result in an in-plane compressive strain in the FeGaB of positive magnetostriction and lead to the magnetic hard axis in the [100] axis.

The strain-mediated VCM in FM metals is drawing increased attention mainly because of its potential for room-temperature magnetization switching via electrical methods without the assistance of a magnetic field, which will be discussed in Section 4.2 in detail. In addition, this mechanism is much more effective for thick metallic samples of $10^2$ nm compared with ones based on charge [76]. The electric field-induced change in the magnetic properties follows the variation tendency of the piezo strain in the FE substrates under an external voltage. When compare with the magnetic propertiess in strain- and charge-mediated VCM, we find that the coercivity, magnetic anisotropy, and magnetization switching are mostly studied in a strain-mediated mechanism, which are also different from the interests in $T_C$ and magnetic moment for a charge-mediated mechanism.

### 3.2.2 Strain mediation of antiferromagnetic metals

Giant (tunneling) magnetoresistance stands out as a seminal phenomenon in the emerging field of spintronics. In a spin valve or a MTJ, the antiferromagnetic layer is used as a fundamental static-supporting material to establish a reference magnetization direction. Recently, except the assistant role of AFM materials as a pinning layer, many attempts have been made to control the antiferromagnet as a functional layer in the emerging field of AFM spintronics due to its ignorable ferromagnetic stray field and invulnerability to perturbations in the external magnetic field [243–245]. The rapid development of AFM spintronics also stimulates the area of voltage control of antiferromagnetic materials. One of the most typical examples is the FeRh alloy. The $Fe_{50}Rh_{50}$ shows a ferromagnetic to antiferromagnetic phase transition at $T^* \approx 350$ K and a total magnetic moment of ~2.9 $\mu_B$ [246–248]. Remarkably, during the ferromagnetic to antiferromagnetic transition, the volume of the lattice changes by about 1% [246], providing the possibility of VCM in FeRh through strain transfer.

The magnetization dependence on temperature for epitaxial FeRh grown on BTO substrates are displayed in Fig. 14, where polarization states of the heterostructures are marked in the FE loop in Fig. 14a. A voltage applied on the BTO backside of 21 V could clearly increase the $T^*$ by around 25 K, manipulating the phase of FeRh without the assistance of heating or cooling. During this process, the modulation of magnetization could reach $\sim 5.5\times10^5$ A/m, corresponding to a huge ME coupling coefficient $\alpha$ of $1.6\times10^{-5}$ s/m. It is worth noting that this value is much larger than some other systems such as $TbPO_4$ ($3\times10^{-10}$ s/m) [249], LSMO/BTO ($2.3\times10^{-7}$ s/m) [64], and CoFeB/PMN-PT ($2\times10^{-6}$ s/m) [77]. Despite the magnetization, the voltage-controlled FeRh phase transition between FM and AFM phases also causes a significant electroresistance effect in the similar system of FeRh/PMN-PT [86]. The relative change is called a giant electroresistance effect, which can be observed at all the temperatures presented here as shown in Fig. 14c. The in-plane strain modification in FeRh (left) depends strongly on the lattice variation along the PMN-PT *c*-axis (right), affirming that the changes in the strain state in FeRh during polarization reversal of PMN-PT is the reason for the electroresistance effect (Fig. 14d). With the sample quality of FeRh improved on BTO, the full phase transition can be modulated by an electric field with the value of 2 kV/cm. Therefore, a larger electroresistance, around 22%, can be achieved in epitaxial FeRh/BTO heterostructures in comparison with FeRh/PMN-PT [250].

The potential for the application of antiferromagnets in high-density memory technologies is attracting further investigations into this topic, where the manipulation of antiferromagnetic moments is central to the research. Considering that the antiferromagnetic moments are not sensitive to the external magnetic field, the exchange coupling, voltage, and current become three candidate methods to drive the partial rotation of antiferromagnetic moments in various materials, IrMn, CuMnAs, and $Mn_2Au$, etc. [243]. The previous studies on the phase transition of FeRh films driven by the strain effect, originally induced by the voltage-dependent ferroelectric polarization, could attract increasing research on the tuning of antiferromagnetic moments by electrical methods [86,250].

### 3.2.3 Strain mediation in ferromagnetic/ferroelectric oxide heterostructures

With the rapid advances in deposition technologies, such as oxide molecular beam epitaxy and PLD systems accompanied by reflection high-energy electron diffraction, epitaxial FM oxides can be deposited on FE substrates with atomic level precision, which would construct the base for VCM via strain in the oxide system [117]. The role of strain was illustrated in the thin films of 20–50 nm-thick LSMO ($x$ = 0.3) and LCMO ($x$ = 0.3) on PMN-PT(001) crystals [26]. The magnetization intensity of LSMO ($x$ = 0.3) displays a butterfly shape when sweeping the electric field, which just tracks the dependence of the piezoelectric strain along the [100] direction in PMN-PT on the electric field as displayed in Fig. 15a and b. The ME coupling coefficient is calculated to be around $\sim 6\times 10^{-8}$ s/m near room temperature in this system. This work attributes the enhanced magnetization at the polarized state to the ferromagnetic-to-paramagnetic phase transition induced by the piezoelectric strain during ferroelectric polarization switching. A large modulation of magnetization was observed in a 40 nm-thick LSMO ($x$ = 0.33) grown on the BTO substrate based on the temperature-dependent phase transition and the resulting lattice variation in BTO [64], where the coupling between ferroelectric and ferromagnetic domains was used to explain the magnetization variation.

Oxides in a spinel structure with magnetic character can also be manipulated by an electric field via strain. For the system of polycrystalline $Fe_3O_4$ film deposited on a PMN-PT (011) substrate, the MA was significantly changed based on both the magnetoelastic and magnetostatic energy contributions as shown in Fig. 16. The electric field-induced modification of MA was demonstrated by both the ferromagnetic resonance frequency and the shape of the magnetization curves [132]. The strain-mediated VCM is observed more widely in spinel ferrimagnetic oxides such as $Fe_3O_4$, CFO, and NFO, where the magnetic moment and anisotropy are dramatically changed under an electric field [132–134,137–139].

Despite the 2-2-type heterostructures, in the BFO/CFO and PZT/CFO 1-3 vertical composite films, where the magnetic $CoFe_2O_4$ nanopillars are embedded in a FE BFO or PZT matrix, an electric field could also modulate the spin configurations of the systems [139,140]. For the strength of the ME interaction between the FE matrix and the magnetic nanopillars, $\alpha$ was estimated to be approximately $1\times10^{-7}$ and $4\times10^{-7}$ s/cm in BFO/CFO and CFO/ PZT, respectively. Here, the origin of the magnetization reversal driven by the electric field is the intimate lattice interaction among the different magnetic nanopillars, induced by three-dimensional hetero-epitaxial growth. Thus, the voltage applied on the ferroelectric matrix would change its lattice through the inverse piezoelectric effect in the process of ferroelectric polarization reversal, which consequently modulates the MA of the magnetic pillars by magnetostriction.

Voltage control of magnetism mediated by strain transfer has been found in various oxide systems, reflected by the dramatic and reversible modification in magnetization under electric fields. Nevertheless, the mechanisms of magnetic properties changes caused by strain engineering are still vigorously pursued and two different origins are proposed: i) the electronic phase transition induced by electrical means and ii) the magnetic anisotropy modulated by electrical means. The electronic phase transition induced by strain under an electric field is demonstrated by studies on FM materials like LSMO [26,64], LCMO [129], and PCMO [251], which has a strong tendency of phase separation. In contrast, the modifications of MA with external voltages were generally accepted in CFO [137,138] and $Fe_3O_4$ [132–134]. Except the FM/FE heterostructures, magnetic modulation under electric fields is also observed in many multiferroic thin films like $ErMnO_3$, SMO, and $EuTiO_3$, where the origin of magnetoelectric coupling is also attributed to the strain [252–255].

### 3.3 Exchange coupling

The exchange coupling is another mechanism for VCM. The exchange coupling or exchange bias (EB) effect is widely observed at the interfaces between various ferromagnets and antiferromagnets, reflected by a shift of the magnetization curve away from the origin. Then if the coupling between FM and AFM layer can be manipulated by external voltage, the magnetic propertiess of the heterostructure are modulated subsequently. If the single-phase multiferroic (antiferromagnetic and ferroelectric) materials such as $Cr_2O_3$ [81–84], $YMnO_3$ [85,256], $LuMnO_3$ [257], and BFO [20,28,158,258,259] are used to control the magnetism by electrical means, the exchange coupling between the AFM and FM order at the interface would bridge the external voltage and the magnetic properties. While the exchange spring in AFM material could transfer the effect of external voltage to the AFM/FM interface and result in the magnetic modulation.

### 3.3.1 Voltage control of exchange coupling in multiferroic heterostructures

The electric field-controlled exchange bias was initially found in perpendicularly magnetized [Co/(Pt or Pd)]/$Cr_2O_3$ heterostructures [81–84]. The EB in the [Co/Pd]/$Cr_2O_3$(0001) heterostructure was reversibly manipulated at 303 K. The antiferromagnetic single domain state in $Cr_2O_3$ is reversed by the combination of magnetic and electric fields. Subsequently, the orientation of the uncompensated spins at the $Cr_2O_3$(0001) interface which pinned the [Co/Pd] hysteretic curve was switched [81]. Figure 17a demonstrates a large isothermal voltage control of the $H_{EB}$ [83]. Once the $|EH|$ reaches a critical threshold ($E$ and $H$ are isothermally applied axial electric and magnetic fields, respectively), a significant electrically controlled magnetic switching is observed. Note that the voltage control of exchange bias in the $Cr_2O_3$/$(Co/Pt)_3$ multilayer also needs the assistance of heating and cooling. The voltage control of $H_{EB}$ was also reported in NiFe/$YMnO_3$ [85], where the $H_{EB}$ is suppressed to 0 by applying a voltage on the bottom across the $YMnO_3$. As the arrow in Fig. 17b shows, when increasing the bias voltage, the magnetization will decrease and even reverse its sign.

BFO is a classic multiferroic material whose ferroelectric $T_C$ is around 1100 K and antiferromagnetic Néel temperature is around 640 K [38,146,260]. Considering the importance of multiferroic BFO in VCM, we need to introduce some basic knowledge of this material first. Although the spin configuration of $G$-type antiferromagnetic (oppositely aligned ferromagnetic planes of the {111}) BFO is fully compensated at the interface of (001) orientation, the exchange bias coupling between the FM and AFM layers could still be generated by the Dzyaloshinskii-Moriya coupling and ferroelectric polarization [261,262]. An abundant ferroelectric domain structure is found in BFO because the eight possible polarizations along any one of the eight degenerate [111] directions are able to form three types of FE domain walls [146,158]. By changing the orientation of the film growth, strain state in film, and substrate, the domain walls of 71°, 109°, and 180° can be effectively controlled [263,264]. Taking advantage of the AFM domain reversal induced by the ferroelectric polarization reversal, the voltage control of AFM domains in BFO at room temperature was demonstrated by Zhao *et al.* [158]. This result establishes the foundation of voltage control of magnetic properties through BFO.

In the BFO/LSMO heterostructure, the exchange bias field and coercivity are switched between two different states via the reversal of FE polarization in a reversible way, which can be reflected by the magnetoresistance in Fig. 18 [28]. Although both strain and charge may contribute to the changes in magnetic properties modulation under electric fields, the voltage control of exchange bias here is mainly attributed to modulation of canted interfacial magnetism (spin degree of freedom) in BFO under electric field sweeping [265]. Subsequently, the $H_{EB}$ is reversibly manipulated between two distinct states with opposite signs based on the FE reversal of BFO in BFO/LSMO without assistance of any magnetic field [258]. The relevant AFM-FE order in BFO and the modification of exchange couplings at the interface induce the bipolar modification.

Although multiferroic materials with both ferroelectric and antiferromagnetic orders, e.g., BFO, are the typical FE layer for VCM based on exchange bias, it is an overgeneralization to attribute all the manipulation of the magnetism by electrical means in these systems to exchange bias. For example, in the LCMO ($x$ = 0.5)/BFO system, where the LCMO ($x$ = 0.5) is near the FM–AFM phase boundary, the variation in the carrier due to FE polarization switching would drive the FM–AFM phase transition in LCMO ($x$ = 0.5) and should play a pivotal role in the modification of magnetic properties under electric fields [214]. Additionally, in the CoFe/BFO system, the mechanical strain generated by ferroelasticity under an electric field could also modulate the preferential direction of the magnetic domain structures and the resulting magnetization [20]. Except the exchange coupling effect, ferroelastic strain is also thought to contribute to the voltage-induced magnetization switching in devices or heterostructures based on BFO. To exclude the influence from ferroelastic strain, a $LuMnO_3$ single crystal was chosen as the substrate to grow ferromagnetic $Ni_{81}Fe_{19}$ film. $LuMnO_3$ with both FE and AFM orders exhibit a 180 ° FE domain, which is not ferroelastic. Thus, the voltage-induced magnetization switching reported in this system could only be caused through the exchange coupling [257].

In the VCM based on exchange bias, multiferroic materials play a very important role. The coercivity, exchange bias field, and even the magnetization switching are manipulated by the electric field based on the associated ferroelectric and antiferromagnetic domains. Such a magnetoelectric effect in the multiferroic/ferromagnetic heterostructures also generates a novel device for magnetoelectric random access memory (MeRAM) [266]. In MeRAM, external voltages can be applied to control the interfacial exchange coupling and switch the magnetization of the FM layer, dramatically reducing the energy consumption. Although the corresponding mechanism is still controversial [261,267,268], these phenomena pave the way for integrating multiferroics into modern memory devices with potentially high performance characteristics.

### 3.3.2 Voltage control of exchange springs in antiferromagnetic alloys

The exchange bias and the magnetic moment can also be manipulated in the systems without multiferroic materials. In the multilayer of Ta (4)/Pt (8)/[Co (0.5)/Pt (1)]$_4$/Co (0.5)/Pt (0.6)/IrMn (3)/$HfO_2$ (2) (units in nanometers) [70], an electric field is applied through an ionic liquid. The electric double layer built up on the surface of the multilayer can be used to manipulate the electron carriers in the multilayer. In such a configuration, similar to the mechanical spring for force transfer (Fig. 19a), the IrMn exchange spring transfers the influence of the electric field to the interface of the magnetic Co/Pt and AFM IrMn and thus the magnetism is modulated. The $H_{EB}$ and $H_C$ of [Co/Pt]/IrMn in Fig. 19b show that a positive and negative $V_G$ would increase and suppress the interfacial exchange interaction, respectively. The electric field effect on exchange coupling is supposed to be prominent within the length of the exchange spring (~6 nm for IrMn [94]). Thus, the voltage control of EB would become weaker as the thickness of the AFM IrMn increases. A similar effect is observed in AFM FeMn but with a longer depth of ~15 nm [69]. Recent progress in the voltage control of the exchange spring in AFM metals through an ionic liquid provides a novel thinking for realizing the exchange-coupling-mediated VCM.

### 3.4 Orbital reconstruction

The modification of carrier density, strain transfer effect, and exchange coupling are the three classic mechanisms responsible for voltage control of magnetic properties. The corresponding physical natures of these mechanisms are the modulations of lattice, charge, and spin by an electrical way, respectively. Lattice, charge, and spin are well known as the members of the four degrees of freedom in strongly correlated electron systems. However, from the integrity of the physics, the orbital, as another important degree of freedom, has received little attention in previous studies about electric-field manipulation. Since the magnitude and the anisotropy of electron transfer between atoms are closely related to the orbital occupation, the orbital occupation plays an vital role in determining the electronic structure and magnetism [269]. At the interface between two dissimilar oxides or a metal and an oxide, the charge transfer and orbital hybridization/reconstruction of two proximate metal atoms or ions through $O^{2-}$ would induce an interfacial covalent bond, leading to a series of fancy electronic structures and phenomena [270,271]. For FE field-effect transistors, the polarization reversal associated with a movement of charged ions, which is known as the essential characteristic in FE materials, can be used to manipulate the orbital reconstruction at the interface, thus providing a promising approach for the orbital-mediated VCM.

3.4.1 Electrically induced orbital reconstruction in ferromagnetic metals

The chemical bonding or orbital hybridization at the interface between FM metals and dielectric/ferroelectric oxides has a great influence on the magnetism of FM metals, so the performance of the FM metals can be manipulated through the electric field applied via the oxides. In the heterostructures of MgO (10 nm)/Fe (0.48 nm), the perpendicular anisotropy of Fe changes significantly, as observed in the representative magnetic hysteresis loops under the application of bias voltages of ±200 V, which are obtained from Kerr ellipticity, $\eta_k$, measurements (Fig. 20a) [23]. As the voltage decreases from +200 V to –200 V, a large enhancement of the perpendicular anisotropy occurs. The modulation of MA can be ascribed to the electric field-induced variation of electron occupancy in the Fe layer. This is also supported by the results reported by Kyuno *et al*. [272] where the MA of 3*d* FM metals was susceptible to the occupancy of 3*d* orbitals at the interface of the 3*d* FM-metal/noble-metal. An electric field of 200 V makes a change in electron filling of $2\times10^{-3}$ electrons per Fe atom at the surface [23], which is high enough to induce a remarkable change in the anisotropy energy of 4 meV per surface Fe atom, resulting in an enhancement of the $3z^2–r^2$ ($m_z = 0$) orbital energy (Fig. 20b).

VCM accompanied by the orbital hybridization was also proposed in the heterostructures of an ultrathin FM metal layer and a FE oxide layer, where the interfacial spin states are easily controllable via the ferroelectric polarization. In the theoretical predication of Duan *et al.* [273], the magnetic moment of the Fe atom in the Fe/BTO heterostructures tends to show an antiparallel alignment with the moment emerging in Ti atoms. As the polarization switches between two states, Ti atoms move toward or away from the center of the oxygen octahedrons, resulting in a difference in the distance between Fe and Ti. As a consequence, for the interfacial 3*d* orbitals of Fe and Ti atoms, the degree of hybridization via the O-2*p* orbitals varies as the Fe-Ti distance changes, as shown by the local density of states (DOS) for each orbital in Fig. 21a. In this way, the orbital-related magnetism can be modulated through the reversal of the FE polarization of BTO. Subsequently, in the multilayers of Au/Co/Fe/BTO/LSMO, Radaelli *et al.* [274] demonstrated that the magnetism at the interface was changed during the reversal of the FE polarization. Based on the reversal of the FE polarization, the magnetic states at the interface can be modulated. Both the Fe–*L* and O–*K* edge EELS confirm that Fe ions at the Fe/$TiO_2$ interface show an enhanced oxidation state [274]. Correspondingly, the comparison between XMCD for $P_{up}$ (polarization upward) and $P_{down}$ (polarization downward) at 300 K in Fig. 21b also shows a suppression of the typical dichroic signal of $FeO_x$ for the case of $P_{down}$.

Following the initial work, the Fe/BTO multilayers were epitaxially prepared on Pt/MgO substrates by Yang *et al*, [275] demonstrating an enhancement of room-temperature magnetism in comparison to a pure Fe film, which is ascribed to the hybridization of Ti and Fe orbitals. Then, the theoretical calculations indicated that the interfacial magnetization is also controllable in the Ni/BTO superlattices with the help of polarization reversal [276]. However, the mechanism for ME coupling at the interface varies as the Ni atoms replace the Fe atoms, that is, the moments of Ni atoms and the Ti atoms play an important role in the coupling of Ni/BTO and Fe/BTO, respectively. Later, theoretical investigations on FE control of magnetization in systems with other 3*d* transition metals, such as Co/BTO and $Co_2MnSi$/BTO, were carried out as well [131,277–279]. Furthermore, the manipulation of magnetization via polarization reversal was predicted in a multiferroic tunnel junction of $Co_2FeSi$/BTO/$Co_2FeSi$ in 2014 [280], on the basis of the multiple orbital hybridizations of Co/Ti or Fe/Ti. Unfortunately, in metal systems, the orbital occupancy of the FM layer is difficult to directly characterize, which places a limit on further insights. Even so, the differences in magnetic properties under various electric fields can be attributed to the changes in interfacial bonding and orbital hybridization based on previous work by predecessors.

### 3.4.2 Electrically induced orbital reconstruction in ferromagnetic oxides

Variations in magnetism induced by the orbital modulation under an electric field in FM metals are mainly demonstrated by theoretical methods or indirect charge transfer. The high-quality FE/FM oxide heterostructures provide an ideal arena for investigation of the relationship between orbital and magnetic behaviors under electric fields [281]. In the BTO/LSMO heterostructure, Ti ions are moved upward and downward to LSMO in polarization upward and downward ($P_{up}$ and $P_{down}$) states, respectively, as observed in the images taken in the high angle annular dark field mode [282]. When the BTO is in a state of $P_{up}$ and $P_{down}$, the orbital hybridization at the interface is suppressed and enhanced due to the Ti displacement, respectively, which is associated with an enhancement and a reduction in the occupancy of the Mn-$x^2$–$y^2$ orbital (Fig. 22a) [41]. The in-plane magnetoresistance (pMR) in the up-polarized BTO is much higher than that of down-polarized BTO because of the enhanced occupancy of in-plane orbitals in the former (Fig. 22b). Consistent with the change in the pMR value, the X-ray linear dichroism (XLD) in Fig. 22c reflects that the FE switching from $P_{up}$ to $P_{down}$ alters the preferred orbital from $x^2$–$y^2$ to $3z^2$–$r^2$, which further confirms the mechanism of the ME coupling based on orbital reconstruction [122]. Moreover, since the Ti-O-Mn bond at the interface can be switched on and off via the FE polarization, it can be adopted for the manipulation of the bulk performance as an orbital switching [282].

Almost at the same time, XLD and XMCD in the work of Preziosi *et al.* [283] demonstrated that the electrical manipulation of the orbital anisotropy and magnetization of Mn atoms can be nonvolatile in LSMO/PZT heterostructures. The results of XAS, XMCD, and XLD of the Mn $L_{2,3}$ edges, respectively indicate that the carrier density, the spin moment, and the orbital occupancy of Mn $e_g$ orbitals can be modulated with the help of the reversal of the FE polarization. The layer sequences of heterostructures used in these two works are opposite, but the electrical modulations of magnetic properties (e.g., $T_C$, MR, and XMCD) in these two works are both closely and clearly related to the orbital occupancy [41,283]. These results are explained by the polar distortions of the $MnO_6$ octahedra caused by the switching of the ferroelectric polarization, which are also supported by the theoretical calculations [284,285]. For example, as demonstrated by the calculation of Chen *et al.* [285], taking advantage of the nanoscale ferroelectric polarization at the interface of FE oxides/FM manganites, the reversible manipulation of the orbital population can be realized and a remarkable variation in orbital occupation can exceed the value of the bulk materials, which is as large as 10%.

Besides, in the heterostructures of BTO/(NFO/BTO)$_n$ on the STO substrate[286], as the thickness of the NFO varies, a significant change in the Ni moments occurs while the Fe moment remains unchanged. This is confirmed by the characterization of the Ni and Fe ions via XMCD as displayed in Fig. 23. The magnetic moments for both Ni and Fe are reduced as the layer number decreases. At the same time, when the NFO layer thickness decreases, the average magnetic moment of the Ni ions is found to be reduced, leaving that of the Fe ions unsusceptible. In this view, it is the interfacial Ni ions that contribute to the reduction of the ferromagnetic moments, so the reduction of the FM properties via orbital hybridization can be ascribed to the increasing ME coupling related to the interfacial Ni ions.

Here, the orbital reconstruction or hybridization at the interface between dissimilar oxides provides a good stage for VCM via orbital degree of freedom. By the ferroelectric movement, the intensity of the interfacial covalent bond and related orbital occupancies are modulated electrically. Thanks to XLD technology, we can characterize the orbital information in epitaxial oxides experimentally. Now the voltage controls of magnetism are successfully realized by all four degrees of freedom. Compared with the other three mechanisms mentioned before, an orbital mechanism is relatively rare. However, as the interfacial orbital reconstruction and hybridization are commonly demonstrated in various systems, it is expected that this type of mechanism will be observed in an increasing number of systems.

## 3.5 Electrochemical effect

The electrochemical effect or the redox reaction during VCM is widely observed in systems with an ionic liquid or other high oxygen mobility materials (e.g., $GdO_x$) used as a grid, and it becomes an important mechanism in the family of VCM. Under the effect of an electric field, anions and cations in the ionic liquid separate from each other and move to opposite electrodes, while the electric double layer (EDL) forms at the ionic liquid/sample interface where the ionic liquid is in contact with the materials, that is, the ions with opposite charges get paired at the ionic liquid/sample interface. Since the distance of the paired charges is very small, the EDL can offer a very large electric field, which might induce the migration of oxygen ions (vacancies) and cause the modulation of various properties including magnetism. In addition, gadolinium oxide ($GdO_x$) is a solid-state electrolyte with high oxygen mobility, which can play as a reservoir of oxygen ions under the electric field, promoting the studies on the VCM via the electrochemical effect in FM metals as well. In this section, we will focus on this emerging field.

### 3.5.1 Interfacial oxidation of metals under electric fields

In the FET-type VCM devices, the dielectric layer separates the ferromagnetic layer from the top gate electrode. In metal/oxide multilayer structures, the insulating oxide layer (usually MgO or $AlO_x$) is supposed to act as the dielectric gating, and a static electric field is built up under an external voltage. The modulation of metal magnetism caused by an electric field is generally considered to result from changing the charge carrier density and electron occupancy. However, at the metal/oxide interface, the interfacial oxidation or hybridization between the metal and oxygen ions is inevitable during the film deposition and fabrication [287]. First-principles calculations predict that in the MgO/Fe structure, the formation of interfacial $FeO_x$ plays an essential role in the magnetic properties of the heterostructure [218]. In the experiment of Bonell *et al.* [73], it was shown that the partial oxidation of Fe atoms in the MgO/FeCo structure could be manipulated by an electric field in a reversible way, inducing an effective modulation of the interfacial magnetism. Unlike a conventional electrostatic mechanism, the observation of VCM via interfacial oxidation can be defined as an electrochemical effect.

When the oxygen mobility of the oxide is high enough, a more controllable and extensive electric field effect can be observed [42,43,74,180]. In a $GdO_x$/Co structure where is $GdO_x$ is known as an ionic conductor, $O^{2-}$ can be driven toward or away from the Co layer under a relatively small gate voltage and thus the interfacial oxidation state is changed, as shown in Fig. 24 [42]. Using the XAS and XMCD spectra, the migration of $O^{2-}$ and the relevant variation of magnetism are detected clearly. The magnetic properties of the ferromagnetic Co layer such as magnetic anisotropy, coercivity, Curie temperature, and domain wall propagation can be altered concomitantly [43,180]. Particularly, the MAE can be modified by approximately 0.7 erg/cm$^2$ under a voltage of several volts [43] and it was confirmed very recently that magneto-ionic motion can reach beyond the interface limit [288]. In FET-type devices where the electrolyte offers a large electric field, a moderate redox reaction can be induced at the interface with a small gate voltage in similar structures with PMA like $HfO_2$/Ni/Co heterostructures [22,62]. The electrochemical effect in a metal/oxide system allows for nonvolatile operation of magnetism and a significant reduction of the energy consumption.

### 3.5.2 Magnetic phase transition of oxides under electric fields

The electric field produced by EDL was previously applied in the voltage control of conductivity in two dimensions, the transition between metal and insulator, and superconductivity [173–177]. The EDL can induce remarkable changes of carrier density in the two-dimensional conductive system such as LAO/STO [289], graphene [290], and $MoS_2$ [175]. By sweeping the voltage, a nonvolatile metal–insulator transition is observed, offering the possibility of a room-temperature operation. In some other EDL-gated systems, e.g., $SmCoO_3$ [173], $NdNiO_3$ [291], and $Ca_{1-x}Ce_xMnO_3$ [292], the conductive behavior of metal–insulator transitions was also modulated by electric fields. Until recently, the magnetic properties of both FM metal and oxide systems were controlled by EDL gating. Nevertheless, the origin for ionic liquid gating is under intense debate and two mechanisms are proposed to date: electrostatic doping and electrochemical reaction.

The electrostatic doping mechanism is based on the capacitor model in the traditional understanding, where only the carrier density ($\Delta n_S \sim CV_G/e$) of the system is modulated by the electric field. Making use of the electrolyte of EMIM-TFSI, Dhoot *et al.* [169] investigated the doping variation caused by the electrostatic field in LCMO devices as shown in Fig. 25a. The electron doping of $2\times10^{15}/cm^2$ caused by a positive $V_G$ drove a phase transition from a FM metal to an AFM insulator. Yamada *et al.* [109] found the electrical manipulation of ferromagnetism in the magnetic semiconductor $Co:TiO_2$ at room temperature through ionic liquid gating, accompanied by a large-density electron accumulation of $>10^{14}/cm^2$. The application of a $V_G$ of several volts drove a transition from a paramagnetic state with a low carrier density into a FM state with a high carrier density as shown in Fig. 25b. In addition, EDL gating has also been used for the modulation of magnetic properties in $Pr_{0.5}Sr_{0.5}MnO_3$ [293] and $Sr_2IrO_4$ [294].

Although the electrostatic doping mechanism successfully illustrates the variations in carrier density near the interface between ionic liquids and FM materials in many cases, it cannot account for the nonvolatility of the electric field in the whole bulk of the films, which is often observed in other situations [121,122,178]. Therefore, another mechanism is proposed which insists that the electric field effect is realized by an electrochemical reaction: the electric field produced by the ionic liquid would lead to a redox reaction in the channel material, which can be reflected by the formation, migration, and annihilation of oxygen vacancies [176]. The electrochemical mechanism based on oxygen vacancies offers an alternative explanation for the nonvolatile and in-depth electric field of EDL. Then the question comes: which is the dominant mechanism, electrostatic doping or electrochemical mechanism, in the VCM system with ionic liquid gating. It is preferred to say that this is strongly dependent on the magnetic system. For details, the electrostatic doping mechanism dominates in the magnetic material with low oxygen diffusion or small electronegativities, while the electrochemical mechanism is the opposite.

For instance, the FM phase transition in manganite films was reversibly tuned under the electric field of EDL via the formation and the annihilation of oxygen vacancies as demonstrated by Cui *et al.* [121]. In the electronic phase diagram of LSMO, positive and negative voltages can drive the sample to a ferromagnetic insulating (FI) phase with a low doping level and antiferromagnetic metallic/insulating (AFM/AFI) phase with a high doping level, respectively (Fig. 26a). As displayed by the magnetoresistance and magnetization (Fig. 26b) under $V_G$ = +3 V, an insulating and magnetically hard (HMI) phase formed under the electric field. This novel phase is related to the oxygen migration under the electric field and it randomly nucleates and grows in the whole thin film according to the Fourier filter process as shown in Fig. 26c.

Except for manganites, the reversible manipulation of charge transport, metal–insulator transition, and magnetism through ionic liquid gating were also reported in another SRO system [295]. In this work, under a small gate voltage, the crossover temperature of the metal–insulator transition can be modulated in a continuous and reversible way in a temperature range of 90–250 K, while the onset of magnetoresistance varies from 70 to 100 K. The magnetic response to the external gate voltage is attributed to the diffusion of oxygen vacancies between the oxide layer and the ionic liquid in a reversible way. The role of oxygen vacancies in voltage control of material properties is also investigated in other non-magnetic systems. For instance, the formation and the migration of oxygen vacancies are demonstrated in $VO_2$ and $WO_3$ by secondary ion mass spectrometry and *in-situ* X-ray diffraction [176,296,297].

The nonvolatile character of ionic liquid gating based on oxygen vacancies also guarantees many *ex-situ* measurements, making it possible to achieve some manipulations by voltages which has been thought to be difficult for a long time. One concrete example is that the EDL gating can be adopted in the electrical modulation of the orbital occupancy associated with the magnetic anisotropy in LSMO as shown in the sketch of Fig. 27a [122]. According to the XLD signals in Fig. 27b, a positive voltage enhances the strain-favored occupancy of the orbital no matter whether the initial strain is tensile or compressive. In contrast, a negative voltage suppresses the preferred occupancy. The enhancement or suppression of orbital occupancy in one direction will increase or decrease the magnetic anisotropy in the corresponding direction [122,128]. In this work, *ex-situ* measurements about the orbital occupancy are guaranteed, taking advantage of the ignorable relaxation of the gating effect of the EDL even when the voltage has been removed.

As mentioned above, the ion migrations (e.g., $O^{2-}$) rather than simple carrier-density modulation are involved in the electrochemical mechanism. Due to the ultrathin thickness of the EDL around 1 nm, a huge electric field is generated at the interface where the ionic liquid contacts the magnetic material and it can be used to modulate the electronic phase of manganites as thick as 20 nm in a rather large region [121]. However, the core issues that limit the application of ionic liquid gating are the slow reaction speed and its integration with the conventional semiconductor industry. For the latter issue, ionic gel provides a possible solution [298], while the former issue still needs improvement. In addition, we also note that the origin of the electric field effect in ionic liquids is under intense debate, which calls for stronger evidence of oxygen ion migration during the process of VCM.

### 3.6 Comparison of five different mechanisms

As we summarized above, the five different VCM mechanisms have their respective characteristics but also some common points. We summarize the characteristics of these five different mechanisms in Table 5. Actually, these five different mechanisms present some cooperative and competitive aspects in many situations, rather than working separately. For example, the mechanisms of strain, carrier modulation, and orbital reconstruction can explain the voltage controls of magnetism even though the material systems are the same, such as LSMO/BTO [41,64,65,213] and Fe/BTO [19,190,273,299]. Meanwhile, both exchange coupling modulation and carrier density variation are thought to be the origin of the changes in magnetic properties of manganites/BFO under electric fields. The debate between electrostatic doping based on charge and electrochemical reaction based on oxygen vacancies exists during VCM via ionic liquid gating. Thus, it is vital to investigate the relationship between various mechanisms and their combined effect in VCM. We establish the correlation between the five different mechanisms, and the reported cooperation or competition is marked by the lines in Fig. 28a. In general, the strain-, charge-, spin-, and orbital-mediated VCM coexist at the interface in many cases. If the ferroelectric material also possesses the ferroelastic property and the antiferromagnetic property, such as BFO, a simultaneous existence of these four degrees of freedom becomes a reality during the interfacial coupling. In contrast, it seems that the novel electrochemical reaction is independent of the system constituted by the other four mechanisms, but its connection with charge degree should also be recognized.

The differences between voltage controls of magnetism based on various mechanisms can mainly be reflected by the different characteristic thicknesses in FM materials. Figure 28b summarizes the characteristic thicknesses of all five mechanisms and their possible superposition with each other [63]. It is noteworthy that for the carrier density modulation, only the part near the interface can be modulated by the electric field or voltage, with the effective modulated thickness according to the typical screening thickness ($\lambda$) of the FM layer expressed as [300]:

$$\lambda \approx (\varepsilon\hbar^2 / 4me^2)^{1/2}(1/n)^{1/6} \tag{1}$$

where $\varepsilon$ denotes the permittivity, $\hbar$ stands for the Planck constant, $m$ and $e$ are the electron mass and charge, respectively, and $n$ is the carrier density. For ferromagnetic metals, $\lambda$ is commonly less than 0.2 nm [216], while it can increases to about 1 nm for ferromagnetic semiconductors [301]. However, the situation varies for other mechanisms. For the case of strain modulation, the limit in the thickness of the FM layer is no longer the screening thickness, which can be larger than $10^3$ nm sometimes. And the thick FE substrate (>$10^2$ μm) can induce a more remarkable strain, which favors the electrical control. While for the charge-mediated case, where the thickness of the FE layer sometimes reduces down to several nanometers, the mechanical clamping or constraint from lattice distortion (i.e., clamping effect) under FE polarization becomes weaker and the influence on the FM layer is reduced. Note that the strain effect caused by lattice distortion relies on the difference between states with and without polarization of the FE layer, so the positive and negative polarization shows little change for the case of strain modulation [71]. In this view, strain-mediated and carrier-mediated control of magnetism follows the dependence of strain and polarization on gate voltages, respectively.

To distinguish the different mechanisms many attempts have been made in experiment. In the LSMO/PZT system, STEM was used by Spurgeon *et al.* [302] to reveal the contribution of the lattice distortions and the surface charge accumulation in the FE layer. Meanwhile, EELS and polarized neutron reflectometry reveal the existence of a charge-transfer screening region with a thickness of around 2 nm at the PZT interface where the magnetization is affected, suggesting that the FM layer's thickness is vital in the determination of the VCM mechanism.

According to the physical image of interfacial chemical bonding, the electric field effect based on orbital reconstruction should be limited to the one to two monolayers or unit cells. This is similar to the charge-mediated mechanism, which means that these two mechanisms are commonly combined together. For the samples sensitive to the carrier density variation such as manganites near the phase boundary, e.g., LSMO ($x$ = 0.2) and LCMO ($x$ = 0.5), the charge-mediated mechanism plays a dominant role. However, for the manganites in the middle of the FM phase region, e.g., LSMO ($x$ = 0.33), the contribution from the modulation of carrier density is relative small. For example, using a control experiment of BTO/LSMO ($x$ = 0.33) on (110) STO substrate without an interfacial Ti-O-Mn covalent bond, Cui *et al.* [41] demonstrated that the obvious electrical modulation of magnetoresistance in the case of a (001) heterostructure is mainly caused by interfacial orbital reconstruction rather than carrier density modulation.

In BFO-based multiferroic multilayers, the exchange coupling-mediated VCM

relies on the interfacial exchange bias field ($H_{EB}$) and the magnetostatic exchange length of the magnetic layer (in the range of 1 to 10 nm) [303]. When the thickness of the FM layer is below the magnetostatic exchange length, the atomic exchange interaction is dominant so that local magnetic moments (spins) show a tendency for parallel alignment along the direction of thickness. In consequence, the $H_{EB}$ of the whole magnetic layer on average decreases as the thickness increases. For instance, in LSMO/BFO heterostructures [304], the average $H_{EB}$ decreases to almost zero as the LSMO thickness increases to about 30 nm.

For a multiferroic heterostructure where more than two mechanisms coexist during the interfacial coupling, the contribution of each mechanism depends closely on the thickness of the magnetic layers. Early research on the dependence of the coupling between mechanisms of strain and charge on thickness can be referred to in the previous review [305]. Recent studies on ME coupling based on both strain and charge have been experimentally demonstrated in various multiferroic-based multilayers [71,306–309], e.g. $Fe_{0.5}Rh_{0.5}$ [71], $L1_0$-ordered FePt [307], $Ni_{0.79}Fe_{0.21}$ [306], Mn-doped ZnO [308], or the Fe thin film [309] deposited on PMN-PT.

Some studies also focus on the reason for the contradictory mechanisms in ionic liquid gating manipulation. Yuan *et al.* [179] described a “phase diagram” of the EDL gating effect to distinguish the contribution of the electrostatic and electrochemical mechanism: electrostatic doping is commonly found if the work frequency of the electric field is high and the operation temperature is low, while the electrochemical reaction plays a dominant role in the situation of a low-frequency electric field and a high temperature. In addition, Ge *et al.* [178] found that the water contaminated in the ionic liquid plays a vital role in gating experiments. A larger modulation rate can be obtained with the ionic liquid containing more water, while almost none is obtained with a baked ionic liquid (without water). Nevertheless, there are various factors that influence the EDL mechanism, i.e., the strain effect and the oxygen mobility, *etc.*, which call for many further investigations.

Overall, there are strong cooperation and competition between different mechanisms especially when the thickness of FM layer decreases to several nanometers. To figure out which mechanism is the dominant factor in different situations needs careful analysis. On the other hand, some charged defects, e.g. oxygen vacancies, introduce some extrinsic factors to the VCM, whose behaviors under electric-field might play a vital role in VCM and need to be clearly revealed in the microscopic point of view.

## 4. Applications of voltage control of magnetism in spintronics

In the continuing exploration for mechanisms under voltage control of magnetism, many practical devices in the field of VCM have been developed and more applications are on the fast track. Considering the evolution of the critical role of the electric field in VCM, typical performance characteristics are chosen to illustrate the trends in this development direction. Initially, electric field-assisted switching was widely used in the tunneling junction, which is the most basic unit in MRAM. With the assistance of an electric field, a prominent decrease in energy consumption in MTJ was obtained and four-state memory was also realized in the multiferroic tunnel junction (MFTJ), which is expected to accelerate commercial products with MRAM. Subsequently, electric field-driven magnetization reversal was popular and continuing efforts toward an innovative materials system and structure design have made it possible for magnetization switching to be purely electrically driven with no auxiliary magnetic field involved, advancing the progress of device miniaturization. More recently, electrical control of current-induced magnetization switching implies a dual modulation by voltage and current, which exhibits great potential to lower critical current density and speed up the practical application of three-terminal memory devices. So far, a relatively comprehensive understanding has been built about the applications of VCM and it emphasizes the underlying mechanisms of VCM at the same time.

## 4.1 Electric field-assisted switching in the tunneling junction

The rapid development in burgeoning Internet technologies, especially big data and cloud computing, calls for innovative storage technology so as to accommodate this urgent need. Among several new classes of nonvolatile memory technologies, MRAM has achieved great progress considering infinite endurance and low power consumption. Considering that MTJ is the most basic unit in the promising nonvolatile and high-density MRAM, an optimization of the MTJ with the assistance of an electric field is a feasible way to promote MRAM toward practical applications. Based on large amounts of efforts, low-power-consumption MTJ and four-state memory in the multiferroic tunnel junction (MFTJ) have been successfully achieved with the utilization of electric fields in some prototype tunneling junction devices.

### 4.1.1 Significant decrease in energy consumption

Spin-transfer torque allows for high-efficiency electric current control of magnetization in ferromagnetic nanostructures, especially in magnetic tunnel junctions, which are highly promising in practical applications for memory and logic devices. However, the current for magnetization switching in MTJ through the spin-transfer torque is still too large (around $10^6$–$10^7$ A $cm^{-2}$), causing an obstacle in further development. Hence, an approach to reduce the spin-transfer torque switching current and the energy consumption in MTJ is desirable.

Wang *et al*. [46] reported on electric field-assisted reversible switching in perpendicular magnetic anisotropic CoFeB/MgO/CoFeB magnetic tunnel junctions, and the gated MTJ device is shown in Fig. 29a. The current density used for spin-transfer torque-induced magnetization switching can be effectively reduced. In the MTJ structure, the thicknesses of the top and bottom CoFeB layers are both less than 1.6 nm, which enables the electric field to penetrate into the integrated films and modulate the ferromagnetic properties significantly. The TMR and coercivity dramatically depend on the relatively small bias voltages ($V_{bias}$) between the top and bottom CoFeB electrodes. Owing to the different thicknesses of the top and bottom CoFeB electrodes, the $H_C$ of these two layers are 12 mT and 2.5 mT at the ground state, respectively. When a voltage is applied to the MTJ, the $H_C$ varies, that is, the $H_C$ of the top and bottom CoFeB becomes 11.5 mT and 7.2 mT at $V_{bias}$ = 890 mV, and reduces to 13.7 mT and 2 mT at $V_{bias}$ = −890 mV, as shown in Fig. 29b. Both the $H_C$ of the top and bottom CoFeB varies monotonously with the applied bias voltage, although the variation tendencies are the opposite. Besides, the TMR can be modulated step by step under a specific bias pulse procedure.

In a similar experiment in thick CoFeB MTJs with in-plane anisotropy, the switching field and TMR show no observable difference between the case of the positive and negative bias voltage [46]. This observation demonstrates that the voltage control of magnetization switching in MTJ originates solely from the interfacial magnetic anisotropy. The study of magnetic anisotropy energy and the relative change under electric fields reveals that the perpendicular anisotropy energy linearly depends on the electric field and the slope of anisotropy energy change is ~50 fJ/V m. The energy barrier of the MgO layer is also greatly reduced under the bias voltage;

therefore, STT-induced switching can occur at a small current density with a magnitude of $10^4$ A $cm^{-2}$. This investigation of electric field-assisted switching in PMA CoFeB MTJs opens the path to ultralow energy STT-based memory and logic devices.

Though the current-induced switching magnetic fields and spin-transfer torque are generally used for switching the magnetization of MTJ, the large current density required would still inevitably cause metal migration and high-energy consumption in electronic devices. At the same time, switching energies of 40–80 fJ/bit have been obtained in electric field-controlled MTJ with perpendicular magnetic anisotropy [21–24,48,88,310–312]. However, the write energy values are still a few orders of magnitude larger than those for the transistors with the same size (~fJ), which are widely used in volatile semiconductor memory technologies in current integrated circuits.

Specifically, the energy consumption due to the Joule heating for the current-induced magnetization switching (CIMS) is expressed as $E_J^C = RI^2 t_{sw}$ [47], while the electric-induced magnetization switching (EIMS) is expressed as $E_J^E = (E_c t_{MgO})^2 t_{sw} / R$ [47], where *R*, *I*, $t_{sw}$, and $t_{MgO}$ are the MTJ's resistance, the current through the MTJ, the sustained time of magnetization switching, and the thickness of the MgO tunnel barrier, respectively. Hence, EIMS consumes less Joule energy when *R* is high, while in contrast, CIMS consumes more Joule energy. As a result, increasing the thickness of MgO provides a route to solve this problem.

Recently, Grezes *et al*. [48] demonstrated that the energy of EIMS fell to ~6 fJ/bit for switching times of 0.5 ns in the nanoscale MTJ with a core structure of CoFeB/MgO/CoFeB when the junction diameter decreases to 50 nm as shown in Fig. 30. More importantly, high resistance-area products with a thicker MgO layer, which reduce Ohmic dissipation, are the key to realize this low power consumption operation. A similar result was also acquired by Kanai *et al.* [47] where a small comparative energy of 6.3 fJ/bit could also switch the MTJ with thick MgO barrier layer of 2.8 nm. These two independent works illustrate that 1 order of magnitude improvement compared to previous reports [21–24,48,88,310–312] is made possible by a high resistance-area tunnel junction with very low Ohmic loss. Meanwhile, in MTJ a combination of spin-transfer torque and electric fields has been demonstrated to behave a lower switching time and energy than spin-transfer torque alone as well as a higher switching reliability than pure electric fields [313]. It is then concluded that electric fields helps low energy and high efficient switching by magnetic fields or current in the MTJ.

Following the extensive research on the modulation of magnetism by electric fields, VCMA coefficient was proposed in the junctions composed of heavy-metal/ferromagnet/insulator [87,314], providing a quantitative evaluation for the capability of VCM and switching energy. Here VCMA can be expressed as VCMA = $\beta E_I$ = $\beta E_{ext}/\varepsilon$, where $E_I$ and $\varepsilon$ are respectively the electric field and dielectric constant of the insulator, $E_{ext}$ denotes external electric field, while the VCMA coefficient $\beta$ offers the standard to judge the ability of VCM in the junctions via the magnetic anisotropy manipulation [315]. It is commonly accepted that to attain a switching energy < 1 fJ/bit, a write voltage < 1 V, and $\beta \geq$ 200 fJ/Vm [316] associated with a large PMA is required [10,317]. To achieve a larger VCMA, lots of attempts have been made in various systems. For the commonly used PMA Ta/CoFeB/MgO junction, a $\beta$ value of ~50 fJ/V m was attained by the use of suitable annealing process [313] or the enhancement of PMA with a butter layer [49]. Note that Cr/ultrathin Fe/MgO heterostructures show a strong PMA up to 2.1 mJ/m$^2$, accompanied by a remarkable increase of $\beta$ up to 290 fJ/V m [50]. The VCMA provides an effective way to the voltage-induced magnetization switching, whose energy consumption is comparable to same-sized CMOS transistors with the utilization of an electric field. The development of VCMA is bound to advance the nonvolatile memory technologies based on MTJ toward practical applications.

### 4.1.2 Promise for high-density storage via four-state memory

Electrical control of magnetism is also attempted in MFTJs, where a FE (or multiferroic) tunnel barrier is sandwiched between two FM electrodes or a FM and a normal metal [124,125,318–322]. Similar to TMR obtained in traditional MTJ, which is extremely sensitive to the spin-dependent electronic properties at the interface between the FM and dielectric layers, the interaction between FE and FM layers can also be studied through transport measurements in the MFTJ. Assisted by the electric field, four distinct resistance states resulting from the TMR as well as the tunnel electroresistance (TER) effects, as illustrated in Fig. 31a, are successfully achieved in MFTJ.

Four resistance states in a MFTJ of Au/$La_{0.1}Bi_{0.9}MnO_3$/LSMO were first reported by Gajek *et al.* [320], but no obvious variation of the TMR with the FE polarization was observed. Subsequently, artificial MFTJs were designed and a MFTJ of LSMO/BTO (1 nm)/Fe was applied by Garcia *et al.* [321]. It was demonstrated that a TMR of −19% and −45% could be detected in this structure with the FE layer polarized under downward and upward states, respectively (Fig. 31b). To be specific, the ferroelectric polarization of BTO was reversed under voltage pulses produced by conductive-tip atomic force microscopy and the TMR was then measured by sweeping the magnetic field and collecting the current under different ferroelectric states at 4.2 K. Considering that the amplitude of variation in TMR during the ferroelectric polarization switching can reach 450%, the MFTJ provides an efficient method to control the spin polarization of the tunnel current locally and it exhibits great potential in nonvolatile and low-consumption devices. Afterwards, even the sign of the TMR was reversed in the MFTJ of Fe/PZT (3.2 nm)/LSMO, and the TMR values were +4% and −3%, when the polarizations were pointed to LSMO and Co respectively (Fig. 31c) [124]. In spite of the unimpressive TMR value, the relative variation with the switching of ferroelectric polarization could reach 230%. Recently, by inserting a LCMO layer with a doping level close to the FM–AFM phase transition into MFTJ, the difference between TMRs at two polarization states could reach 500% [125]. A similar resistive switching caused by the doping level change and resulting phase transition under FE polarization switching has also been demonstrated in first-principles calculations [322]. Thus, it is the orientation of the ferroelectric tunnel barrier that manipulates the spin polarization of current and it makes the MFTJ a popular candidate in multivalued storage.

At the same time, the temperature windows for the realization of four-state memory in MFTJ have been gradually increased in the past few years. However, almost all the work related to ferroelectric control of spin polarization as mentioned above can only be performed below room temperature. It is clear that further efforts toward room-temperature operation of MFTJ should be vigorously pursued so as to accelerate its practical applications. In particular, the use of at least one LSMO ferromagnetic electrode limits the operating temperature [323]. Moreover, both TMR and TER of the MFTJ are limited at higher temperature in consideration of the thermally activated inelastic tunneling processes through defects in the ferroelectric tunnel barrier [125,324,325]. Although a four-state memory at room temperature with LSMO/$Ba_{0.95}Sr_{0.05}TiO_3$ (3.5 nm)/LSMO tunnel junctions was reported by Yin *et al.*

[326], the resistance contrasts were extremely small (<1%) and no apparent variation of the spin-dependent signal under different ferroelectric polarization orientations was observed. Thus, innovative systems with sole transition metals or their alloys applied as bottom and top electrodes should be investigated to realize an efficient ferroelectric control of spin polarization at room temperature in MFTJ.

So far, by the combination of magnetic and electric fields, the MFTJ devices exhibit robust four-state memory, which provides a promising route to the realization of high-density information storage. Considering that the MTJ and corresponding TMR are widely applied in the read head, sensors, and MRAM, the investigations of VCM in MTJ and MFTJ based on oxides have strong potential in the emerging field of low-power and nonvolatile devices in spintronics. Furthermore, they are bound to accelerate exciting improvements in application-oriented oxide electronics.

### 4.2 Electric field-driven magnetization reversal

Magnetic field-free switching is appealing in the development of modern storage technology because it will contribute to device miniaturization and the integration level. However, it is a great challenge to break the limits of 90° switching magnetization without the assistance of a magnetic field or spin-transfer torque. An alternative method to replace the external magnetic field is the introduction of an electric field [327,328]. The manipulation of magnetization reversal by a pure electric field rather than a magnetic field has been reported by some groups [51,52,59], which illuminates the direction of magnetic field-free switching in the future.

#### 4.2.1 Magnetization switching assisted by a small magnetic field

The strain-mediated magnetoelectric effect has been widely utilized in VCM through a strong coupling between ferroelectric and ferromagnetic phases. The concomitant electrical manipulation of magnetic anisotropy paves the way for magnetic field-free magnetization switching. In a Co/PMN-PT multiferroic heterostructure, the magnetic easy axis of the Co layer is rotated from 0° to 90° when the polarization state changes under poling electric fields of +8 kV/cm and –8 kV/cm as illustrated in Fig. 32a [59]. As a result, nonvolatile and reversible 90° rotation of magnetization was achieved without an external magnetic field at room temperature, which overcame the weakness of volatile 90° rotation of magnetization in the previous report [20]. Moreover, as displayed in Fig. 32b, a reversible 180° magnetization switching can be realized under a stimuli sequence of ±5 kV/cm together with a very small auxiliary field of 0.5 mT, and calculations according to phase-field simulations also supported this reversal. Based on the strain-mediated magnetization rotation and corresponding change of anisotropic resistance, nonvolatile memory devices can be designed by using an electric field instead of a magnetic field.

Another way to realize magnetization switching under electric fields is to import the exchange bias into the system. Voltage control of exchange bias mediated by piezoelectric strain was demonstrated in FeMn/NiFe/FeGaB/PZN-PT(011) heterostructures by Liu *et al.* [329]. The exchange bias set by antiferromagnetic FeMn and NiFe/FeGaB ferromagnetic layers was shifted up to 218% under the competition between the uniaxial anisotropy induced by an electric field and the unidirectional anisotropy via the magnetoelastic interaction. With the application of a suitable magnetic field, near 180° deterministic magnetization switching can be obtained [329]. A similar result was also reported by Xue *et al.* [327] and it implies that AFM/FM/FE multiferroic heterostructures are prototypes to tune exchange coupling by electric fields and constitute an important step toward a practical MRAM at room temperature.

### 4.2.2 Magnetization switching realized under no magnetic field

Despite the fruitful results that have been achieved in the electric field dominant magnetization switching, generally, an auxiliary magnetic field is still indispensable so as to fulfill the reversal of magnetization, because VCM does not break the time-reversal symmetry. Naturally, efforts continue to be devoted to pure modulation by an electric field and certain representative research has appeared.

Among the relevant reports [31,330,331], we can see that in an exchange-coupled heterostructure of CoFe/BFO [51], a 180° switching of in-plane net magnetization was driven purely by voltage. Moreover, due to the features of room-temperature switching and the perpendicular geometry of the voltage applied (Fig. 33a), it is appealing in those potential applications with high-density technical demands. Upon poling the BFO film with a PFM tip, 180° switching of polarization in BFO was realized and the resultant 180° switching of Dzyaloshinskii-Moriya vector at the surface further reversed the local in-plane magnetization [332]. Specifically, the reversal of exchange-coupled magnetic domain in the upper CoFe ferromagnetic layer was directly observed by XMCD and PEEM under zero magnetic field as shown in Fig. 33b. Combined with the ferroelectric loop (*P*–*V*) and resistance versus voltage (*R*–*V*) loops (left panel of Fig. 33c), it was demonstrated that the magnetization reversal is accompanied by the switching of the polarization. At the same time, a comparison between the *R*–*V* loops and resistance values from magnetoresistance curves (*R*–*H*) in the right panel of Fig. 33c illustrates a clear coupling between the FE poling and the magnetization switching. In addition, a comparable GMR signal obtained in the spin valve structure, highlighted by dotted lines as visual guides, verifies the success of the 180° magnetization switching under a single electric field.

Electric field-induced magnetization switching has also been proposed in magnetostrictive memory devices based on magnetic spin-valve multilayers grown on ferroelectric substrates [333]. In 2014, Li *et al.* [80] tuned the TMR of MTJ at a zero magnetic field with the magnetization of the CoFeB-free layer in MTJ controlled by voltage via the (011) PMN-PT substrate. Both the TMR and the electroresistance suggest that the electric field-induced strain can control the magnetization rotation of one FM layer in MTJ. More importantly, the TMR can be tuned via electric field-induced strain at 0 mT and the modulation can be up to 15%. Thus, for the first time, electrical manipulation of TMR at a zero magnetic field is realized at room temperature and this is significant for electric field-tuned spintronics with ultralow energy consumption.

At the same time, more attention has been paid to avoiding the use of magnetic fields in integrated devices. Full reversal of perpendicular magnetization purely driven by electric fields was predicted in a three-dimensional multiferroic nanostructure by phase-field simulations [63], which shows great potential for revolutionizing spintronic devices. Very recently, via the combination of exchange-biased systems and ferroelectric materials, Chen *et al.* [52] demonstrated a reversible electrical magnetization reversal in CoFeB/IrMn/PMN-PT multiferroic heterostructures without a biased magnetic field at room temperature. Noteworthily, it was predicted recently that a shyrmion core reversal controlled by voltages can be realized in the PMA CoFeB/MgO/CoFeB MTJ without an external magnetic field [334], offering an alternative to field-free switching via VCM. Also, the voltage control of interfacial Dzyaloshinskii-Moriya interaction was detected at the Au/Fe/MgO system through spin waves [335]. All these progress in the field undoubtedly accelerates the wider integration of electric fields in magnetic applications.

## 4.3 Electric field modulation combined with current in spintronics

### 4.3.1 Spin–orbit torque

In the past 5 years, spin–orbit torque (SOT), which is a new scheme for spin-transfer torque, has attracted the interest of many groups [198,336–339]. Spin–orbit torque can be observed in devices with only one ferromagnetic layer. A conventional device is based on multilayers where heavy metal (HM), a ferromagnetic layer, and oxide form a sandwich structure, as shown in Fig. 34a [340]. The HM could be Pt, Ta, Hf, and W or other heavy metals, which have a large spin-Hall angle. The ferromagnetic materials, which exhibit a perpendicular magnetic anisotropy (PMA), like ultrathin Co, CoFeB, or Co/Pt multilayers are often chosen as the FM layer [341–343]. The capping oxide layer can be MgO, $AlO_x$, $HfO_2$, and $Ta_2O_5$, to enhance PMA and protect the FM metal from air oxidation.

The Hall devices are often used in spin–orbit torque research. When the in-plane current is applied to the device, the spontaneous spin Hall effect (SHE) in HM will generate a spin-polarized current near the HM/FM interface by dividing the electrons up and down according to the spin direction. The spin current then diffuses into the FM layer, and the spins, which are perpendicular to the magnetic moment, can be absorbed by the FM layer, producing a spin torque to the magnetic moment. A small external field should be applied to break the symmetry and then realize deterministic switching. The spin–orbit torque, mainly originating from SHE, is called damping-like torque ($\boldsymbol{\tau}_{\mathrm{DL}}$), and the orientation of $\boldsymbol{\tau}_{\mathrm{DL}}$ is determined by the magnetic moment ($\boldsymbol{m}$) and the spin polarization direction ($\boldsymbol{\sigma}$): $\boldsymbol{\tau}_{\mathrm{DL}} = \tau_{\mathrm{DL}}^{0}(\boldsymbol{m} \times \boldsymbol{\sigma} \times \boldsymbol{m})$, where $\tau_{\mathrm{DL}}^{0}$ is the parameter proportional to current density, representing the efficiency with which HM excites the spin current and produces damping-like torque. The current-induced magnetization switching is shown in Fig. 34b [341].

Moreover, when there exists an inversion asymmetry between the top and bottom surfaces of the ultrathin FM layer, the motion of electrons is affected by the Rashba field that can be viewed as an equivalent magnetic field. In FM materials, the *s*-*d* exchange interaction and the Rashba field are involved in spin–orbit coupling of conducting electrons. When the current flows through the FM layer, it will generate a torque to switch the orientation of the magnetic moment. The torque, mainly originating from the interfacial Rashba effect, is called field-like torque ($\boldsymbol{\tau}_{\mathrm{FL}}$). Similar to damping-like torque, the direction and magnitude of $\boldsymbol{\tau}_{\mathrm{FL}}$ is given by $\boldsymbol{\sigma}$ and $\boldsymbol{m}$: $\boldsymbol{\tau}_{\mathrm{FL}} = \tau_{\mathrm{FL}}^{0}(\boldsymbol{m} \times \boldsymbol{\sigma})$, where $\tau_{\mathrm{DL}}^{0}$ is the efficiency where current induces field-like torque.

Although it is possible to apply an external field to devices by fabricating magnetic stacks, there are many disadvantages such as fabrication difficulties, low integration density, and stability. The realization of field-free SOT switching is of great interest to SOT applications [344]. There are three main methods to realize field-free SOT switching: i) antiferromagnetic IrMn, PtMn [345,346]; ii) wedged film [347]; and iii) interlayer exchange coupling [348]. SOT is now investigated in magnetization reversion [198,336–339], domain wall motion [349–351], and nano oscillators [338,352]. Compared to STT, SOT-induced magnetization switching requires lower critical current density and energy dissipation when writing and wiping data in the device. Thus, SOT is promising in the application of spintronics memory devices with high speed and low power consumption, as well as in logic-in-memory devices.

### 4.3.2 Electric field control of spin–orbit torque

In the field of spin–orbit torque, the effective electric field helps reduce the power consumption of the practical spin–orbit torque devices. The realization of voltage control of SOT has also been reported by several groups [56,57,353,354].

In Pt/Co bilayers with an $Al_2O_3$ insulator that isolates the Au gate electrode, the current-induced effective field can be modulated by 4.3% when the applied electric field reaches 2.8 MV $cm^{-1}$ [56]. The measurements are mainly conducted using a heterodyne detection of different harmonics in the Hall voltage. By analyzing the variation of first and second harmonic Hall signals, the effects of the electric field on the magnetic anisotropy and the current-induced spin–orbit torque are demonstrated. The electric field modulation of a spin–orbit effective field is eight times larger than the variation in magnetic anisotropy generated by direct electric gating, indicating dominant role of the electrical control of SOT for the variation of critical switching current. The large electric field effect on the spin–orbit torque comes from the strong spin–orbit interaction in the Pt/Co/$Al_2O_3$ structure. Although the current-induced electric field mainly originates from the spin Hall effect of Pt and the spin–orbit interaction, the Rashba effect makes a considerable contribution to the spin–orbit torque. Considering that the Rashba effect comes from the inversion asymmetry of the top and bottom interface of the Co layer, the screening effect of a 0.6 nm thick Co layer avoids the electric field effect from penetrating into the bottom Pt/Co interface. Thus, the electrical modulation of spin–orbit torque in this system is due to the affected Co/$Al_2O_3$ interface and the interfacial Rashba effect.

Besides the metallic system, an effective modulation of spin–orbit torque has been proven in the system of a topological insulator (TI). In a Cr-doped TI (Cr-TI), the electric field can modulate the spin–orbit torque strength by a factor of four [57]. The magnetization switching can be realized by sweeping the current as well as the electric field, resulting in a change of the magnetic anisotropy and the topological surface state. Although a TI usually shows a low Curie temperature, the voltage control of SOT by changing the electronic structure opens the way to a modulation of spin–orbit torque in ferromagnetic systems. Using ionic liquid gating, a large electric field can be built within a relatively small gate voltage, which enables a strong modulation effect of the electric field in the metallic system, despite the short screen length. Yan *et al*. [58] demonstrate that in a perpendicularly magnetized Pt/Co/$HfO_2$ structure, the critical switching current density can be modulated by a factor of two at a practical temperature (180 K), as shown in Fig. 35. The electric field effect of magnetic anisotropy is the minor reason for the modification effect because it changes less than 10% within the range of applied gate voltages. Via the first and second harmonic Hall measurements, it was found that the field-like torque has a negligible contribution to current-induced magnetization switching and the damping-like torque, which mainly originates from SHE of Pt dominating the current-induced torque. Moreover, there is a considerable electric field modulation on the damping-like torque and the effective spin Hall angle. Thus, the electric field effect on the effective spin-Hall angle of Pt/Co systems mainly contributes to the voltage control of critical switching currents.

These works combine the current-driven method with voltage control in the magnetization switching process, which may advance the understanding of fundamental spin–orbit torque theory and stimulate more research interest in the spin–orbit torque phenomenon. Moreover, the dual application of the electric field and the current is compatible with conventional semiconductor technology. Progress toward low-power consumption and practical memory with logic devices can be accelerated as the critical current density is significantly reduced with the utilization of VCM.

## 5. Conclusion and Outlook

In the review article, we tended toward presenting the latest developments in the area of voltage control of magnetism, whereas it is a rather difficult task to encompass all the points of interest in this rapidly growing field. From the perspective of materials, three kinds of magnetic materials (metals, semiconductors, and oxides) and three representative dielectric gating materials (normal dielectrics, ferroelectrics, and electrolytes) were discussed, associated with their special contributions to VCM performance. The combination of magnetic materials and dielectric gating materials produces four device configurations to realize VCM. Besides the conventional mechanisms on charge, strain, and exchange coupling at the interfaces of heterostructures, the investigation scenarios on orbital reconstruction and electrochemical effect were also elucidated. As a promising method for reducing power consumption in memory technologies, several devices are designed to reflect VCM performance, including electric field-assisted switching in the tunneling junction, (pure) electric field-driven magnetization reversal, and electrical control of spin–orbit torque. Thanks to the efforts made by researchers worldwide focusing on VCM over the past decades, the developments in this area truly stand out and are exciting. Nevertheless, it should be noted that this area is still quite definitely in the infant stage, and a large number of open questions remain to be answered, with respect to both the mechanisms and the practical applications:

(1) The mechanisms of VCM still need to be clarified in detail and extended further. The coupling between different ME mechanisms is a pivotal issue when the thickness of thin films decreases to several nanometers. The cooperation and competition between different mechanisms should be revealed with more evidence of the microscopic nature. It is also important to answer questions concerning which mechanism is the dominant factor in different situations.

(2) The practical applications of VCM require further investigation into: i) enhancement of operating temperature upon room temperature for practical systems; ii) decreasing the switching voltage to a level far below the breakdown threshold of magnetic tunnel junction; iii) switching the magnet (Φ = 10~30 nm) with enough thermal stability ($\Delta = K_u V/k_B T > 60$, where $K_u$ and $V$ are anisotropy constant and volume, respectively, $k_B$ and $T$ parameter Boltzman's constant and temperature, respectively); iv) device preparation and integration for VCM; v) reducing the error rate down to $10^{-15}$.

(3) How can the influence of voltage on magnetic properties be optimized and enhanced? In general, the insulating oxide should be very thick to solve the leakage current problem in the traditional semiconductor industry, but this will significantly reduce the effective electric field. Ionic liquid gating provides a promising way for generating a very large electric field at the interface, whereas its integration in the traditional semiconductor industry needs to be considered and its operation speed should be enhanced.

(4) Finding a room-temperature route to realizing a purely voltage-controlled 180 ° reversal of magnetization in experiment without the assistance of an external field is significant for the practical application of VCM. Based on some prototypical results in single FM layers, this idea needs to be further promoted in microelectronic devices with practical applications.

(5) Electric fields provide a promising way to dramatically reduce current density for the realization of magnetization switching, in both spin-transfer torque and spin–orbit torque-based devices. This is one of the most fascinating research areas in spintronics with strong industrial background.

(6) Charged oxygen vacancies are inevitable in FE, FM oxides, and barrier materials,

which can be modulated by the external voltage to have some influence on magnetic properties. Thus, the role and behavior of charged oxygen vacancies in VCM needs to be clearly revealed in the future. In addition, the behaviors of other defects at the interface with electric field switching also need to be clarified. It is vital to overcome the obstacle of polarization fatigue in practical devices under repetitive electric cycles.

(7) Associated with the emerging research directions in magnetism community, such as skymions and Dzyaloshinskii-Moriya interaction, as well as yet unexplored voltage controlled behaviors, e.g., interlayer coupling, VCM will have a more broad platform to show its flexibility, bringing about unprecedented physical phenomena and extensive application potential.

In the end, we hope that this article can inspire further investigations into voltage control of magnetism and promote developments in this area.

**Acknowledgements**

We thank Y. N. Yan's contribution to this work on electric field control of spin–orbit torque. The authors are grateful to Dr. S. Miwa of Osaka University, Dr. W. G. Wang of University of Arizona, Prof. N. Kioussis of California State University Northridge, X. Li of UCLA for fruitful discussions. The authors acknowledge the support from Beamline BL08UIA in the Shanghai Synchrotron Radiation Facility. This work was supported by the National Key Research and Development Program [grant number 2016YFA0203800], the National Natural Science Foundation of China [grant numbers 51671110, 51231004, 51571128, 51601006, 51322101, and 51202125] and the National High Technology Research and Development Program of China [grant numbers 2014AA032901, and 2014AA032904].

Table 1. Summary of the metal-based heterostructures used for VCM with corresponding device configurations, performances (coupling coefficient $\alpha$ and operation temperature $T$), coupling mechanisms, and results by the voltage control.

| System | Device | $10^8\alpha$ (s/m) | $T$ (K) | Mechanism | Results | Reference |
|---|---|---|---|---|---|---|
| MgO/Fe | FET | - | 300 | Orbital | MA | [23] |
| MgO/Fe | FET | - | Theory | Orbital | MA | [72] |
| $ZrO_2$/MgO/Fe | FET | - | 300 | Charge | MA | [355] |
| Fe, Co | - | - | Theory | Orbital | MA & $M$ | [35] |
| Fe, Co, Ni | - | - | Theory | Orbital | MA | [356] |
| BTO/Fe | FET | 0.3–1.6 | Theory | Charge | $M$ | [190,216] |
| BTO/Fe | FET | - | Theory | Orbital | $M$ | [273,276] |
| Fe/BTO | BG | - | 300 | Strain | $H_C$ | [19] |
| BTO/Fe | FET | 7.3 | Theory | Charge | $M$ | [190] |
| (Co)/Fe/BTO | BG | 0.2 | 300 | Orbital | $M$ | [299] |
| $SiO_2$/MgO/FeCo | FET | - | 300 | Electrochemistry | $M$ | [73] |
| Fe/MgO/FeCo | MTJ | - | 300 | Charge | MA & MR | [87] |
| Co/PMN-PT | BG | - | 300 | Strain | MA & $M$ | [59] |
| Co/BTO/Pt | FET | 0.4 | Theory | Charge | $M$ | [216] |
| $GdO_x$/Co | FET | - | 300 | Electrochemistry | $H_C$, MA, & $M$ | [42,43,180] |
| $GdO_x$/Co | FET | - | 300 | Electrochemistry | Domain | [74] |
| P(VDF-TrFE)/Co | FET | - | 300 | Charge | $H_C$ | [357] |
| Pt/Co/$Cr_2O_3$(0001) | BG | - | 250–290 | EB | EB, $H_C$, & $M$ | [60] |
| [Co/Pt]/$Cr_2O_3$(111) | BG | - | 150–250 | EB | EB, $H_C$, & $M$ | [81,82] |
| [Co/Pd]/$Cr_2O_3$(0001) | BG | - | 303 | EB | EB | [83,84] |
| $HfO_2$/Co/Pt | FET | - | 295 | Charge | Domain | [358] |
| $AlO_x$ or $TaO_x$/Co/Pt | FET | - | 300 | Charge | Domain | [359] |
| EMI-TFSI/MgO/Co/Pt | FET | - | 220–380 | Charge | $T_C$, $H_C$, & $M$ | [61] |
| DEME-TFSI/$HfO_2$/FeRh | FET | - | 100–350 | Electrochemistry | M, & $T_C$ | [360] |
| DEME-TFSI/$HfO_2$/FeMn/[Pt/Co] | FET | - | 10–300 | Electrochemistry | EB, $H_C$, & MR | [69] |
| DEME-TFSI/$HfO_2$/IrMn/[Pt/Co] | FET | - | 10–300 | Electrochemistry | EB, $H_C$, & MR | [70] |
| DEME-TFSI/$HfO_2$/Ni/Co | FET | - | 50–350 | Electrochemistry | $H_C$ | [22] |
| DEME-TFSI/$HfO_2$/[Ni/Co] | FET | - | 10 | Electrochemistry | EB, $H_C$ | [62] |
| Ni/BTO | BG | | 120–350 | Strain | Domain & $M$ | [79] |
| Ni/BTO | BG | 0.05 | 300 | Strain | $M$ & MA | [76] |
| Ni/BTO | FET | - | Theory | Orbital | $M$ | [276] |
| Ni/BTO/Pt | FET | 1.5 | Theory | Charge | $M$ | [216] |
| [Cu/Ni]/BTO | BG | 60 | 300 | Strain | MA | [361] |
| Polyimide/MgO/CoFe | FET | - | 300 | - | MA & $H_C$ | [317] |
| CoFe/BFO | BG | - | 300 | EB & Strain | $M$ | [20] |
| CoFe/(Cu/CoFe)/BFO | BG | 10–30 | 300 | Domain | $M$ & MR | [51] |
| CoFe/BFO/SRO/PMN-PT | BG | - | 100–300 | Strain | EB | [230] |
| CoFe/PZT ($x$ = 0.48) | BG | - | 300 | Strain | $M$ | [75] |
| IrMn/Co/Cu/CoFeB/MgO/PZT ($x$ = 0.5) | BG | - | 300 | Strain | MR & $H_C$ | [362] |
| NiFe/YMO(0001) | BG | - | 2–100 | EB | EB, $H_C$, $M$, & MR | [85] |
| Fe-Ga/BTO | BG | - | 300 | Strain | Domain | [78] |
| FeRh/BTO | BG | 1600 | 325–400 | Strain | FM-AFM & $M$ | [71] |
| FeRh/BTO | BG | - | 300–400 | Strain | MR & FM-AFM | [250] |
| FeRh/PMN-PT | BG | - | 300–423 | Strain | MR & FM-AFM | [86] |
| NaOH/FePt or FePd | FET | - | 300 | Charge | MA & $H_C$ | [18] |
| $ZrO_2$/MgO/CoFeB | FET | - | 300 | - | MA & $H_C$ | [21] |
| CoFeB/PMN-PT | BG | - | 300 | Strain | FMR | [231] |
| CoFeB/PMN-PT | BG | - | 300 | Strain | MA & $M$ | [363] |
| CoFeB/PMN-PT | BG | 200 | 300 | Strain | MA & $M$ | [77] |
| CoFeB/IrMn/PMN-PT | BG | - | 300 | Strain | EB & $M$ | [52] |
| CoFeB/MgO/CoFeB | MTJ | - | 300 | Charge | MA & MR | [88] |
| CoFeB/MgO/CoFeB | MTJ | - | 300 | - | $H_C$ & MR | [46] |
| CoFeB/$AlO_x$/CoFeB/PMN-PT | MTJ | - | 300 | Strain | MA & MR | [80] |
| BTO/$Co_2$MnSi | - | - | Theory | Charge & Strain | $M$ | [278] |
| $Co_2FeSi$/BTO | - | - | Theory | Orbital | DOS | [280] |

Table 2. Summary of the semiconductor-based heterostructures used for VCM with corresponding device configurations, performances (coupling coefficient $\alpha$ and operation temperature $T$), coupling mechanisms, and results by the voltage control.

| System | Device | $10^8\alpha$ (s/m) | $T$ (K) | Coupling | Results | Reference |
|---|---|---|---|---|---|---|
| (Ga,Mn)N/p-GaN/n-GaN | FET | - | 300 | Charge | $M$ & MR | [364] |
| (Ga,Mn)As | FET | - | Theory | Charge | MA | [96] |
| (Ga,Mn)As | FET | - | Theory | Charge | MA & $M$ | [97] |
| $Al_2O_3$/(Ga,Mn)As | FET | - | ≤70 | Charge | $H_C$ & $T_C$ | [98] |
| $Al_2O_3$, $ZrO_2$, | FET | - | ≤140 | Charge | MR & $T_C$ | [99] |
| $Al_2O_3$, $HfO_2$/(Ga,Mn)As | FET | - | ≤90 | Charge | MR & $T_C$ | [100] |
| $SiO_2$/(Ga,Mn)As | FET | - | 10–50 | Charge | MR, $T_C$, & $M$ | [16] |
| $ZrO_2$/(Ga,Mn)As | FET | - | 97 | Charge | MR, MA, & $M$ | [101] |
| $ZrO_2$/(Ga,Mn)As | FET | - | 2 & 5 | Charge | MR, MA, & $M$ | [102] |
| $HfO_2$/(Ga,Mn)As | FET | - | ≤50 | Charge | $T_C$ & $M$ | [33] |
| $HfO_2$/(Ga,Mn)As | FET | - | <80 | Charge | MR, $H_C$, $T_C$, & $M$ | [103] |
| KClO4-PEO/(Ga,Mn)As | FET | - | <200 | Charge | MR & $T_C$ | [104] |
| $ZrO_2$/(Ga,Mn)Sb | FET | - | ≤30 | Charge | $T_C$ & MR | [365] |
| Polyimide/(In,Mn)As | FET | - | 5–50 | Charge | MR, $H_C$, & $T_C$ | [16] |
| $SiO_2$/(In,Mn)As | FET | - | 40 | Charge | $H_C$ | [195] |
| $SiO_2$/(In,Mn)As | FET | - | 51 | Charge | MR & $H_C$ | [106] |
| SiN/$Mn_xGe_{1-x}$ | FET | - | 50 | Charge | MR | [107] |
| PZT ($x$ = 0.8)/Co:$TiO_2$ | FET | - | ≤400 | Charge | $M$ & $H_C$ | [108] |
| DEME-TFSI/Co:$TiO_2$ | FET | - | 300 | Charge | MR | [109] |
| Co:ZnO | FET | - | 300 | Charge | $H_C$ & $M$ | [110] |
| $SiO_x$/Co:ZnO | FET | - | ≤10 | Charge | MR | [111] |
| Fe:ZnO | FET | - | 300 | Charge | $H_C$ & $M$ | [366] |
| DEME-TFSI/MnBiTeSe | FET | - | 2 & 20 | Charge | MR | [112] |
| (Cd,Mn)Te/(Cd,Zn,Mg)Te | FET | - | <3 | Charge | $T_C$ & $M$ | [113] |
| $(BiSb)_2Te_3$/$Cr(BiSb)_2Te_3$ | FET | - | ≤10.4 | Charge | MR, $H_C$, & $T_C$ | [114] |

Table 3. Summary of the oxide-based heterostructures used for VCM with corresponding device configurations, performances (coupling coefficient $\alpha$ and operation temperature $T$), coupling mechanisms, and results by the voltage control.

| System | Device | $10^8\alpha$ (s/m) | $T$ (K) | Coupling | Results | Reference |
|---|---|---|---|---|---|---|
| BFO/LCMO ($x$ = 0.5) | FET | - | ≤300 | Charge | $M$ | [214] |
| LCMO ($x$ = 0.3)/PMN–PT | BG | 3.4 | 10 & 50 | Strain | $M$ | [129] |
| LCMO ($x$ = 0.3)/PMN–PT | BG | - | 210 | Strain | $T_C$ & $M$ | [26] |
| EMIM-TFSI/LCMO ($x$ = 0.2) | FET | - | 50–250 | Charge | $T_C$ | [169] |
| LSMO ($x$ = 0.3)/PMN–PT | BG | 6 | 330 | Strain | $T_C$ & $M$ | [26] |
| PZT($x$ = 0.8)/LSMO ($x$ = 0.2) | FET | 0.08–0.62 | 100 | Charge | $T_C$ & $M$ | [27,34] |
| PZT/LSMO ($x$ = 0.2) | FET | - | 50–300 | Charge | $T_C$ | [172] |
| PZT ($x$ = 0.8)/LSMO ($x$ = 0.2, 0.33, 0.5) | FET | - | ≤350 | Charge | $T_C$ & $M$ | [123] |
| LSMO ($x$ = 0.33)/BTO | BG | 2–23 | 200–300 | Strain | MA & $M$ | [64] |
| BTO/LSMO ($x$ = 0.33) | FET | - | 200–370 | Charge | $M$ | [65] |
| BTO/LSMO ($x$ = 0.33) | FET | - | 10–400 | Orbital | $T_C$ & MR | [41] |
| LSMO ($x$ = 0.175)/PZT ($x$ = 0.8) | BG | - | 4.2–300 | Orbital | $M$ | [283] |
| BFO/LSMO ($x$ = 0.3) | FET | - | 5.5 | EB | EB & $H_C$ | [28,258] |
| DEME-TFSI/LSMO ($x$ = 0.2) | FET | - | ≤400 | Electrochemistry | $T_C$ | [178] |
| DEME-TFSI/LSMO ($x$ = 0.4) | FET | - | 10 | Electrochemistry | MR, $H_C$, & $T_C$ | [121] |
| DEME-TFSI /LSMO ($x$ = 0.54) | FET | - | 10 | Electrochemistry | MA | [122] |
| PCMO ($x$ = 0.4)/PMN-PT | BG | - | 30 & 10 | Strain | $M$ | [251] |
| BFO/LSMO ($x$ = 0.25) | FET | - | Theory | Charge | $M$ | [213] |
| Fe/BTO/LSMO ($x$ = 0.33) | MTJ | - | 4.2 | Charge | MR | [321] |
| Au/LBMO/LSMO ($x$ = 0.33) | MTJ | - | 4 | Charge | MR | [320] |
| Co/BTO/LSMO ($x$ = 0.33) | MTJ | - | 10 | Charge | MR | [319] |
| Co/PZT ($x$ = 0.8)/LSMO ($x$ = 0.3) | MTJ | - | 10 & 50 | Charge | MR | [124] |
| LSMO ($x$ = 0.3)/BTO/LCMO ($x$ = 0.5)/LSMO ($x$ = 0.3) | MTJ | - | 80 | Charge | MR | [125] |
| LSMO ($x$ = 0.3)/BFO/LSMO ($x$ = 0.3) | MTJ | - | 80 | Charge | MR | [324] |
| Co/BTO/LSMO ($x$ = 0.3) | MTJ | - | 10 | Charge | MR | [192] |
| LSMO ($x$ = 0.3)/$Ba_{0.95}Sr_{0.05}TiO_3$/LSMO ($x$ = 0.3) | MTJ | - | 300 | Charge | MR | [326] |
| $CrO_2$/BTO /Pt | - | 1 | Theory | Charge | $M$ | [216] |
| $Fe_3O_4$/ BTO | - | 2 | Theory | Charge | DOS | [131] |
| $Fe_3O_4$/PMN–PT | BG | 6.7 | 300 | Strain | MA & $M$ | [132] |
| $Fe_3O_4$/PZT | BG | - | 300 | Strain | MA | [132] |
| $Fe_3O_4$/PZT | BG | - | 300 | Strain | MA | [133] |
| $Fe_3O_4$/PZN–PT | BG | 10.8 | 300 | Strain | MA & $M$ | [132] |
| $Fe_3O_4$/CFO/PZT ($x$ = 0.47) | BG | 3.3 | 300 | Strain | $M$ | [134] |
| PZT/$Ni_{0.23}Fe_{2.77}O_4$/PZT | - | - | 300 | Strain | $M$ | [135] |
| $Zn_{0.1}Fe_{2.9}O_4$/PMN–PT | BG | 2.3 | 300 | Strain | MA & $M$ | [136] |
| CFO/PMN-PT | BG | 3.2 | 300 | Strain | MA & $M$ | [137] |
| CFO/BTO | BG | - | 250–350 | Strain | MA | [138] |
| CFO-BFO | Nano. | 10 | 300 | Strain | $M$ | [139] |
| CFO-PZT ($x$ = 0.47) | Nano. | 40 | 300 | Strain | $M$ | [140] |
| BTO/NFO/BTO | - | - | 300 | Orbital | $M$ | [286] |
| SRO/STO/SRO | MTJ | - | Theory | Charge | $M$ | [212,215] |

Table 4. Summary of the dielectric materials used for VCM with corresponding category and performances, i.e. dielectric constant ($\kappa$), ferroelectric Curie temperature ($T_C$), saturated polarization ($P_S$), and capacitance per unit area ($C$).

| Materials | Category | $\kappa$ | $T_C$ or $T_M$ | $P_s$ (μC/cm$^2$) | $C$ (μF/cm$^2$) | Reference |
|---|---|---|---|---|---|---|
| MgO | Dielectric | 9.9 | - | - | - | [168] |
| $Al_2O_3$ | Dielectric | 8.9–11.1 | - | - | - | [152] |
| $SiO_2$ | Dielectric | 3.9 | - | - | - | [152] |
| $Si_3N_4$ | Dielectric | 7 | - | - | - | [152] |
| $Cr_2O_3$ | Dielectric | 11.9–13.3 | - | - | - | [367] |
| $ZrO_2$ | Dielectric | 23 | - | - | - | [152] |
| $HfO_2$ | Dielectric | 25 | - | - | - | [152] |
| $Ta_2O_5$ | Dielectric | 22 | - | - | - | [152] |
| STO | Dielectric | 2000 | - | - | - | [152] |
| Polyimide | Dielectric | 2.8–3.5 | - | - | - | [368] |
| BTO | FE | 1200–10000 | 393 | 16–25 | - | [369,370] |
| PTO | FE | 200 | 763 | 35–75 | - | [370–372] |
| YMO | FE | 20 | 913 | 5.5–6.2 | - | [160,370,373] |
| BFO | FE | 87–400 | 1103 | 65 | - | [374–376] |
| PZT | FE | ~500–5000 | 663 | 85 | - | [376,377] |
| (L)BMO | FE | 25–200 | 750–770 | 12 | - | [378] |
| PMN-PT | FE | 2000–4000 | 112–202 | 58 | - | [379] |
| DEME-TFSI | Electrolyte | 14.5 | 180 | - | 9.2 | [174] |
| EMIM-TFSI | Electrolyte | 12.25 | 258 | - | 10 | [169,380,381] |
| $KClO_4$-PEO | Electrolyte | ~10 | 220 | - | 7.4 | [380,382] |
| TMPA-TFSI | Electrolyte | ~10 | 295 | - | ~10 | [383,384] |
| $GdO_x$ | Electrolyte | 14 | - | - | - | [385] |

Table 5. Comparison of five different mechanisms. The conclusions are appropriate to most cases.

| Mechanism | Device | Thickness (nm) | Orientation | Dielectric layer | Magnetic layer |
|---|---|---|---|---|---|
| Charge | FET & MTJ | $10^{-1}$–$10^{0}$ | Any | Ferroelectric & Dielectric | Metals, Semiconductors & Oxides |
| Strain | BG & Nano. | $10^{1}$–$10^{6}$ | Any | Piezoelectric | Metals & Oxides |
| Exchange coupling | FET & BG | $10^{0}$–$10^{1}$ | Any | Multiferroic | Metals & Oxides |
| Orbital | FET & BG | $10^{0}$ | (001) | Ferroelectric | Metals & Oxides |
| Electrochemistry | FET | $10^{0}$–$10^{1}$ | Any | Ionic Liquid & $GdO_x$ | Metals, Semiconductors & Oxides |

**Figure Captions**

**Fig. 1.** Publications per year on voltage control of magnetism according to Web of Science: http://apps.webofkonwledge.com/.

**Fig. 2.** Different device configurations for VCM: (a) field effect transistor, FET type, (b) back gate, BG type, (c) magnetic tunnel junction, MTJ type, and (d) nanostructure, Nano type.

**Fig. 3.** Sketches of the different magnetic properties expected in VCM (a) magnetic anisotropy, (b) coercivity, (c) magnetization, (d) exchange bias, (e) Curie temperature, and (f) magnetoresistance.

**Fig. 4.** (a) Schematic of a VCM device using an ionic liquid as an insulating material. Reproduced with permission [70]. Copyright 2015, Wiley. (b) Sketches of the operating principle for EDL with a positive gate voltage ($V_G > 0$, left column) and a negative gate voltage ($V_G < 0$, right column). Reproduced with permission [32]. Copyright 2016, Institute of Physics.

**Fig. 5.** Schematic of the different magnetic responses to an electric field in heterostructures with different mechanisms: (a) charge, (b) strain, (c) exchange coupling, (d) orbital, and (e) electrochemistry.

**Fig. 6.** (a) Schematic graph of the electrolytic cell containing a FePt or FePd film. (b) Electric field effect on the coercivity in FePt. Reproduced with permission [18]. Copyright 2007, AAAS. Electric field effect on the (c) Hall resistance of Co/Pt and (d)

Curie temperature of the Co/Pt double layer. Reproduced with permission [181]. Copyright 2011, Nature Publishing Group.

**Fig. 7.** (a) Schematic graph of VCM in an (In,Mn)As FET. (b) The dependence of $R_{Hall}$ on the magnetic field under different gate voltages. Reproduced with permission [16] Copyright 2000, Nature Publishing Group.

**Fig. 8.** (a) The electric field-dependent Kerr rotation in PZT ($x$ = 0.2)/LSMO ($x$ = 0.2) at 100 K. Reproduced with permission [27]. Copyright 2009, Wiley. (b) XANES curves of PZT ($x$ = 0.2)/LSMO ($x$ = 0.2) with different polarizations. The inset is the dependence of the absorption at 6549.7 eV on the electric field. Reproduced with permission [34] Copyright 2010, American Physical Society. (c) EELS of the Mn $L_{2,3}$ edge in a LSMO/PZT/LSMO heterostructure. Reproduced with permission [205]. Copyright 2015, Nature Publishing Group. (d) Schematic model of the spin configurations in LSMO ($x$ = 0.2) at the PZT ($x$ = 0.2) interface under different polarizations. Reproduced with permission [34]. Copyright 2010, American Physical Society.

**Fig. 9.** The temperature-dependent resistivity and magnetization curves for LSMO at doping levels (a) $x$ = 0.20 and (b) $x$ = 0.33 with and without $Pb(Zr_{0.2}Ti_{0.8})O_3$ capping layers. (c) The calculated depth profile of the modulation of $e_g$ electronic density under different polarizations. Reproduced with permission [123]. Copyright 2013, American Chemical Society.

**Fig. 10.** (a) The voltage-controlled magnetoresistance in LAO/STO. Reproduced with permission [386]. Copyright 2010, American Physical Society. (b) MFM frequency images for LAO/STO under different gate voltages. Reproduced with permission [220]. Copyright 2014, Nature Publishing Group.

**Fig. 11.** (a) Heterostructure and test setup schematics for $CaRuO_3$/$CaMnO_3$. (b) Spin asymmetry for samples with a bias voltage of 0 and –400 V, respectively. Reproduced with permission [227]. Copyright 2015, American Physical Society.

**Fig. 12.** (a) The loop-like electrical modulation of magnetization and the corresponding polarization current recorded synchronously. (b) Schematic of the polarization orientations for (001) PMN-PT. (c) Correlation between domain switching and distortion. Reproduced with permission [77]. Copyright 2012, American Physical Society.

**Fig. 13.** (a) Sketch for a FeGaB/PZN-PT heterostructure. (b) Magnetization of FeGaB/PZN-PT with various electric fields. Reproduced with permission [242]. Copyright 2009, Wiley.

**Fig. 14.** (a) The dependence of FE polarization on external voltage for FeRh/BTO. (b) The magnetization as a function of temperature for FeRh/BTO under various bias voltages. Reproduced with permission [71]. Copyright 2014, Nature Publishing Group. (c) Electric field-dependent $\Delta\rho/\rho_{min}$ and net polarization. (d) The strain of FeRh (left axis) and PMN-PT (right axis) along their crystalline axes at different electric fields. Reproduced with permission [86]. Copyright 2015, Nature Publishing Group.

**Fig. 15.** (a) In-plane piezoelectric strain as a function of the electric field in a PMN-PT crystal. (b) Magnetization as a function of the electric field for a LSMO ($x$ = 0.3)/PMN-PT (001) heterostructure. Reproduced with permission [26]. Copyright 2007, American Physical Society.

**Fig. 16.** (a) Ferromagnetic resonance absorption and (b) magnetization for $Fe_3O_4$/PZN–PT under different electric fields. Reproduced with permission [132]. Copyright 2009, Wiley.

**Fig. 17.** (a) Magnetization of $[Co (0.6 nm)/Pd (1.0 nm)]_3$/Pd (0.5 nm)/$Cr_2O_3$(0001) at $T$ = 303 K. Reproduced with permission [83]. Copyright 2010, Nature Publishing Group. (b) Magnetization loops of NiFe/YMO/Pt at 2 K with different external voltages. Reproduced with permission [85]. Copyright 2006, American Physical Society.

**Fig. 18.** Voltage control of exchange bias in a BFO/LSMO ($x$ = 0.3) heterostructure. Reproduced with permission [28]. Copyright 2010, Nature Publishing Group.

**Fig. 19.** (a) The sketch for voltage control of an IrMn exchange spring. (b) The $H_{EB}$ and $H_C$ with varying $V_G$. Reproduced with permission [70]. Copyright 2015, Wiley.

**Fig. 20.** (a) Magneto-optical Kerr ellipticity $\eta_k$ for MgO/Fe under different gate voltages. (b) Sketch of the influence of the external voltage on the orbital occupancy of Fe. Reproduced with permission [23]. Copyright 2009, Nature Publishing Group.

**Fig. 21.** (a) DOS of Fe-3*d* and O-2*p* orbitals at the Fe/$BaTiO_3$ interface. Reproduced with permission [273]. Copyright 2006, American Physical Society. (b) XMCD of the Fe $L_3$ with BTO in $P_{up}$ or $P_{down}$ at 300 K. Reproduced with permission [299]. Copyright 2014, Nature Publishing Group.

**Fig. 22.** (a) Schematic for the covalent bond and orbital reconstruction at the BTO/LSMO ($x$ = 0.33) interface in $P_{up}$ and $P_{down}$. (b) The LSMO ($x$ = 0.33) thickness-dependent in-plane magnetoresistance in $P_{up}$ and $P_{down}$ at 9 T. (c) XLD of Mn in states of $P_{up}$ and $P_{down}$ for samples of various LSMO ($x$ = 0.33) thicknesses. Reproduced with permission [41]. Copyright 2015, Wiley.

**Fig. 23.** (a) Fe and (b) Ni $L_3$ XMCD intensity plotted against the NFO thickness and the layer number. The names of samples are determined by the thickness of every layer in BTO/$(NFO/BTO)_n$ (units in nm). Reproduced with permission [286]. Copyright 2014, American Physical Society.

**Fig. 24.** (a) Schematic graphs of $GdO_x$/Co/Pt samples, (b) $R_H$-$H_z$ curves when positive gate voltage (red lines) and negative gate voltages (blue lines) are applied, (c) XAS and XMCD of Co under different electric fields. Reproduced with permission [42]. Copyright 2014, American Physical Society.

**Fig. 25.** (a) Temperature-dependent resistance for $La_{0.8}Ca_{0.2}MnO_3$ of 5 nm at $V_G$ = 0, –3, and +3 V. Reproduced with permission [169]. Copyright 2009, American Physical Society. (b) AHE of Co:$TiO_2$ measured at different $V_G$. Reproduced with permission [109]. Copyright 2011, AAAS.

**Fig. 26.** (a) Electronic phase diagram of $La_{1-x}Sr_xMnO_3$ with a different doping level *x*. (b) Magnetoresistance and the normalized magnetization of the sample at $V_G$ = +3 V. (c) Fourier-filtered images for LSMO (*x* = 0.41) at $V_G$ = +3.0 V. Reproduced with permission [121]. Copyright 2014, Wiley.

**Fig. 27.** (a) Schematic graphs of voltage control of 3*d* orbital occupancy. (b) Normalized XAS and XLD signals of $La_{0.46}Sr_{0.54}MnO_3$ grown on the STO substrate. Reproduced with permission [122]. Copyright 2015, Wiley.

**Fig. 28.** (a) Schematic graphs of the correlations between five different mechanisms. (b) The effective thickness for VCM based on various mechanisms. Reproduced with permission [63]. Copyright 2016, Wiley.

**Fig. 29.** (a) Schematic graphs of the gated MTJ device. (b) TMR with various bias voltages. Reproduced with permission [46]. Copyright 2012, Nature Publishing Group.

**Fig. 30.** Measured (dots) and calculated (dashed-dotted lines) scaling trend of switching energy with junction diameter for over 90% switching probabilities. Reproduced with permission [48]. Copyright 2016, AIP Publishing.

**Fig. 31.** (a) Sketch of the four resistance states in MFTJS. Magnetoresistance curves with different polarization states for (b) Fe/BTO/LSMO (*x* = 0.33) and (c) Fe/PZT (*x* = 0.2)/LSMO (*x* = 0.3). Reproduced with permissions [321] and [124]. Copyright

2010, AAAS and 2012, Nature Publishing Group.

**Fig. 32.** (a) Magnetization of Co/PMN-PT at different polarization states: $P_R^+$ (left) and $P_R^-$ (right). (b) Pulsed electrical operation with $H_{au}$ (top) and the resulting magnetization (bottom) measured at 0.3 mT. Reproduced with permission [59]. Copyright 2014, Wiley.

**Fig. 33.** (a) Sketch of a $BiFeO_3$-based multiferroic heterostructure. (b) XMCD-PEEM images under different external voltages. (c) The left panel: the polarization dependence and resistance dependence on external voltages; right panel: magnetoresistance of corresponding spin valve. Reproduced with permission [51]. Copyright 2014, Nature Publishing Group.

**Fig. 34.** (a) Schematic graphs of the usual spin–orbit torque device. Reproduced with permission [340]. Copyright 2016, Nature Publishing Group. (b) Current-induced switching owing to spin–orbit torque. Reproduced with permission [341]. Copyright 2012, American Physical Society.

**Fig. 35.** Voltage control of current-induced switching in a Pt/Co/$HfO_2$ structure. Reproduced with permission [58]. Copyright 2016, Wiley.

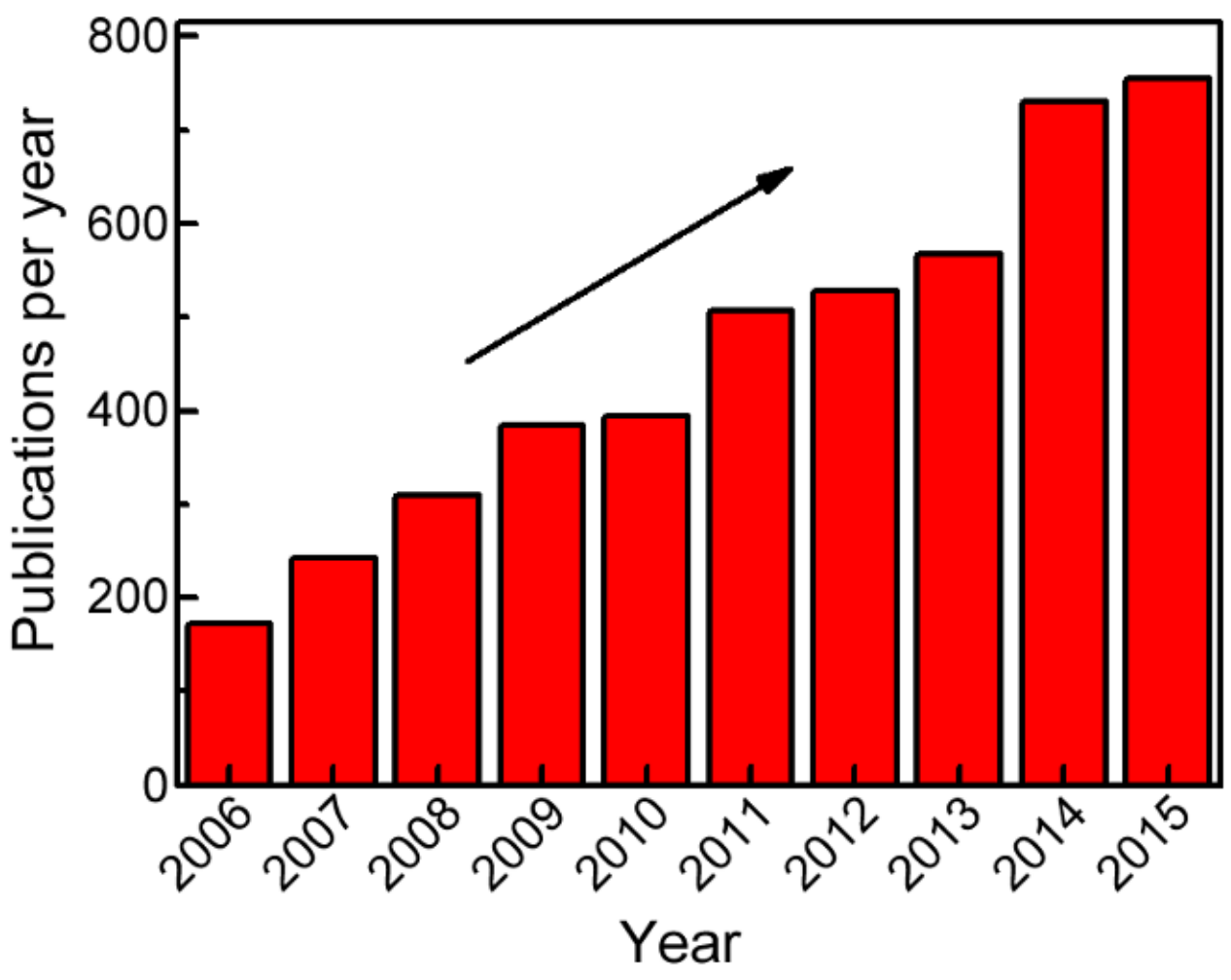

**Fig. 1**

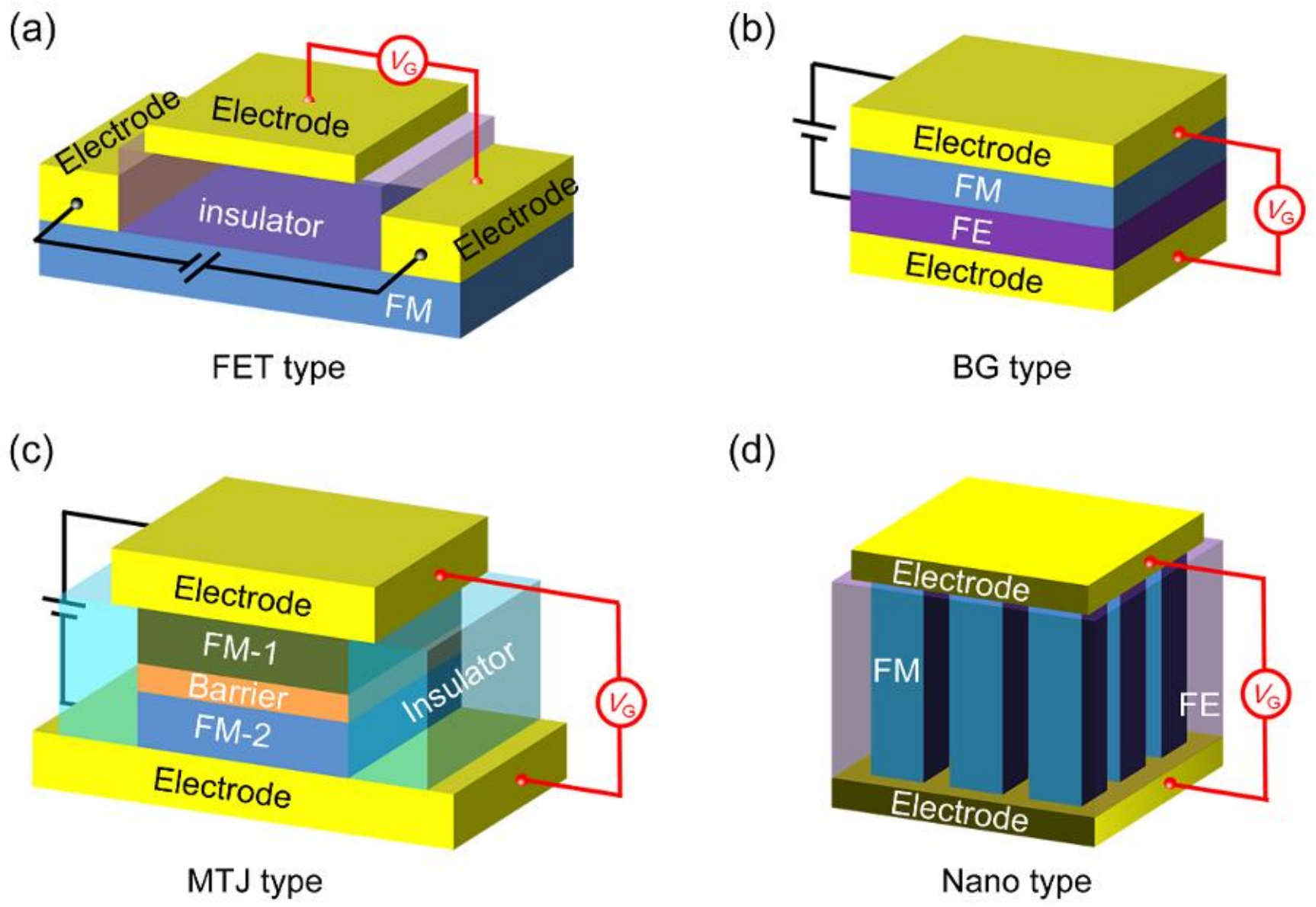

**Fig. 2**

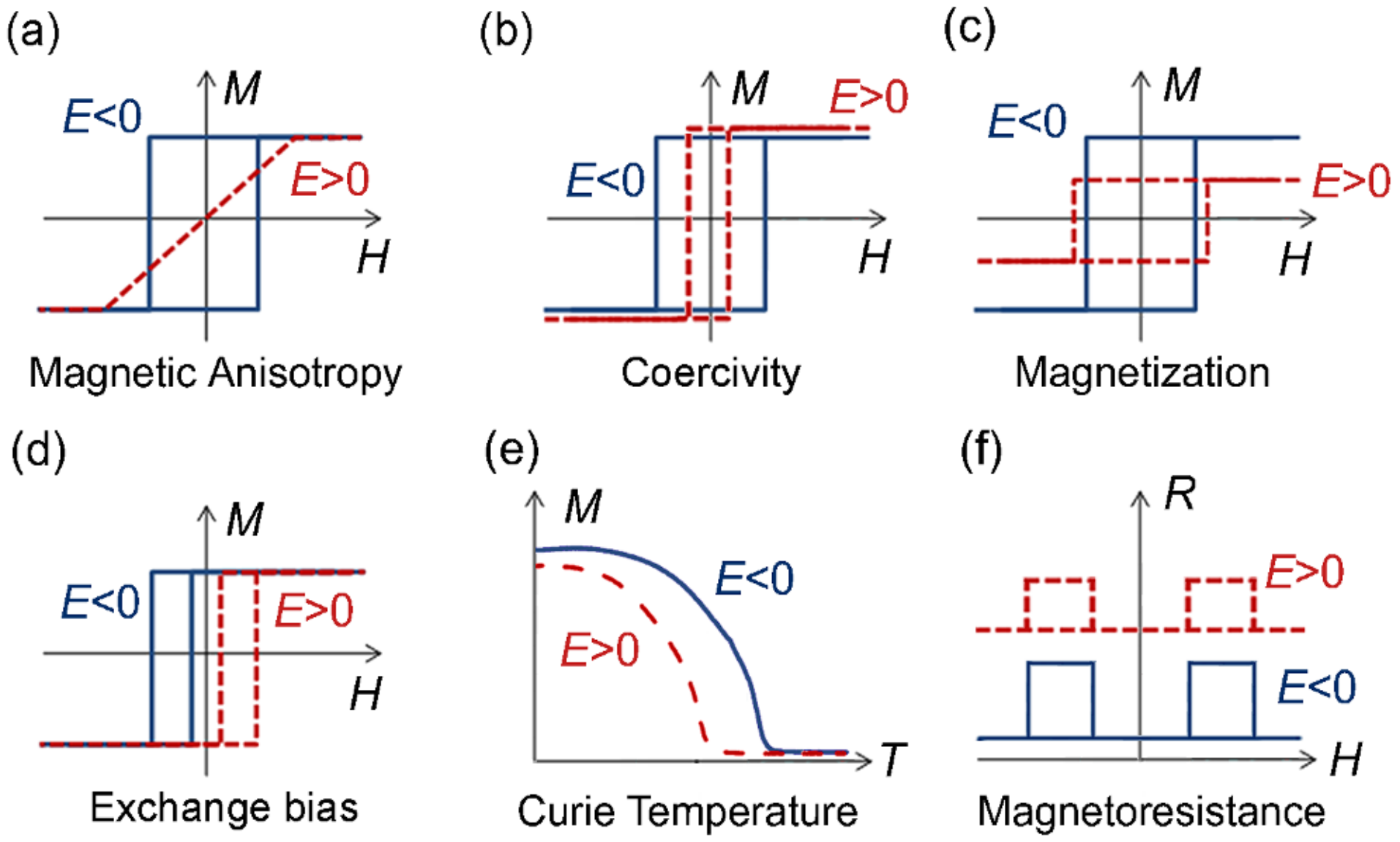

**Fig. 3**

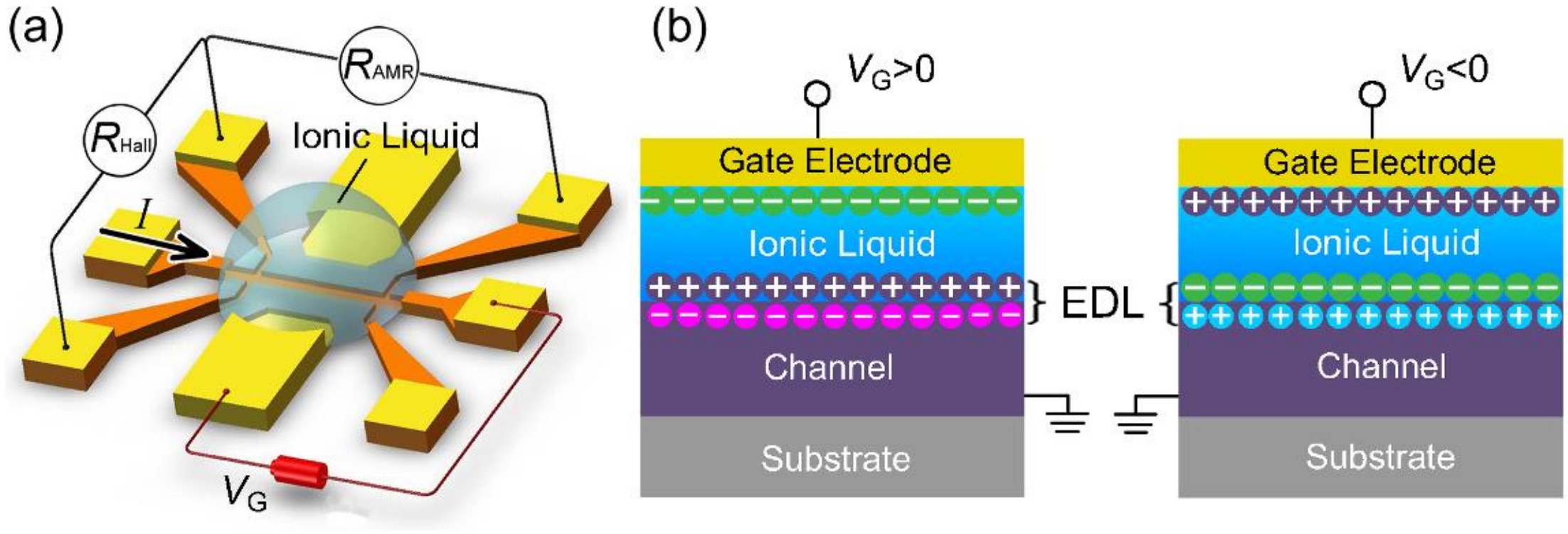

**Fig. 4**

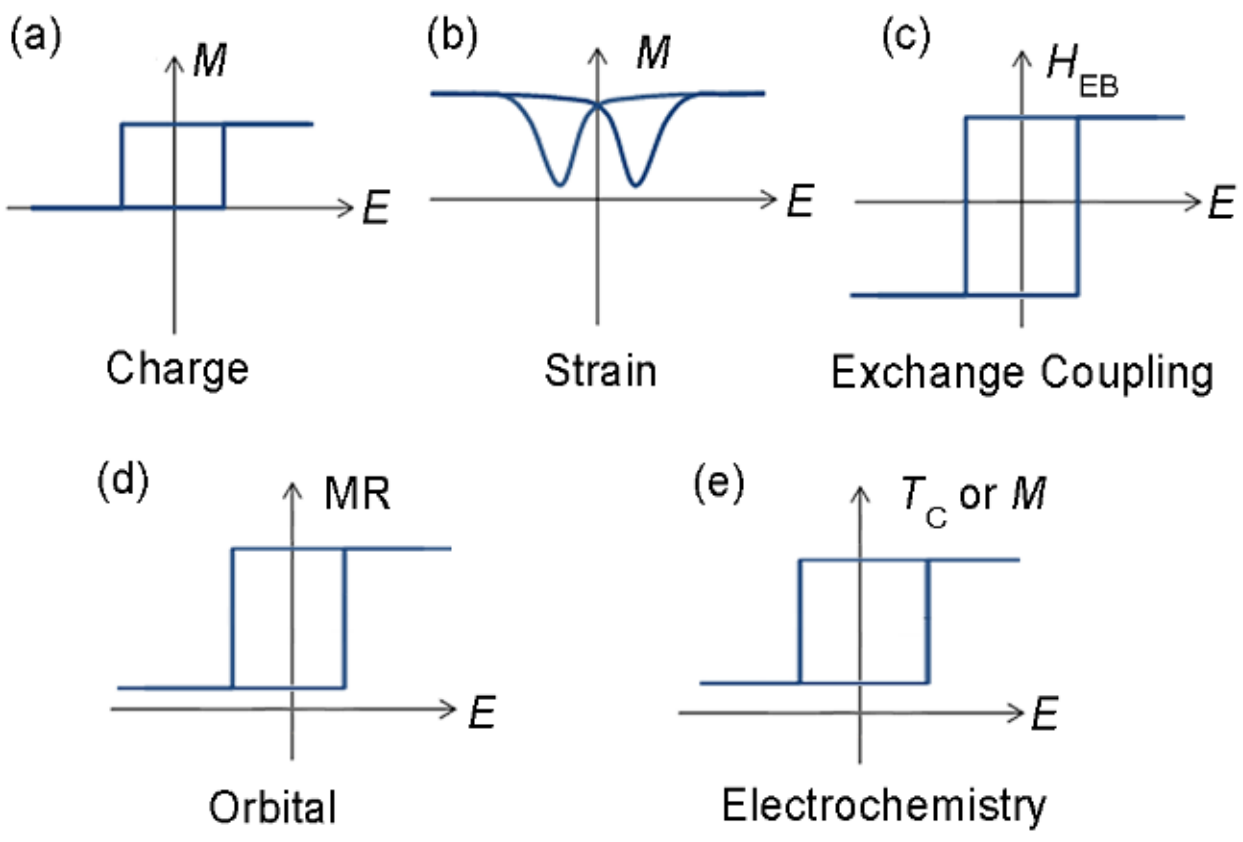

**Fig. 5**

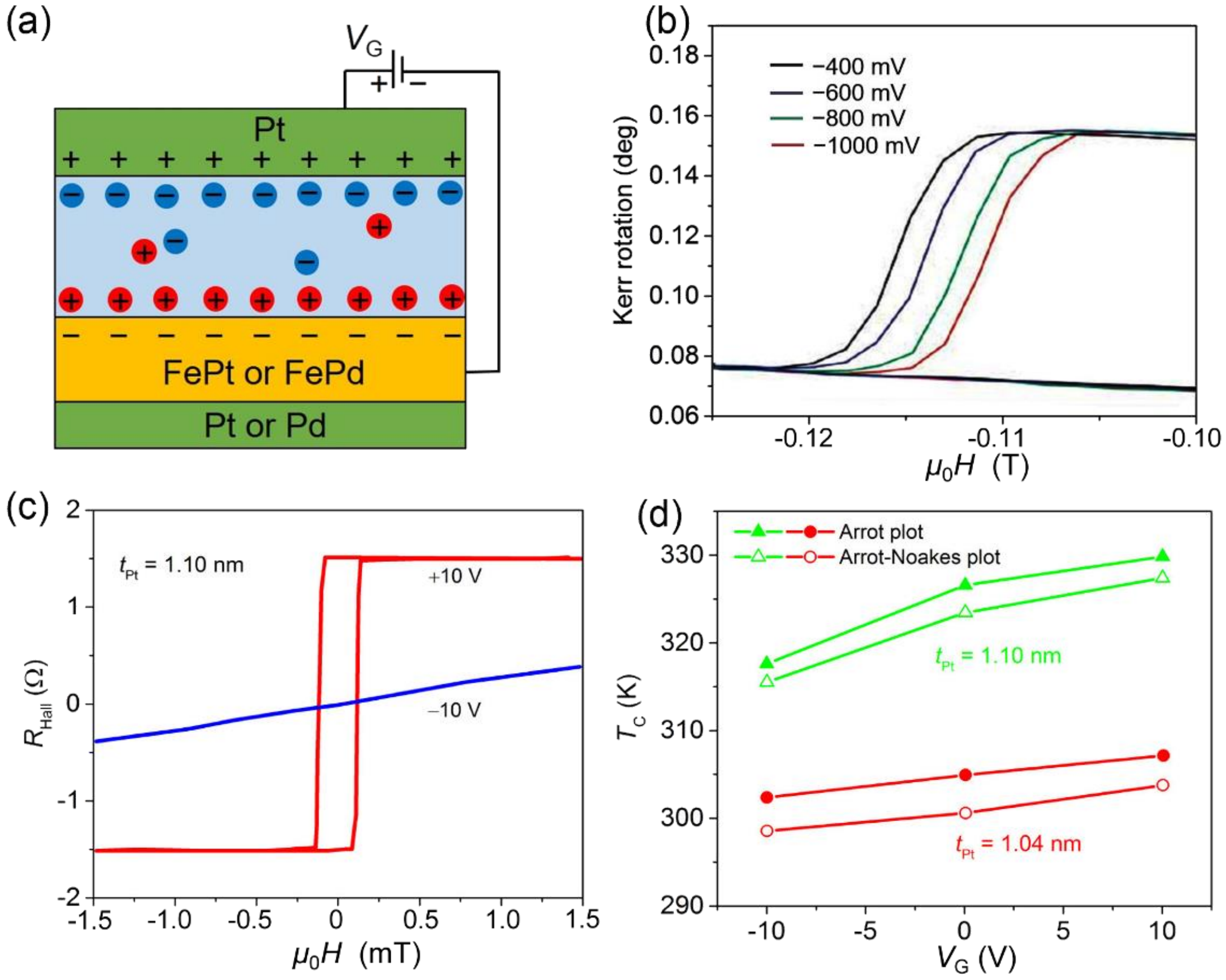

**Fig. 6**

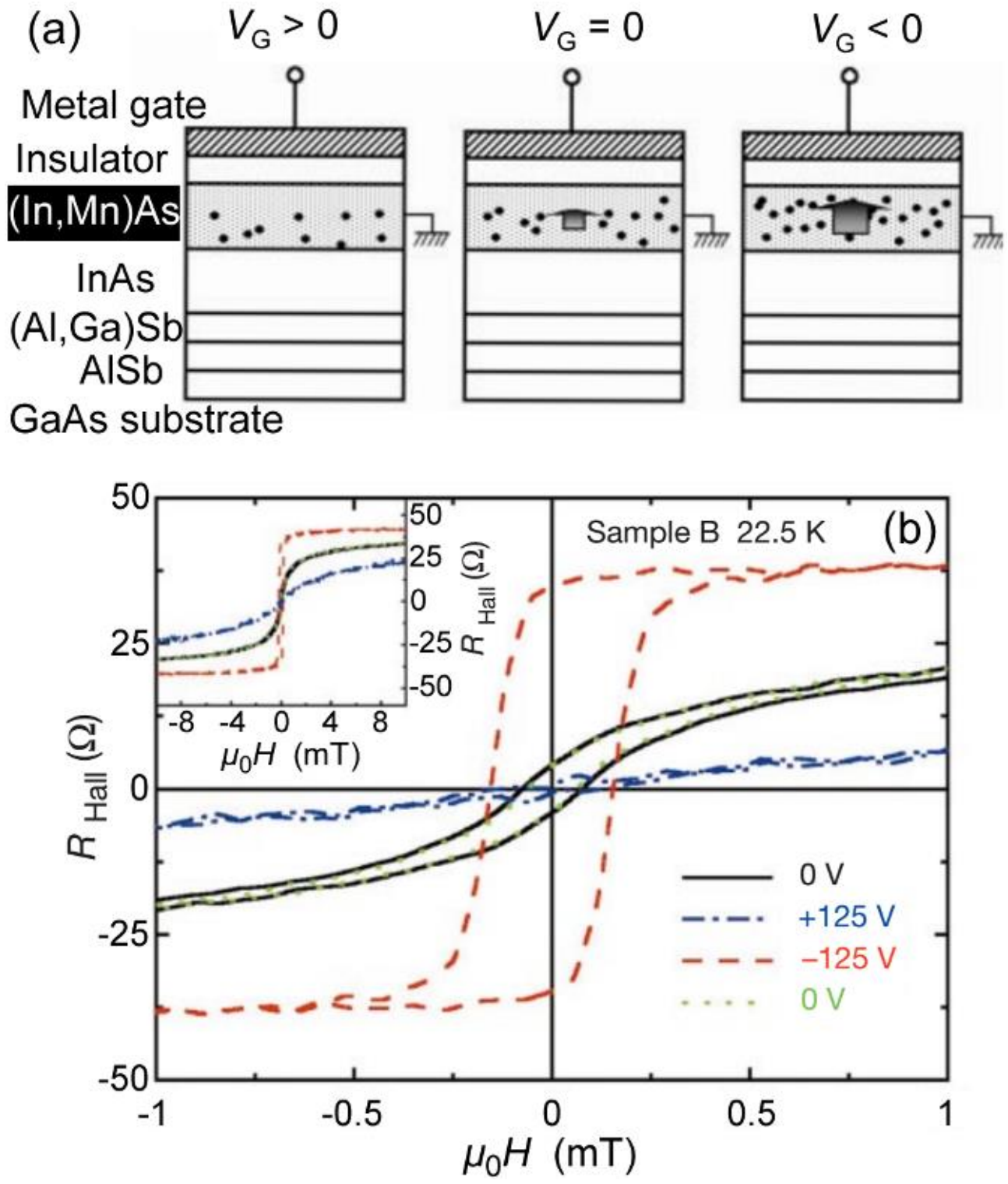

**Fig. 7**

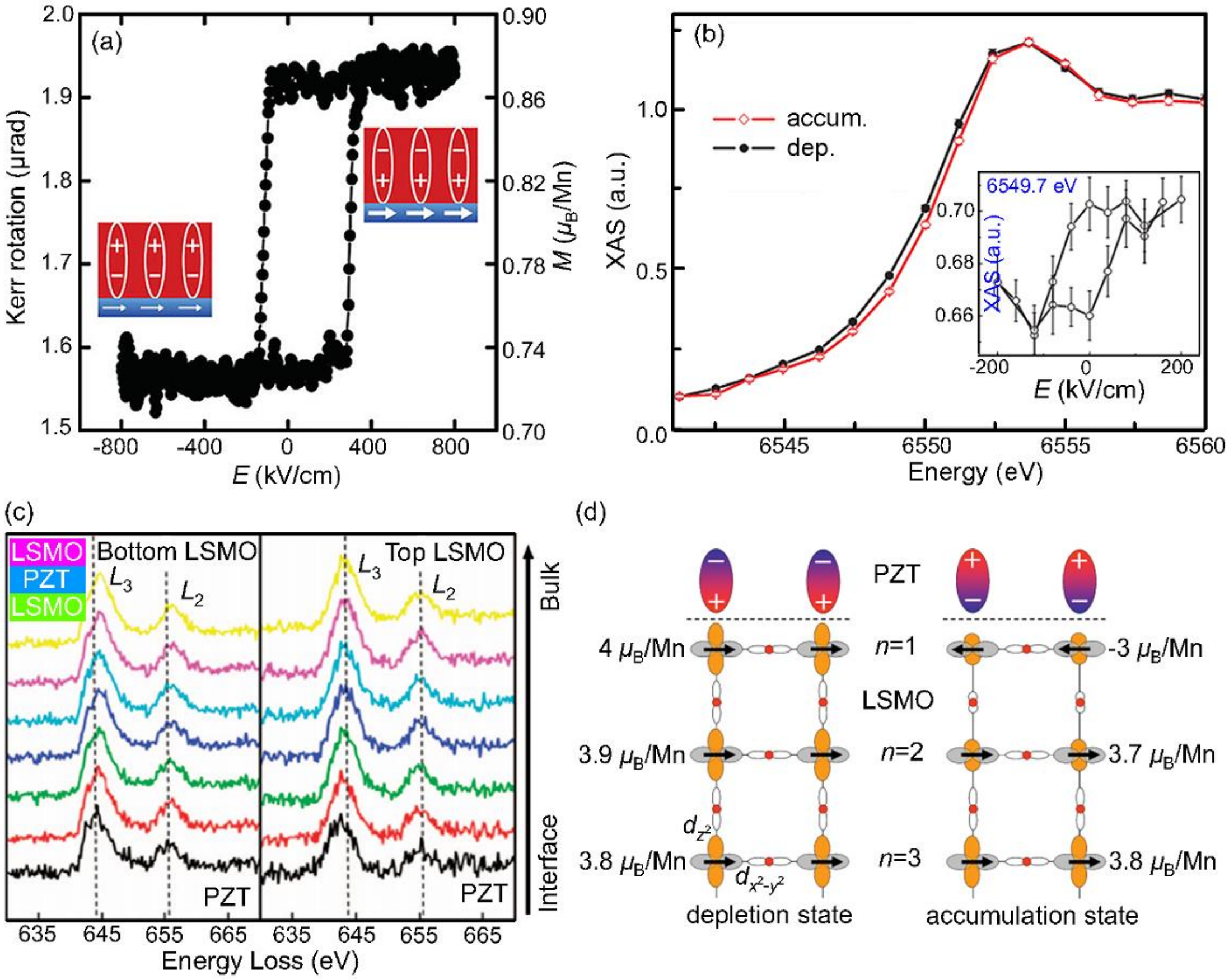

**Fig. 8**

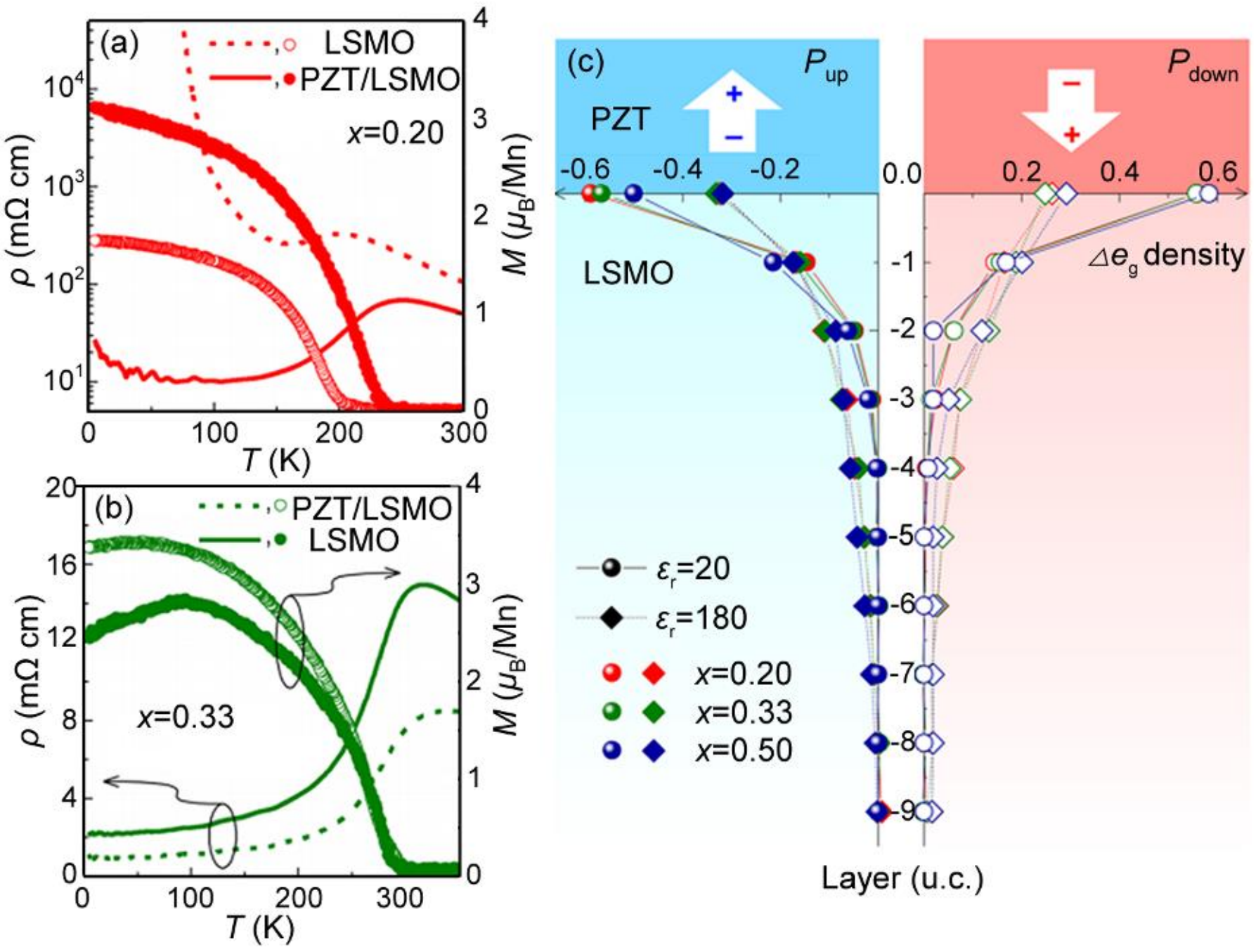

**Fig. 9**

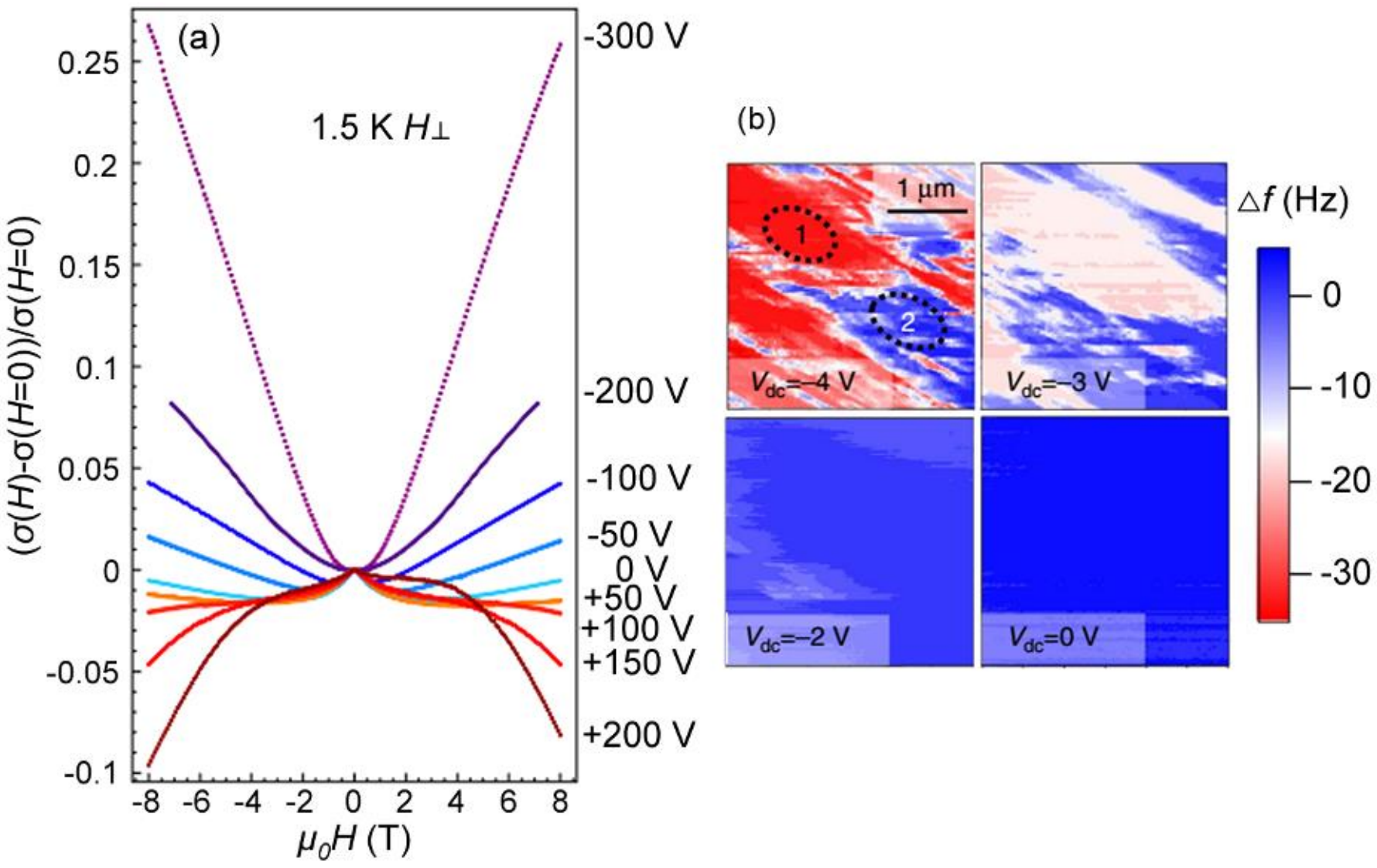

**Fig. 10**

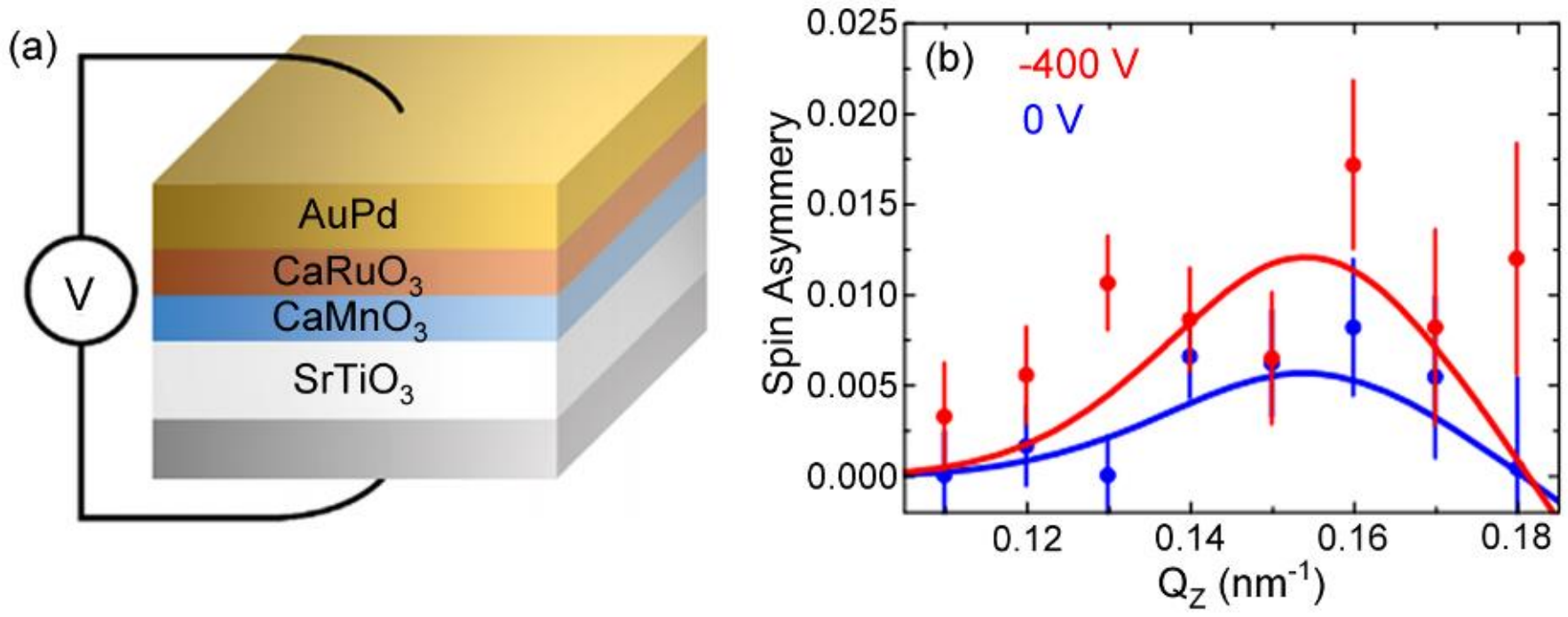

**Fig. 11**

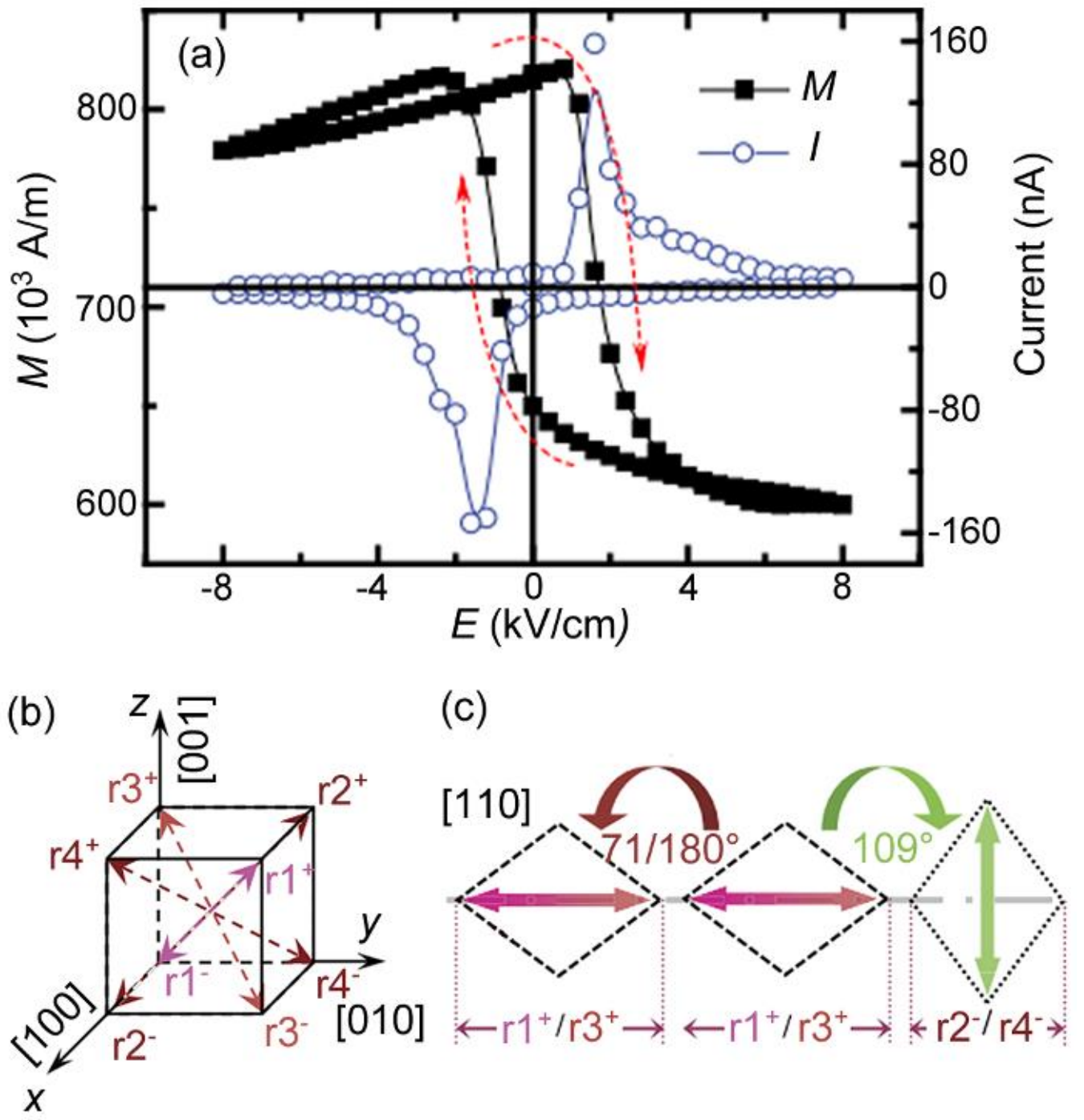

**Fig. 12**

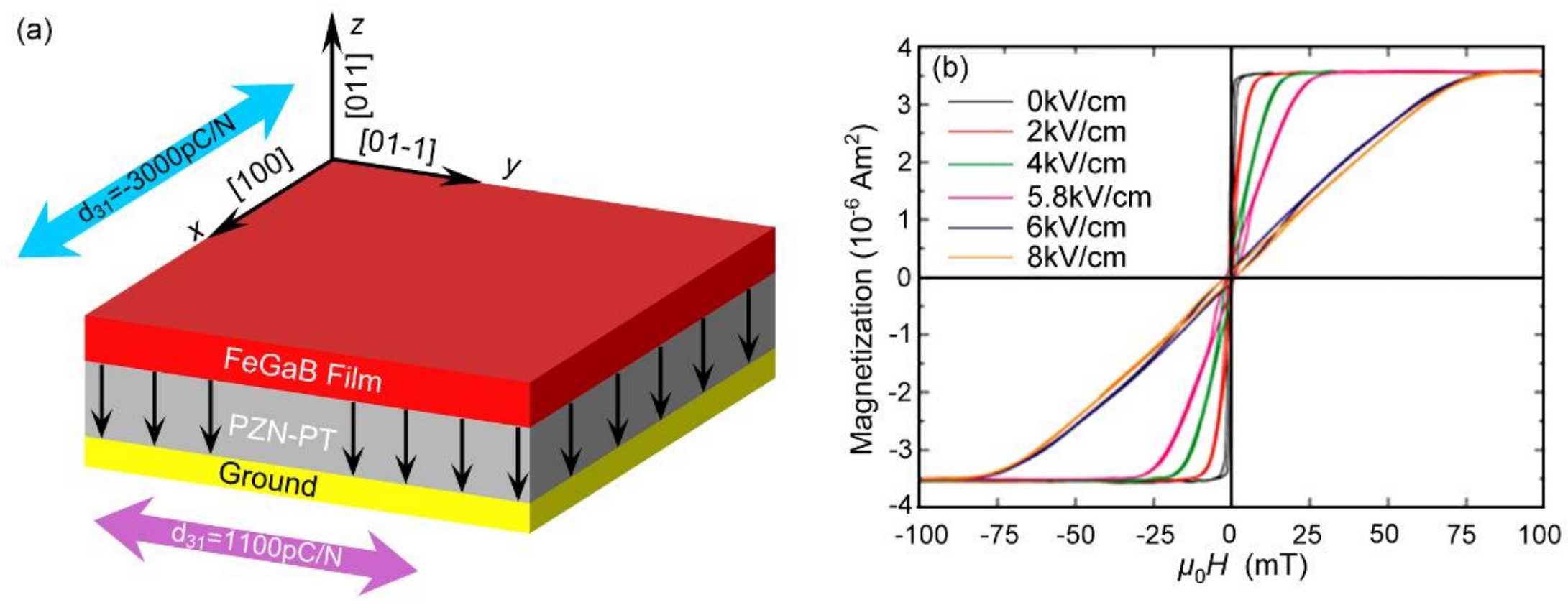

**Fig. 13**

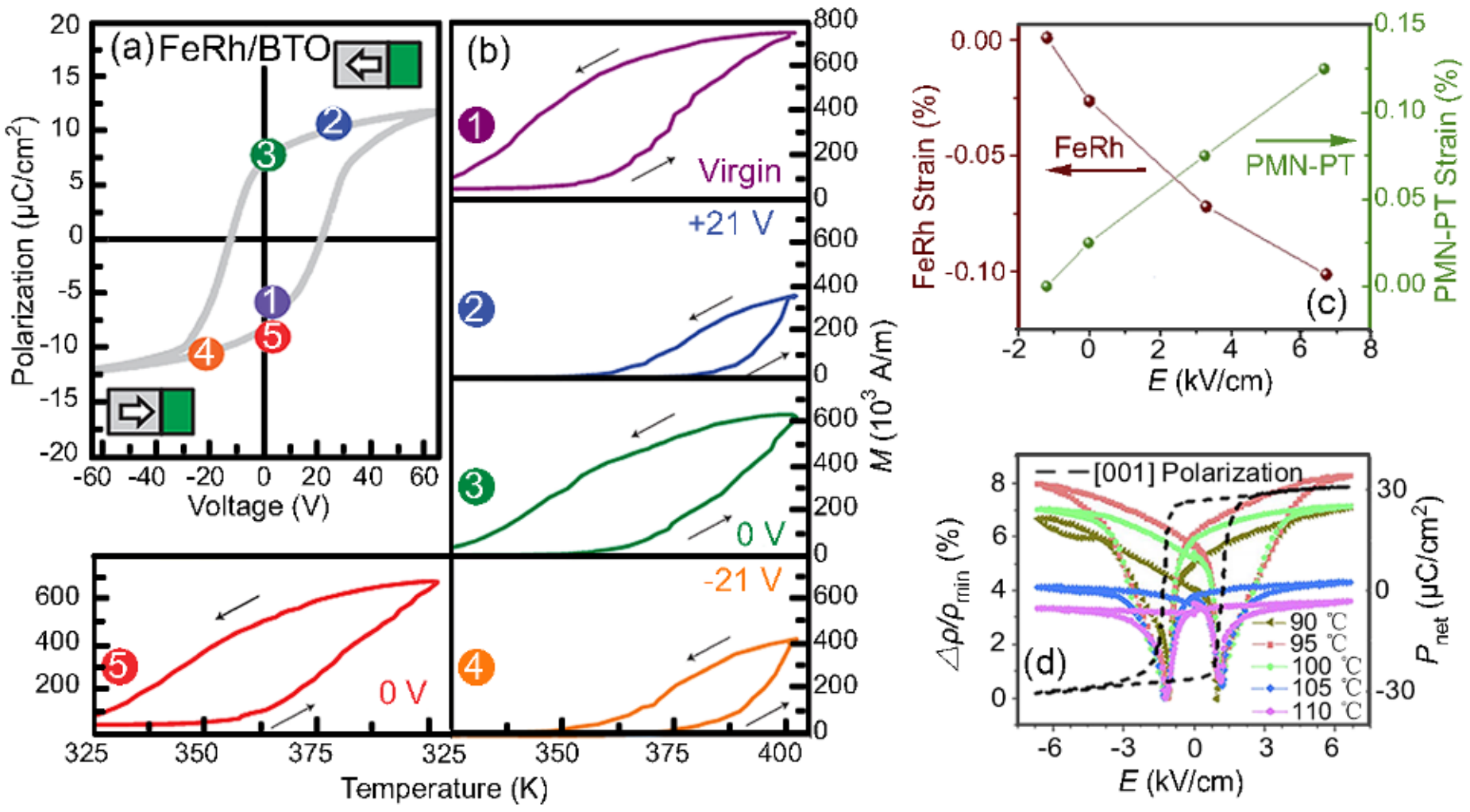

**Fig. 14**

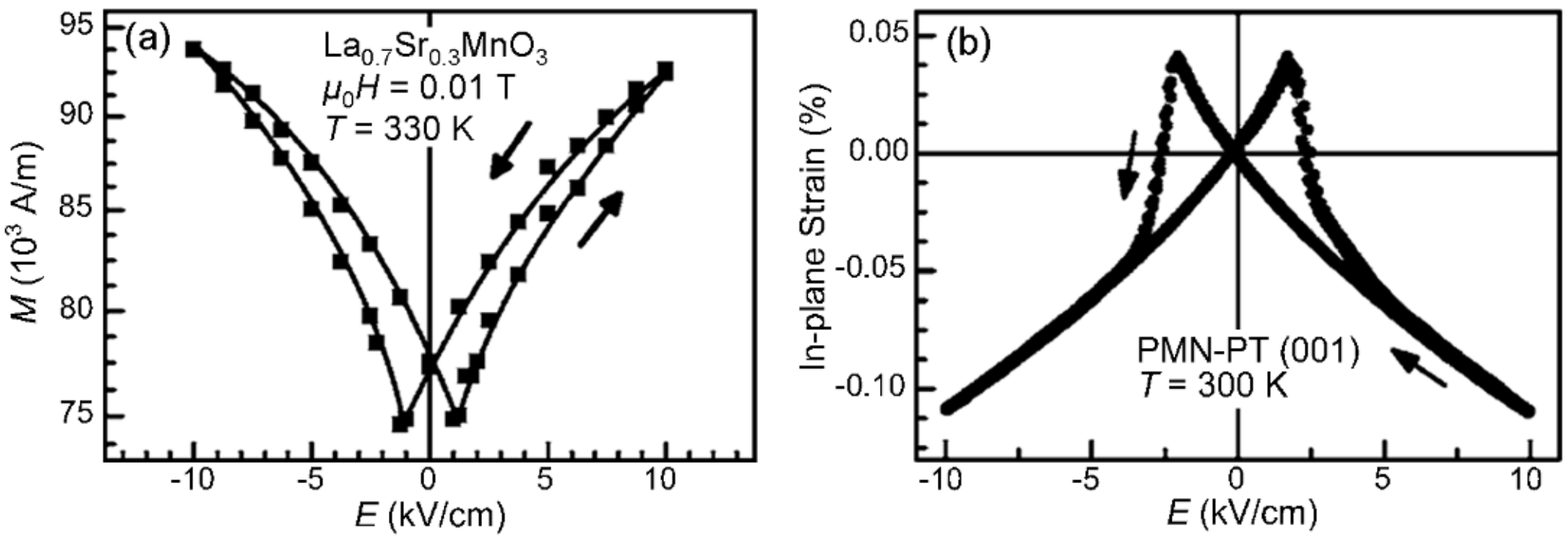

**Fig. 15**

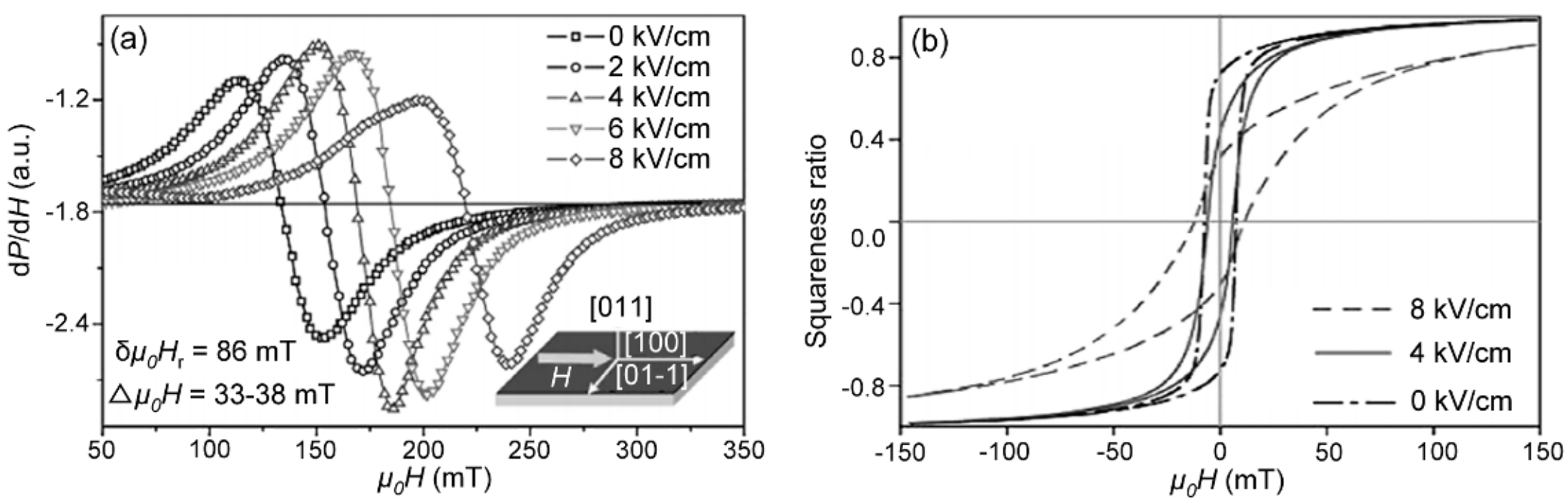

**Fig. 16**

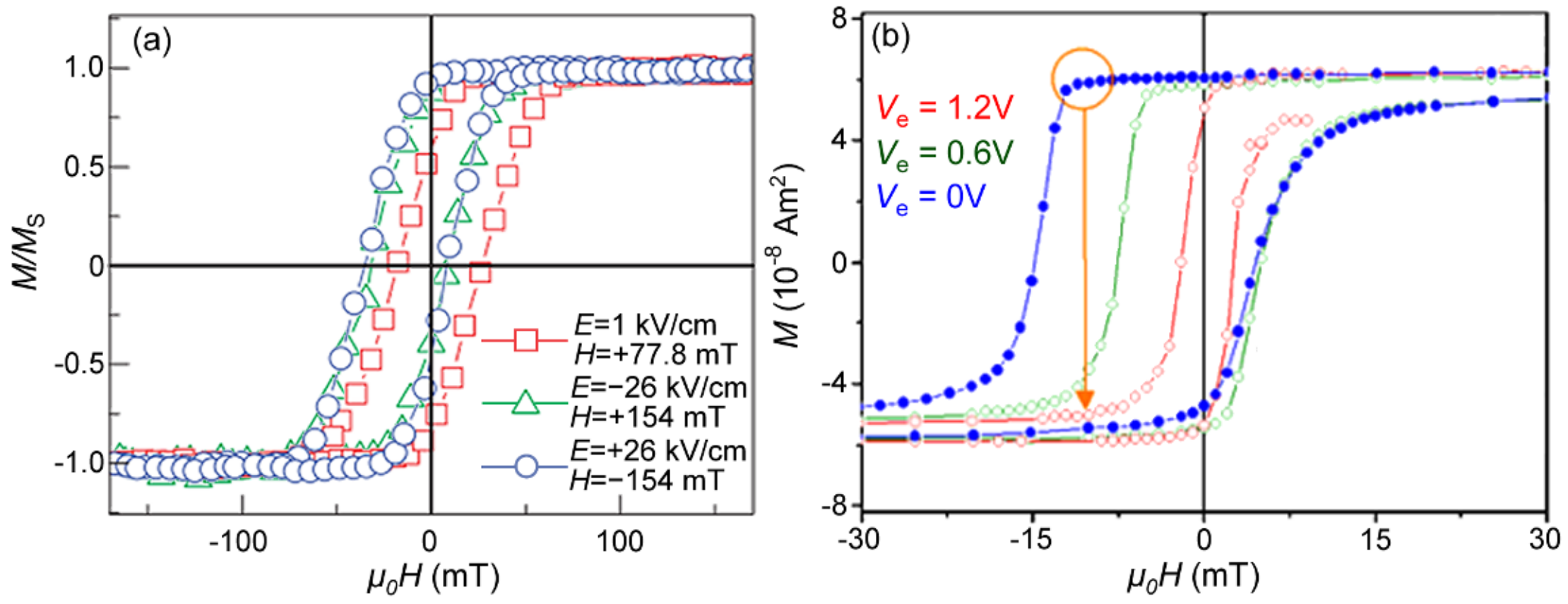

**Fig. 17**

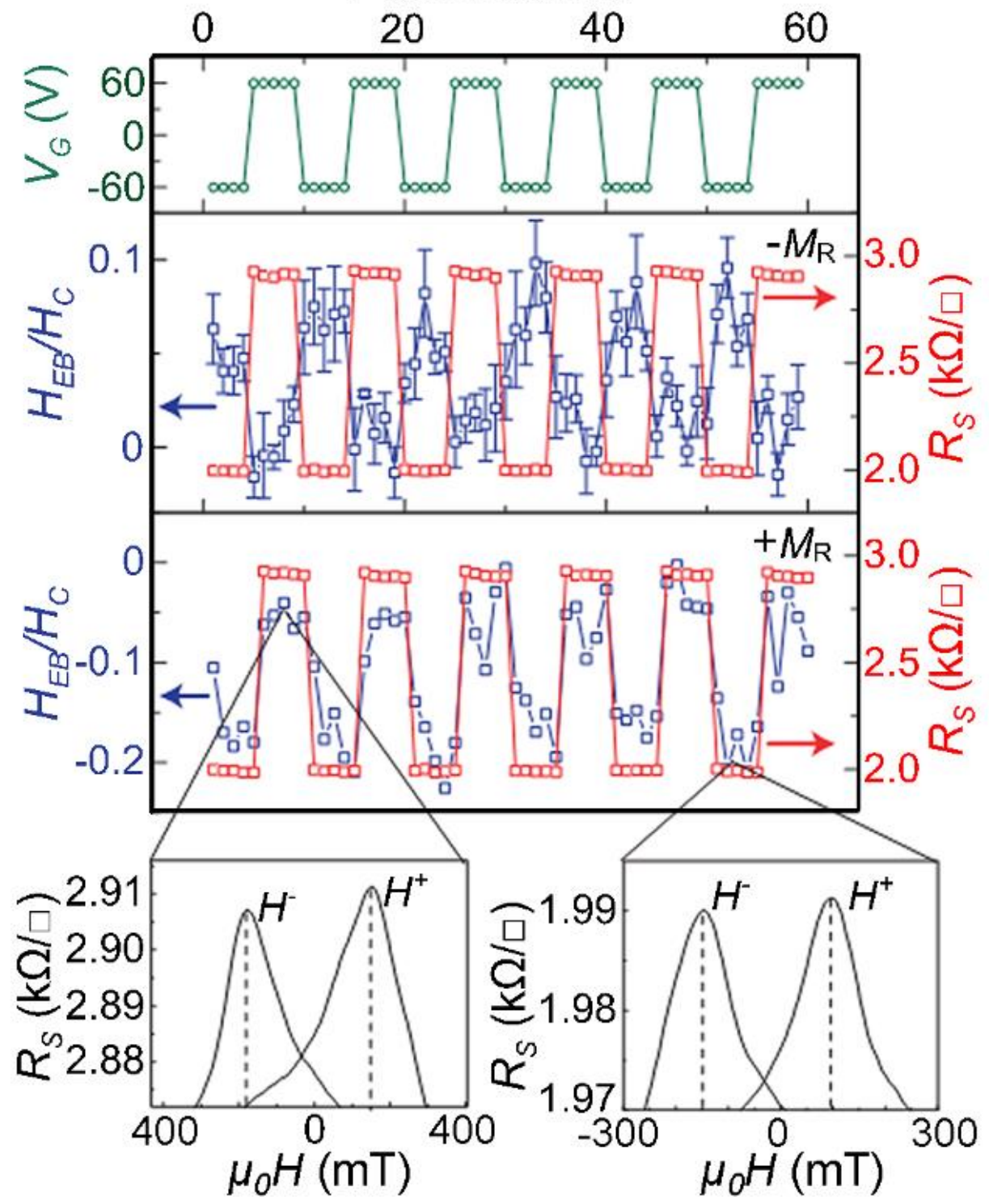

**Fig. 18**

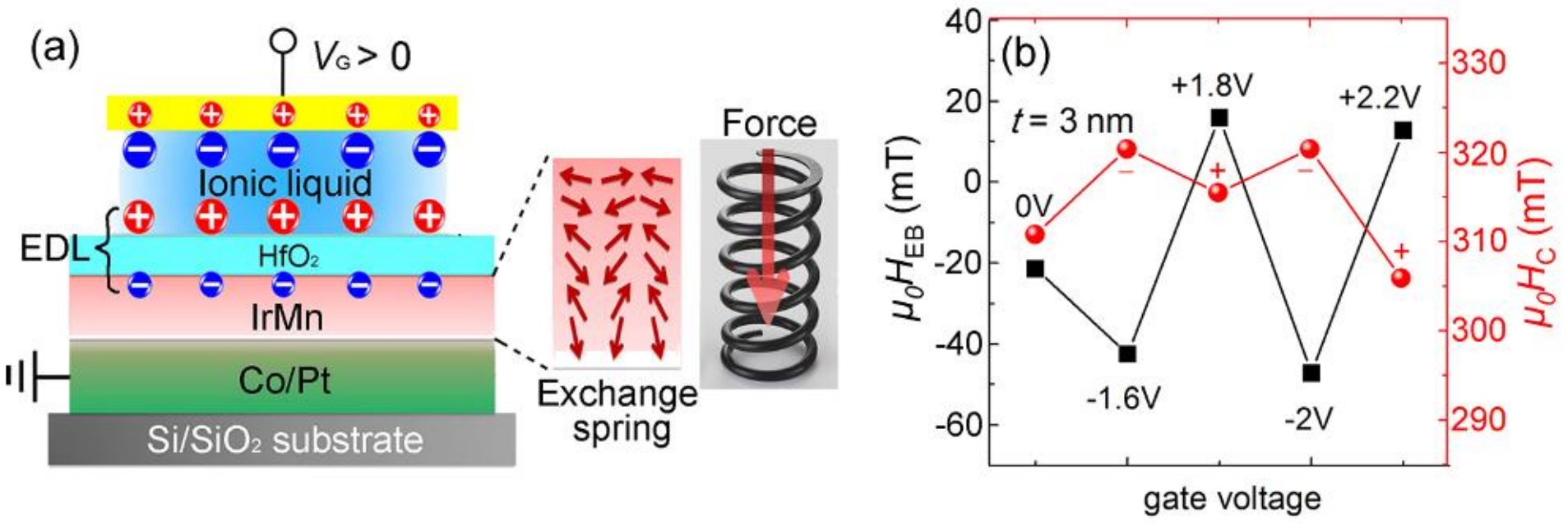

**Fig. 19**

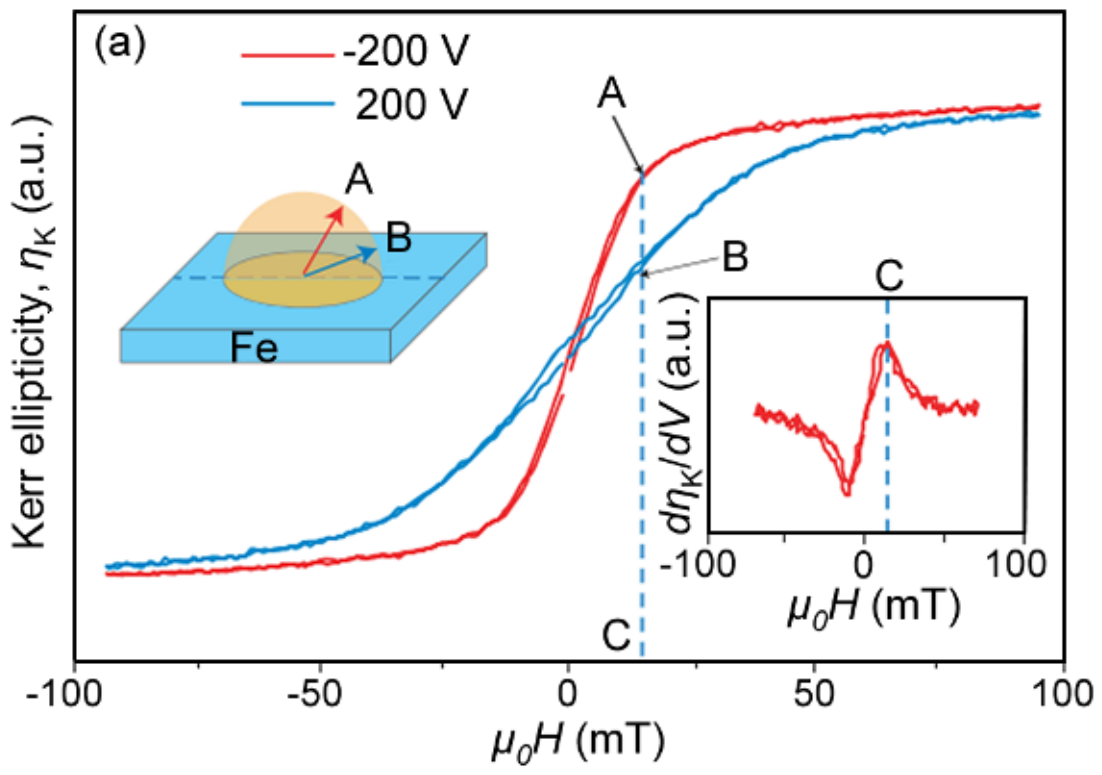

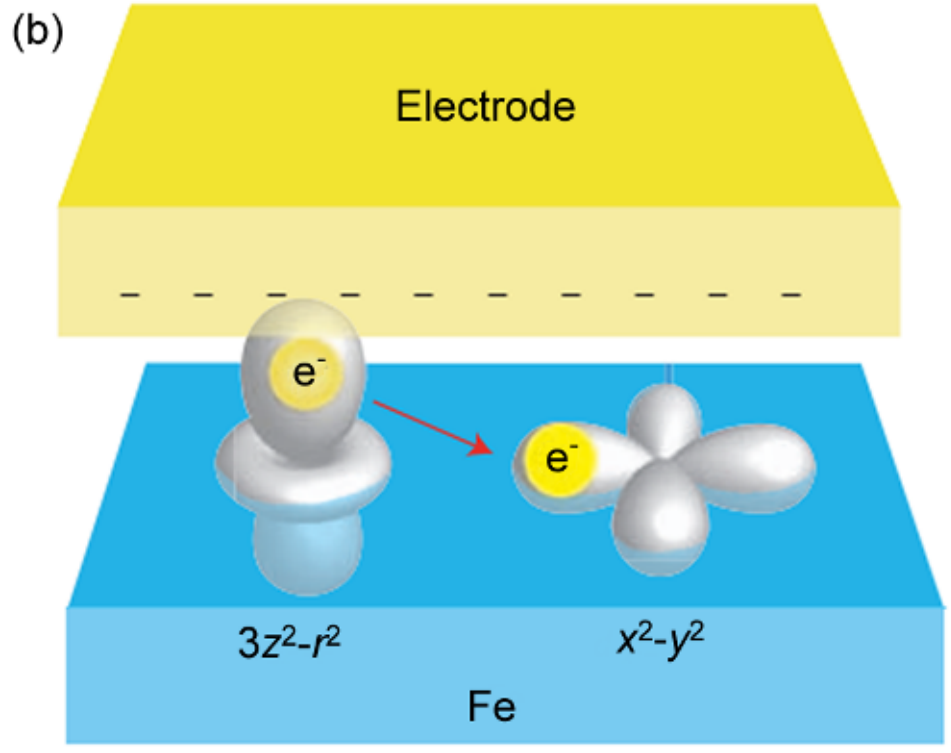

**Fig. 20**

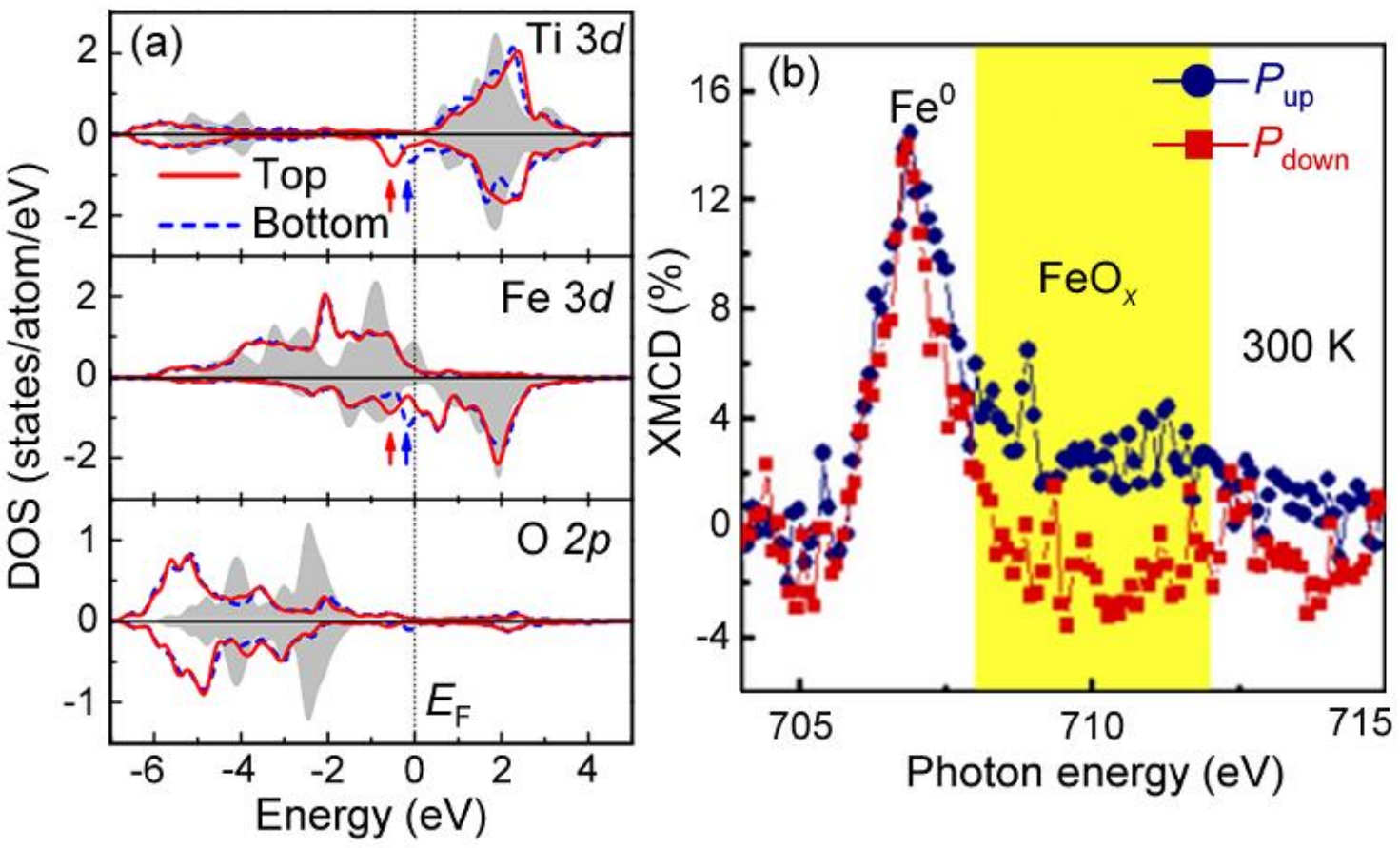

**Fig. 21**

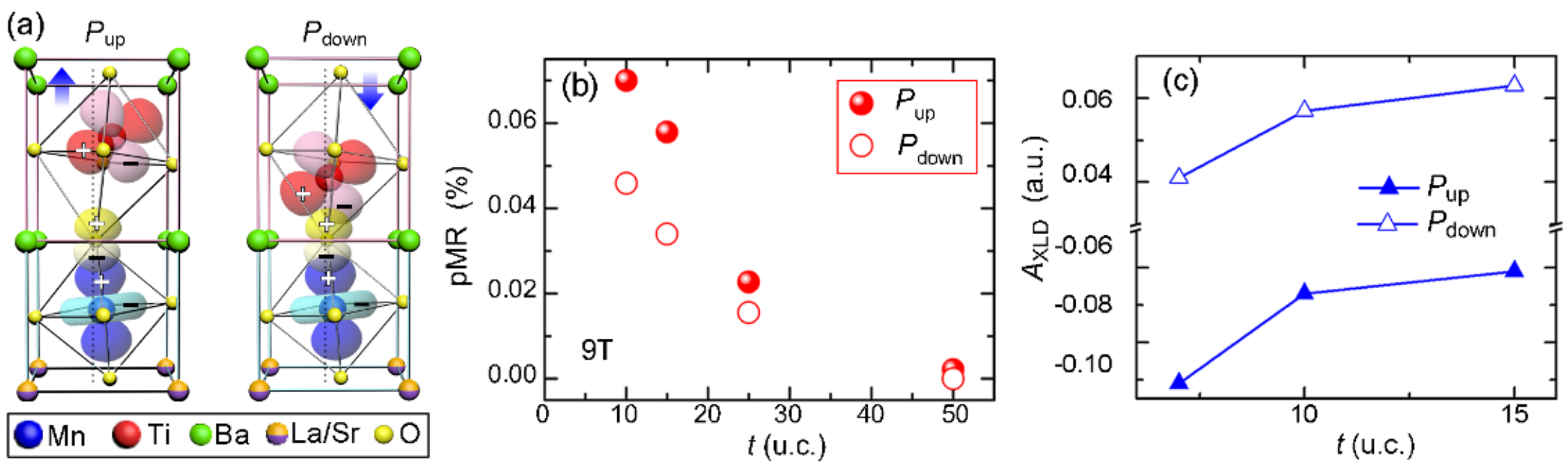

**Fig. 22**

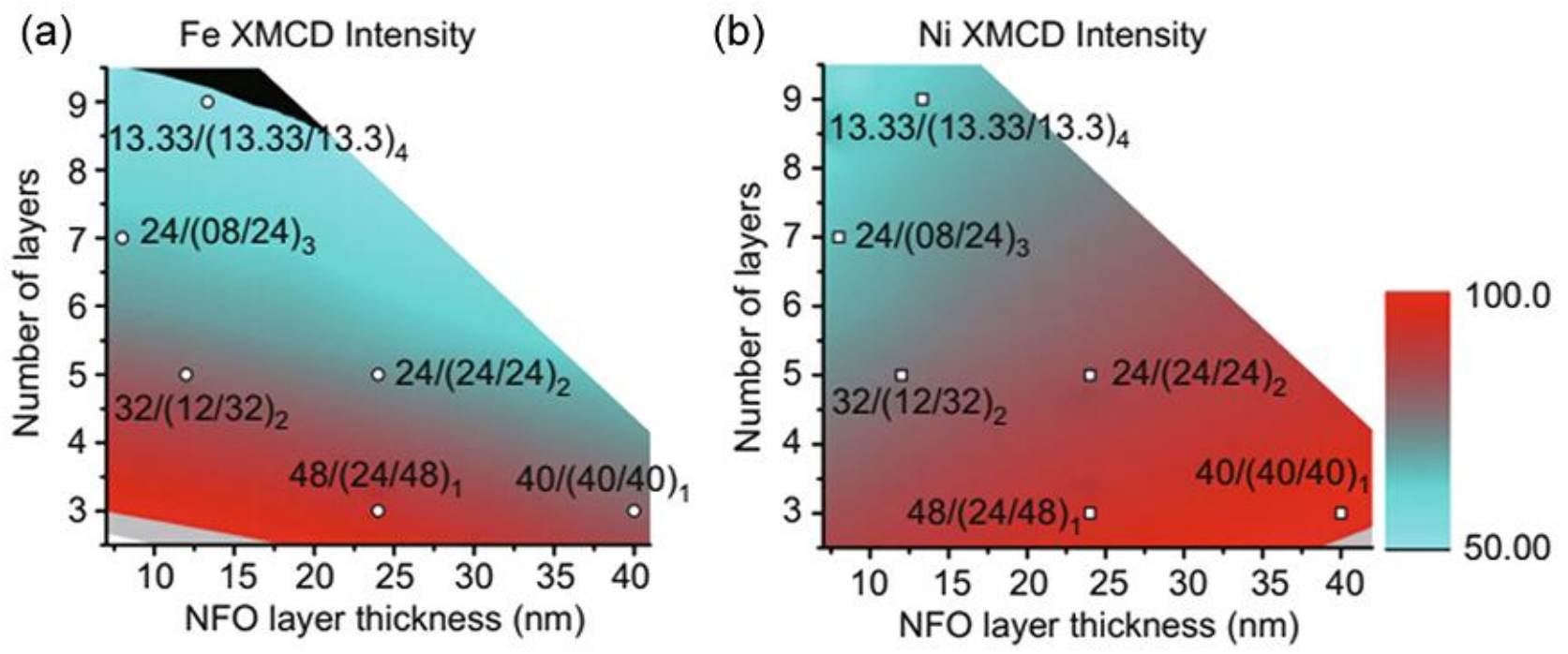

**Fig. 23**

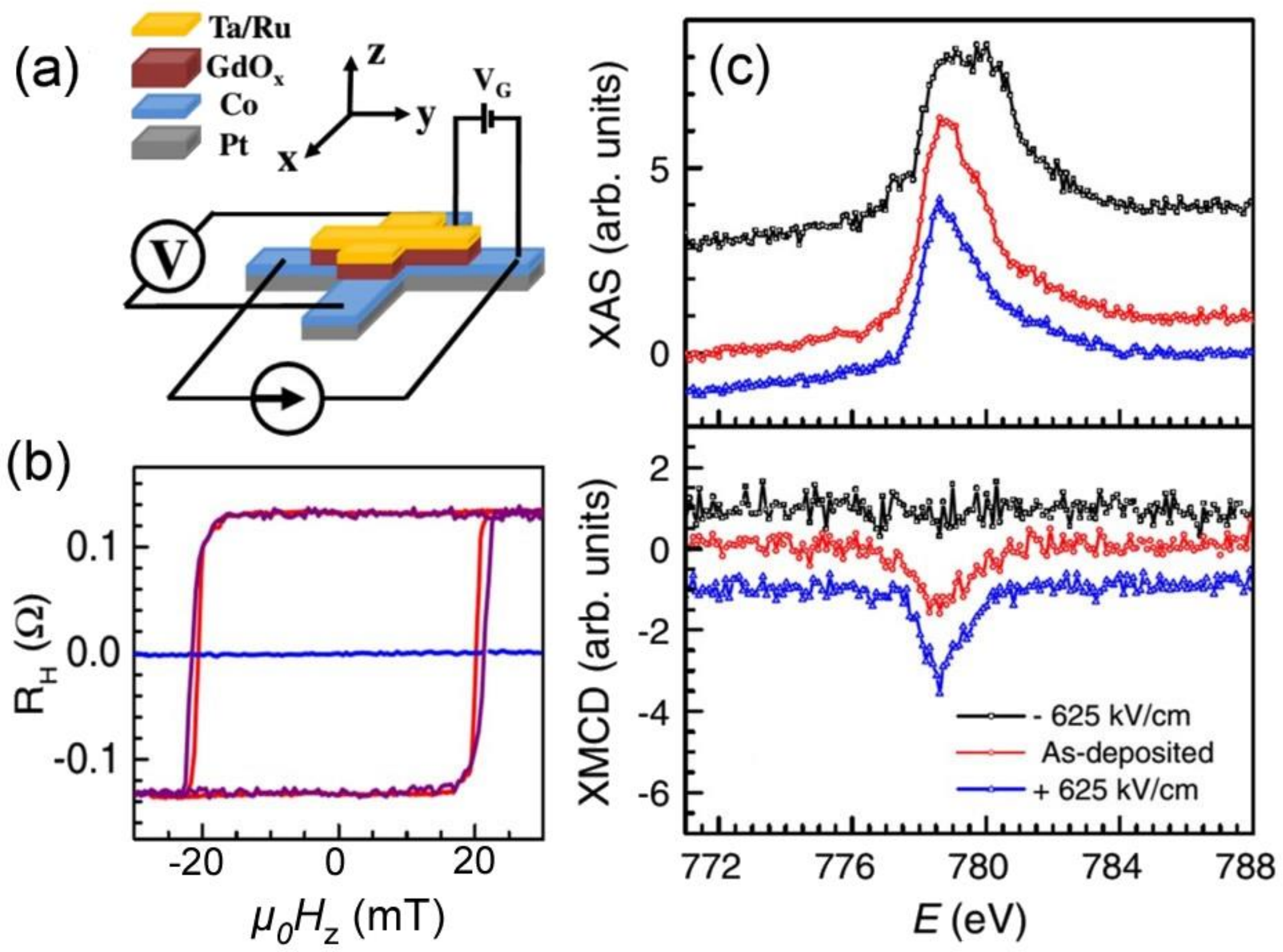

**Fig. 24**

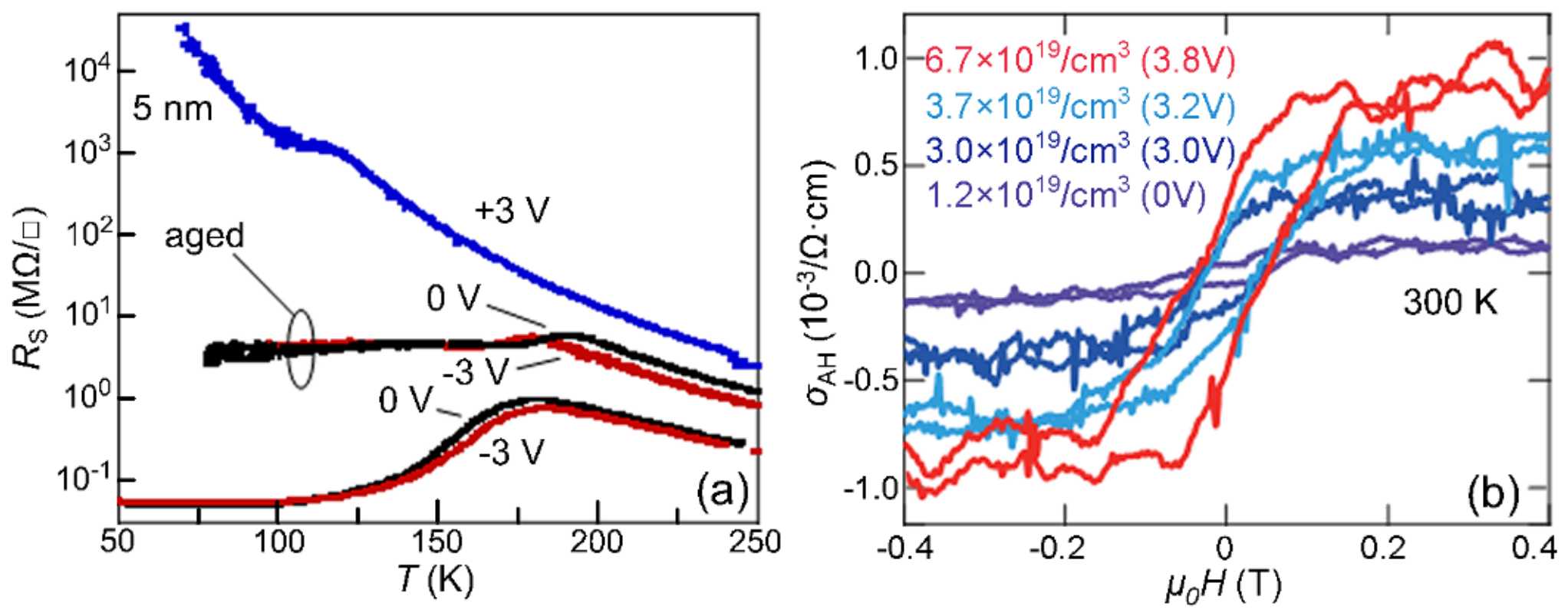

**Fig. 25**

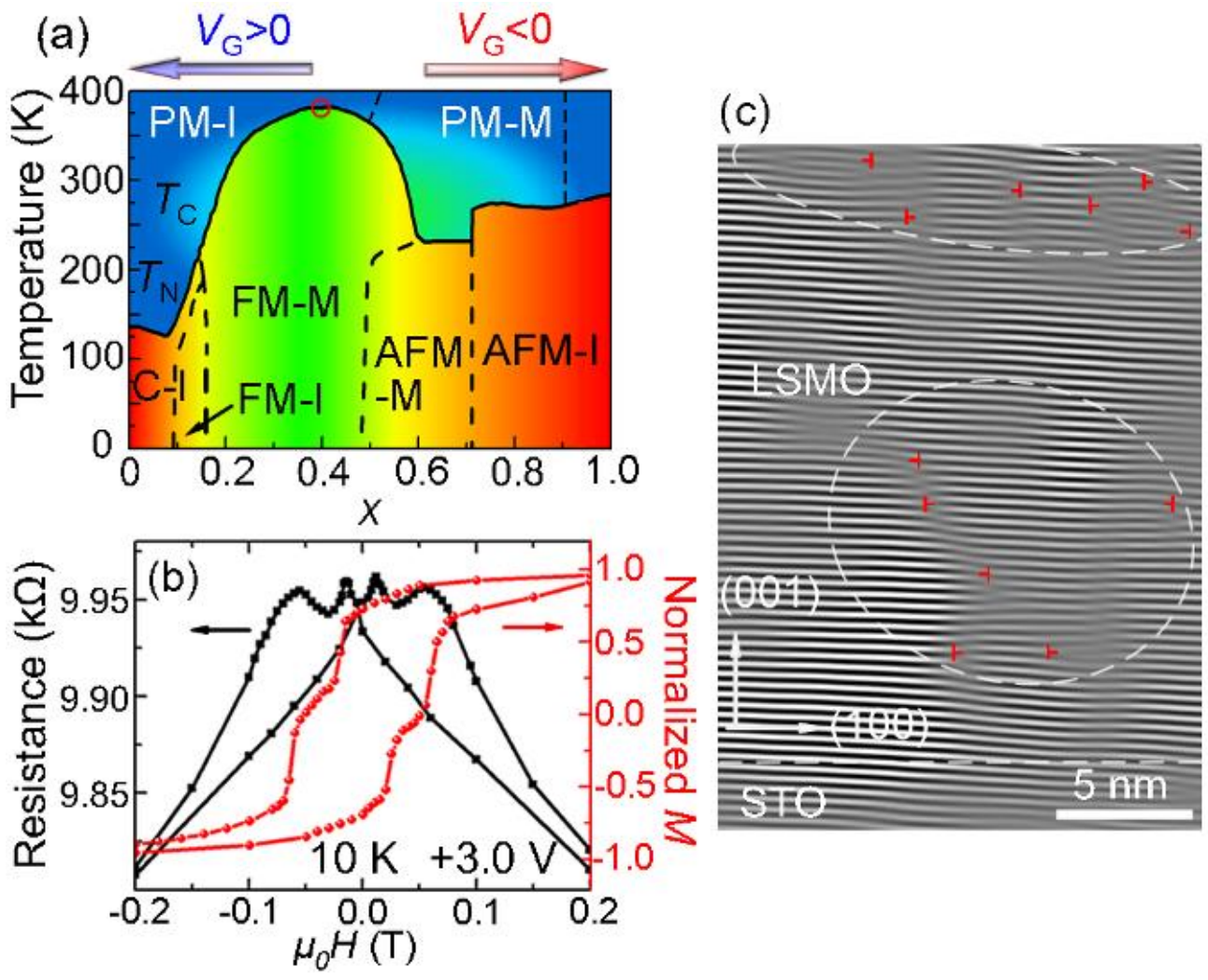

**Fig. 26**

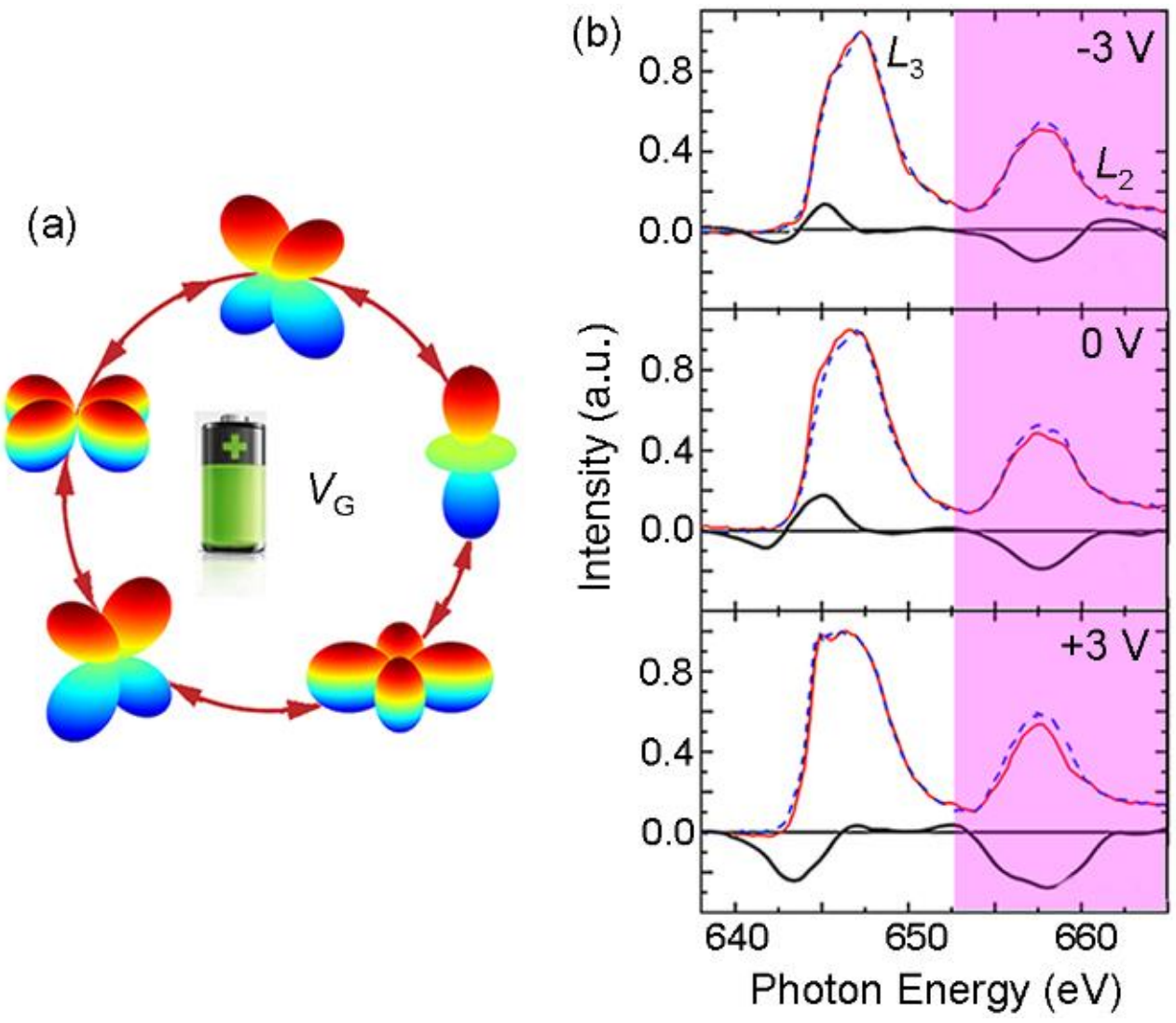

**Fig. 27**

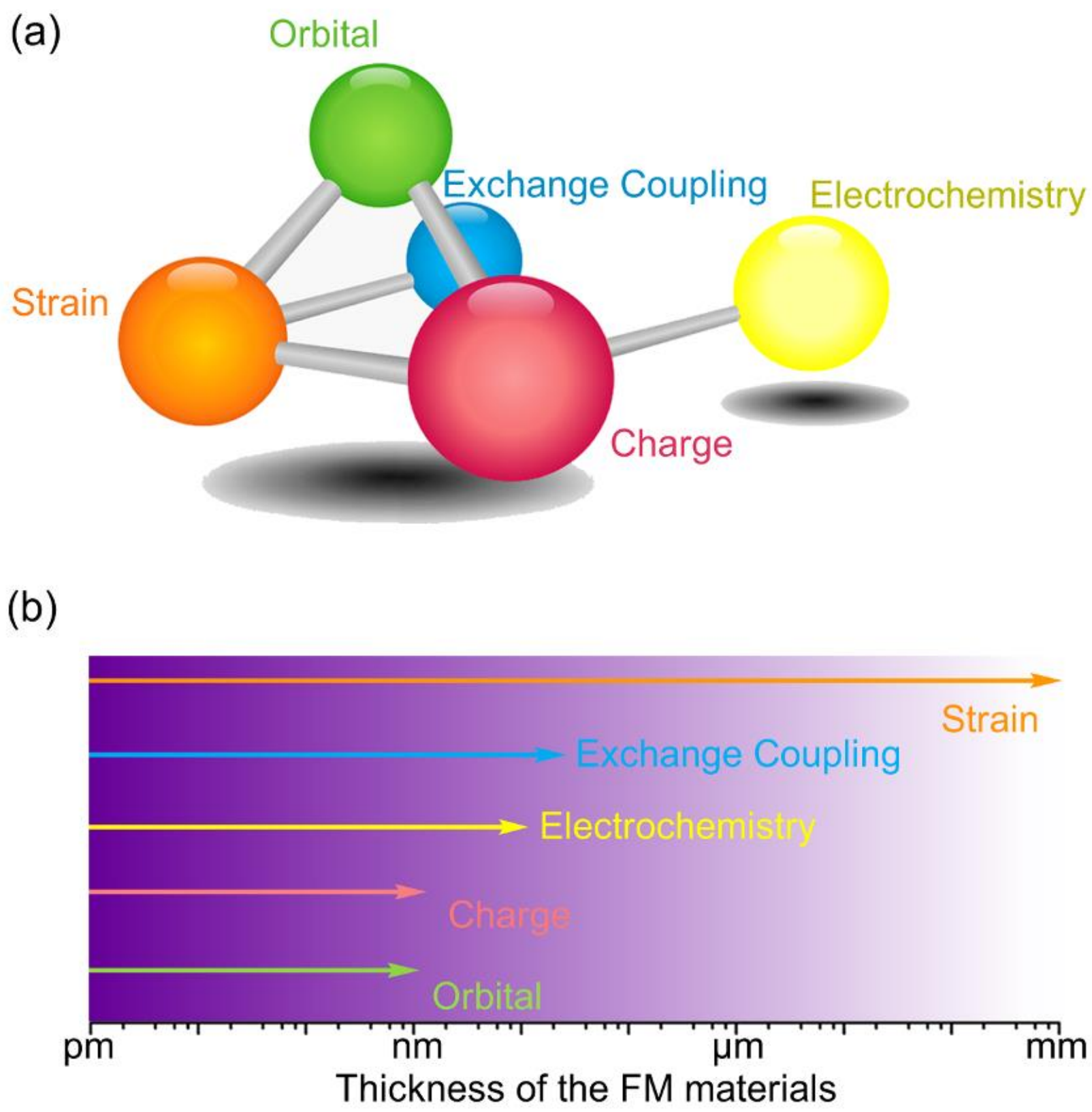

**Fig. 28**

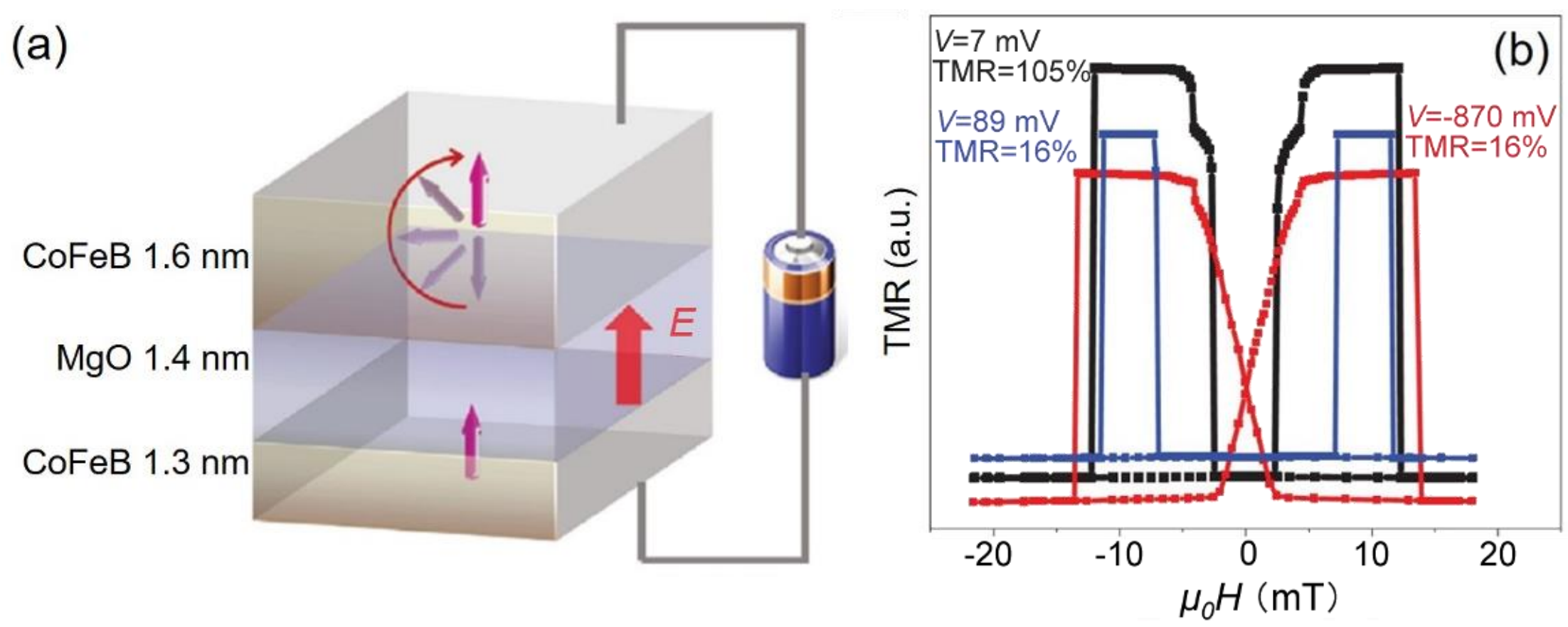

**Fig. 29**

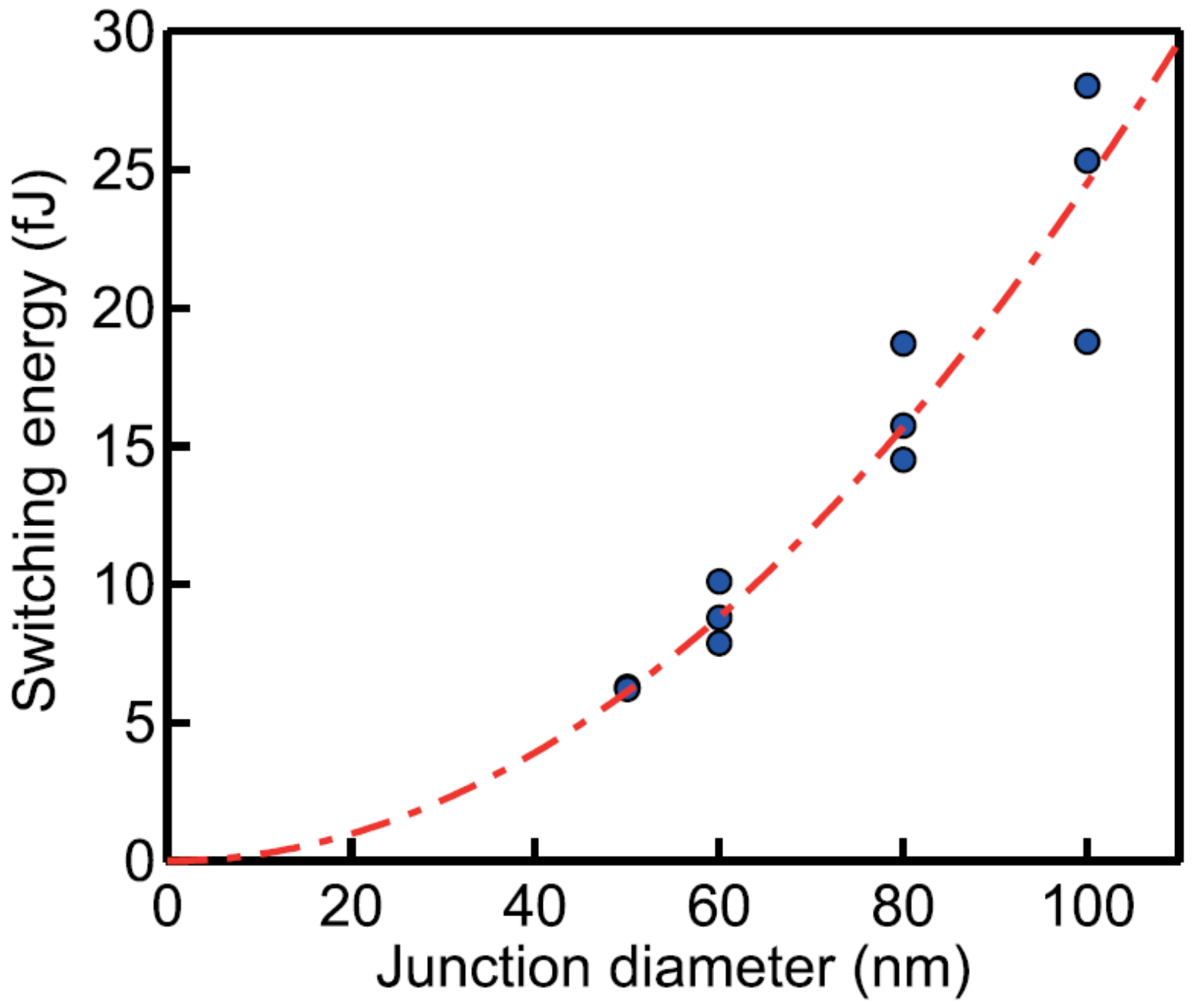

**Fig. 30**

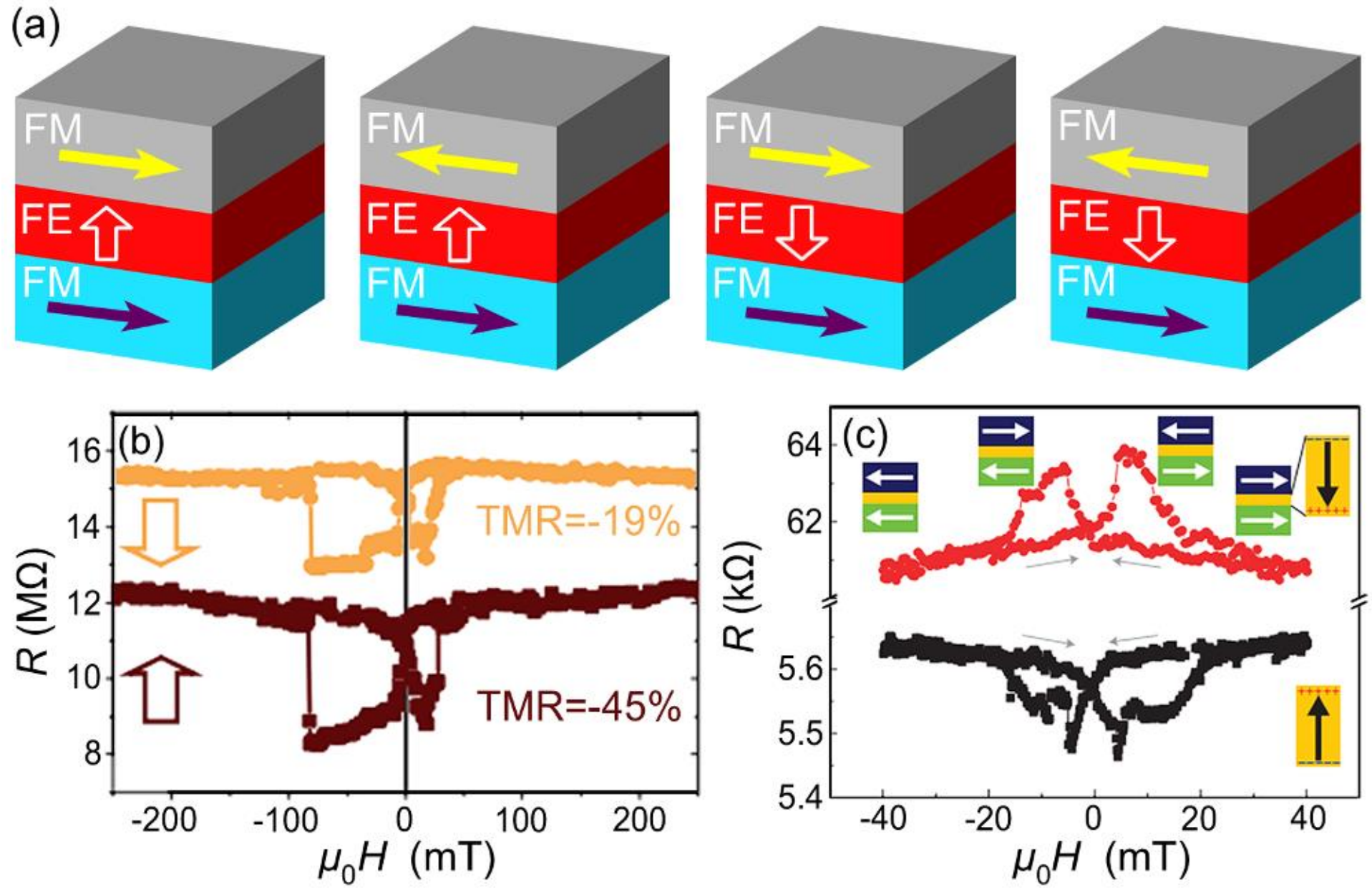

**Fig. 31**

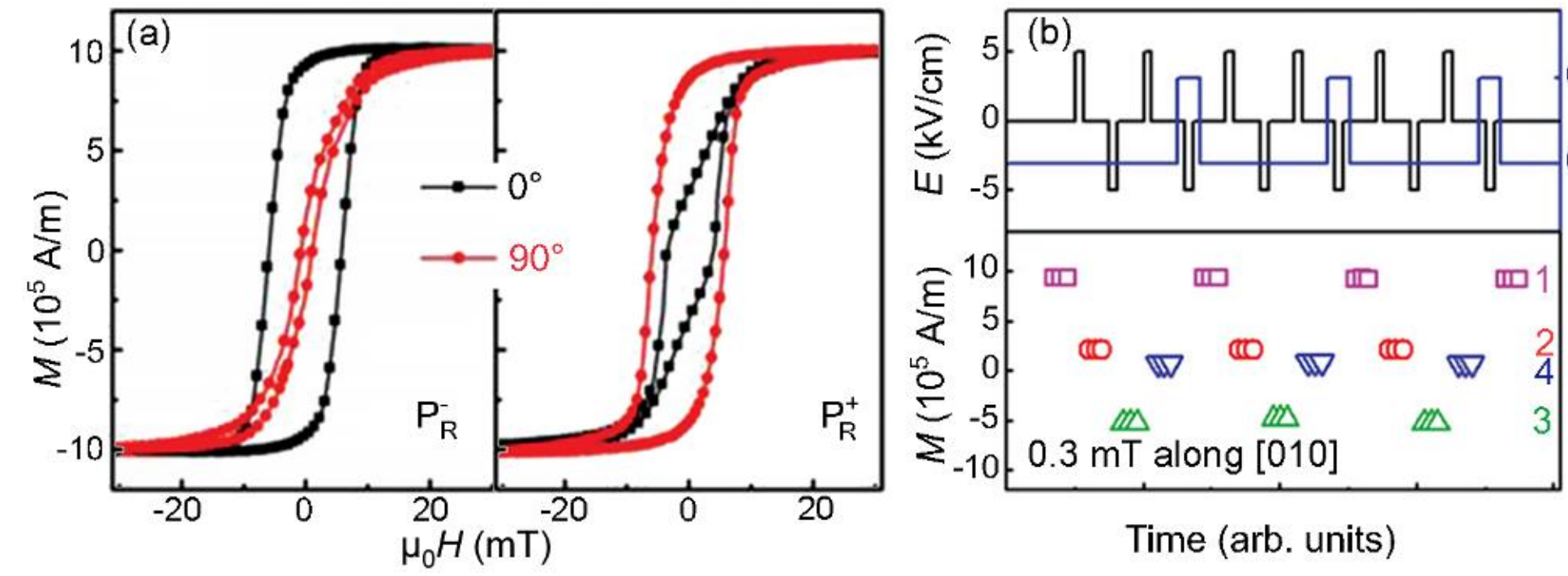

**Fig. 32**

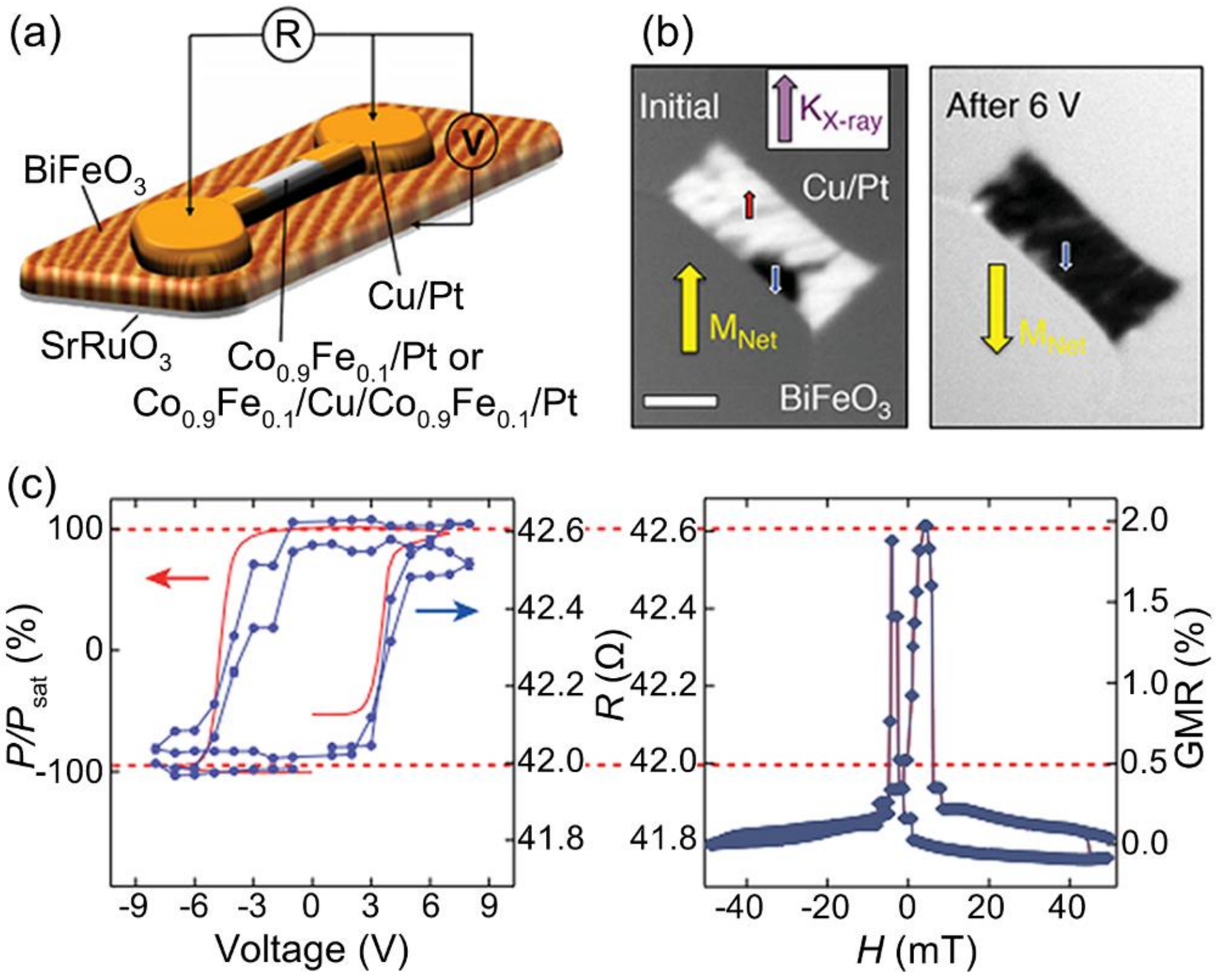

**Fig. 33**

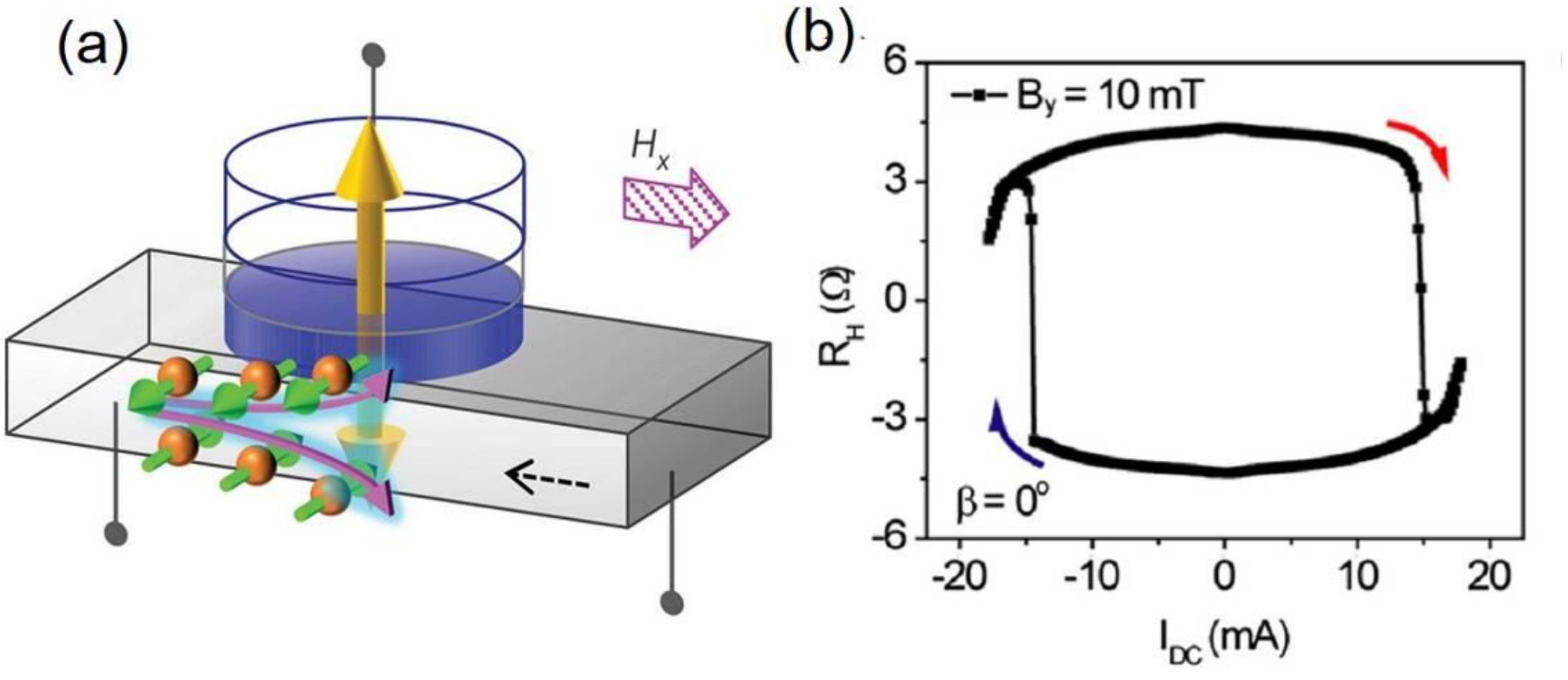

**Fig. 34**

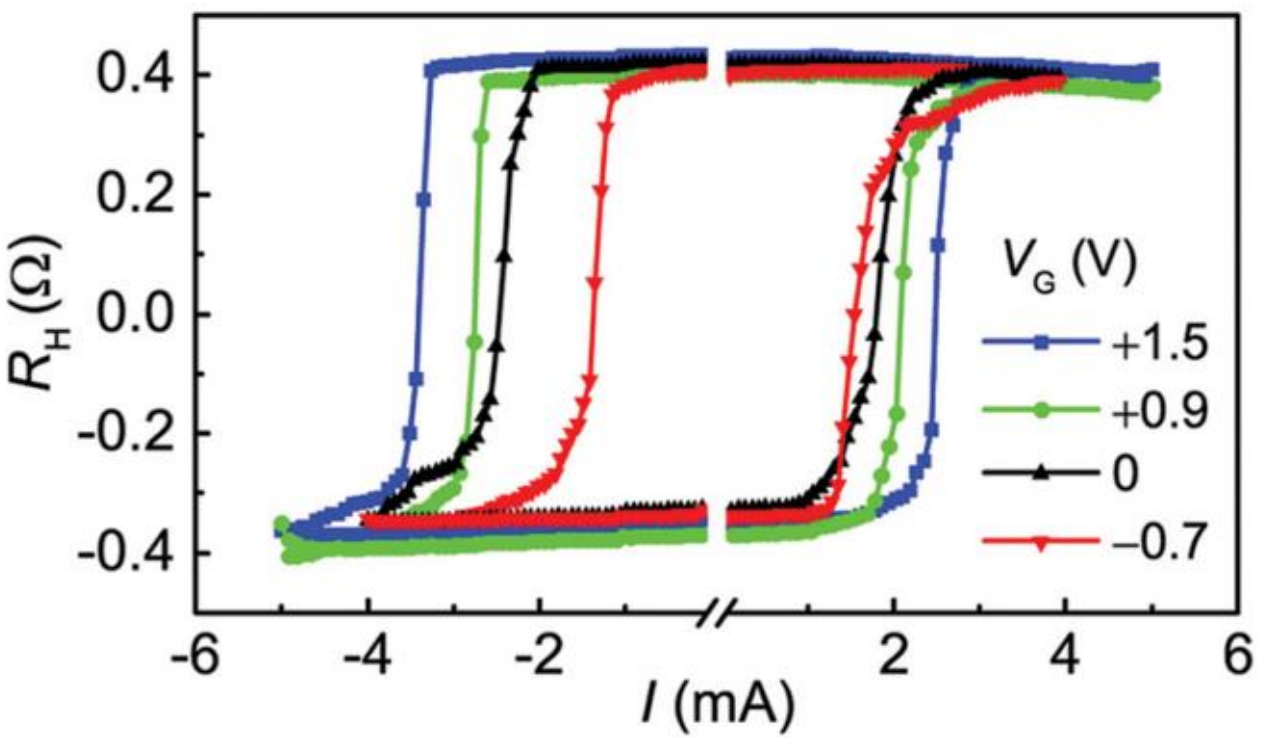

**Fig. 35**